\pdfoutput=1
\documentclass[10pt,twoside]{article}
\usepackage[a4paper,textwidth=130mm,textheight=205mm,centering,headsep=7mm]{geometry}
\usepackage[T1]{fontenc}
\usepackage{amsmath,amsthm}
\usepackage{newtxtext}
\usepackage{newtxmath}
\usepackage{mathrsfs}
\usepackage{graphicx}
\usepackage{booktabs}
\usepackage{enumitem}
\usepackage{tikz-cd}
\usepackage{titlesec}
\usepackage{fancyhdr}
\usepackage[font=small,labelfont=bf,labelsep=period,figurename=Fig.]{caption}
\usepackage{cite}
\usepackage{placeins}
\usepackage[hidelinks]{hyperref}

\titleformat{\section}{\normalfont\bfseries}{\thesection.}{0.5em}{}
\titlespacing*{\section}{0pt}{3.2ex plus 1ex minus .2ex}{1.6ex plus .2ex}
\titleformat{\subsection}[runin]{\normalfont\bfseries}{\thesubsection.}{0.5em}{}[.]
\titlespacing*{\subsection}{0pt}{2.2ex plus .8ex minus .2ex}{0.5em}
\titleformat{\paragraph}[runin]{\normalfont\itshape}{}{0pt}{}[.]
\titlespacing*{\paragraph}{0pt}{1.4ex plus .5ex minus .2ex}{0.5em}
\fancypagestyle{firstpage}{\fancyhf{}%
  \fancyfoot[C]{\small\thepage}}

\numberwithin{equation}{section}

\newtheoremstyle{cmpthm}{1.3ex plus .5ex minus .2ex}{1.3ex plus .5ex minus .2ex}{\itshape}{}{\bfseries}{.}{.5em}{}
\newtheoremstyle{cmpdef}{1.3ex plus .5ex minus .2ex}{1.3ex plus .5ex minus .2ex}{\normalfont}{}{\bfseries}{.}{.5em}{}
\theoremstyle{cmpthm}
\newtheorem{theorem}{Theorem}[section]
\newtheorem{proposition}[theorem]{Proposition}
\newtheorem{lemma}[theorem]{Lemma}
\newtheorem{remark}[theorem]{Remark}
\theoremstyle{cmpdef}
\newtheorem{definition}[theorem]{Definition}

\newcommand{\De}{\Delta}
\newcommand{\Df}{\Delta_\phi}
\newcommand{\cO}{\mathcal{O}}
\newcommand{\ve}{\varepsilon}
\newcommand{\dd}{\mathrm{d}}
\newcommand{\hz}{\hat z}
\newcommand{\hzb}{\hat{\bar z}}
\newcommand{\Plib}{\Phi_{\mathrm{lib}}}
\newcommand{\GFF}{\mathrm{(GFF)}}
\DeclareMathOperator{\cl}{cl}
\DeclareMathOperator{\inter}{int}

\begin{document}

\thispagestyle{firstpage}
\begin{flushleft}
\vspace*{2mm}
{\LARGE\bfseries Liberation in the Heavy Limit\par}
\vspace{5mm}
{\large  Sridip Pal$^{\dagger}$, Gordon Rogelberg$^{*}$}

\vspace{2mm}
{\small
$^{\dagger}$\,[Tata Institute of Fundamental Research Mumbai, sridip@theory.tifr.res.in]\\
$^{*}$\,[Department of Physics, Yale University, gordon.rogelberg@yale.edu]\par}

\end{flushleft}
\vspace{2mm}

\vspace{4mm}

\noindent{\small\textbf{Abstract:} We study CFT four-point functions of identical scalar primaries as their dimension $\De$ grows for fixed $0<z,
\bar{z}<1$, and call them \emph{liberated} if they grow at the same exponential rate as those of a generalized free field (GFF), as is the case for heavy free particles propagating along geodesics in AdS. We show that the correlator is liberated at the crossing symmetric point if and only if the light spectrum is sparse, and that liberation then holds wherever $z$ and $\bar z$ lie on the same side of $1/2$. If the low twist states are also sparse, liberation holds everywhere except in an explicit band around $z+\bar z=1$. In particular, if the primary twist gap lies above $\sqrt{2}\De$, then the correlator is liberated everywhere. For twist cutoffs below $\De$, composites of two spinning generalized free fields fail to be liberated in a region that fills the band near the double lightcone limit. In two dimensions, Virasoro primaries cannot be liberated unless the central charge grows faster than $\De$. The same framework sharpens the universality of large-$c$ torus partition functions of two-dimensional CFTs.\par}

\section{Introduction}
\label{sec:intro}

Let $\mathcal{G}(z,\bar z)$ be the reduced four-point function of identical Hermitian scalar primaries of dimension $\De$, normalized so that the identity contributes $1$. On the real domain $0<z,\bar z<1$, or causal diamond, positivity of the operator product expansion (OPE) and crossing imply
\begin{equation}\label{eq:intro-lower}
 \mathcal G\geq\max\left\{1,\left(\frac{u}{v}\right)^{\!\De}\right\},
 \qquad u=z\bar z,\quad v=(1-z)(1-\bar z).
\end{equation}
We ask when this elementary lower bound determines the leading logarithm of the correlator as the external dimension tends to infinity. More precisely, for a sequence of operators in a CFT or in a family of CFTs, under what spectral assumptions does
\begin{equation}\label{eq:introlib}
 \lim_{\De\to\infty}\frac{1}{\De}\log\mathcal G
 =\max\left(0,\log\frac{u}{v}\right)
\end{equation}
hold, and on what kinematic region?

The benchmark is the generalized free field (GFF), whose reduced correlator is
\begin{equation}\label{eq:introGFF}
 \mathcal G^\GFF=1+u^{\De}+\left(\frac{u}{v}\right)^{\!\De}.
\end{equation}
The term $u^{\De}$ never dominates on this domain. Thus \eqref{eq:introlib} is equivalent to $\log(\mathcal G/\mathcal G^\GFF)=o(\De)$, which we call \emph{liberation}. This allows a relative factor $e^{o(\De)}$, so it does not require the ratio of correlators to tend to one. The name recalls liberation at large spin, discovered in \cite{Komargodski:2012ek,Fitzpatrick:2012yx} and partly proven in \cite{Pal:2022vqc,vanRees:2024xkb}. There, the exchanged spin grows, as probed by the four-point function in the double lightcone limit. Here, instead, the external dimension grows at fixed cross ratios. In both cases, the identity contribution controls the limit of the correlator.

Our first result is a criterion at the crossing symmetric point $z=\bar z=1/2$.
The correlator is liberated at this point, i.e., $\log\mathcal G(1/2,1/2)=o(\De)$, if and only if the number of light exchanged states of dimension at most $E$, weighted by their squared OPE coefficients, grows no faster than $(3+2\sqrt2)^{E}$, up to factors $e^{o(\De)}$. This rate is the inverse of the factor $\rho^{E}=(3-2\sqrt2)^{E}$ with which such a state contributes to the correlator at $z=\bar z=1/2$. We call such a spectrum \emph{sparse}. Liberation at the self-dual point is further equivalent to liberation on both wedges $z,\bar{z}\leq 1/2$ and $z,\bar z\geq 1/2$. See Theorem~\ref{thm:weak-crossing} for the precise formulation.

At first sight, it may seem surprising that the self-dual point should ``know'' so much. The mechanism is elementary, though. Positivity makes $\mathcal G$ increase with $z$ and $\bar z$, so $1\leq\mathcal G(z,\bar z)\leq\mathcal G(1/2,1/2)=e^{o(\De)}$ whenever $z,\bar z\leq1/2$, which is liberation there. Crossing then maps this region to $z,\bar z\geq1/2$. The connection with sparseness also follows because every exchanged state contributes positively to $\mathcal G(1/2,1/2)$, so if the total is not exponentially large, neither is the contribution of any group of light states. Conversely, crossing fixes the average dimension exchanged at this point to $\sqrt2\De$ (in the $\rho$-expansion), in particular, sparseness up to just above $\sqrt2\De$ is enough to bound $\mathcal G(1/2,1/2)$.

Readers familiar with the extended Cardy formula in $2$D CFTs will immediately recognize the similarity of the above with vacuum dominance of the torus partition function $Z(\beta)$ up to subexponential terms for $\beta\neq 2\pi$ \cite{Hartman:2014oaa}
\begin{equation}\label{hks}
    \lim_{c\to\infty}\frac{1}{c}\left(\log Z(\beta)-\frac{c\beta}{12}\right)=\rm{max}\left(0, \frac{\pi^2}{3\beta}-\frac{\beta}{12}\right)
\end{equation}
under an appropriate sparseness condition on the low-lying spectrum. Using our framework, we prove the converse: sparseness is equivalent to $\log Z(2\pi)-\frac{\pi c}{6}=o(c)$, which establishes \eqref{hks} also at $\beta=2\pi$.

The self-dual criterion has a direct application to quantum field theory in rigid AdS. The boundary correlators of a gapped quantum field theory in AdS$_{d+1}$ obey all the conformal bootstrap axioms except the existence of a stress tensor. The flat-space limit takes the AdS radius to infinity in units of the bulk masses. The dimension of an operator dual to a massive particle then grows at fixed cross ratios, and its four-point function should reduce to the scattering amplitude of the particle. The reference \cite{vanRees:2022zmr} used this to derive a dispersion relation, unitarity, and the Froissart--Martin bound for the amplitude without the LSZ axioms. The amplitude is read off from the connected part of the correlator, which is exponentially smaller than the disconnected part. Extracting it therefore requires, at the very least, that the full correlator grow at the same exponential rate as the generalized free one, which is liberation. Their spectral assumption, that every exchanged primary has dimension above $\sqrt2\De$, is a special case of sparseness, so liberation comes for free.

Our second result extends liberation into the mixed wedges, $z\leq1/2\leq\bar z$ and its image, under twist sparseness.
If states of twist below $2\alpha\De$ are also sparse for some $0<\alpha\leq 1$, liberation holds everywhere except in an explicit region around the self-dual line $z+\bar z=1$, henceforth called \emph{undetermined region}. It shrinks as $\alpha$ grows and, in the radial expansion, disappears at $\alpha=1/\sqrt2$. So sparseness and a twist gap above $\sqrt2\De$ give liberation everywhere.

The result is derived using the iterative argument of \cite{Dey:2024nje} (used in the context of analyzing the torus partition function of $2$D CFT for independent left and right inverse temperatures), which we name \emph{the relay}. Starting from a mixed point, where one of $z,\bar z$ is below $1/2$ and the other above, each step bounds the correlator by a low-twist term plus its value at another mixed point, then we repeat. If the steps reach the region $z,\bar z\leq1/2$, where liberation is already known, in a finite number of relays, this proves liberation at the starting point. The undetermined region consists of the points from which they never get there. A quantity conserved \cite{Dey:2024nje} at each step gives its boundary in closed form. See Theorem~\ref{thm:strong-crossing} for details.

In the context of torus partition function of $2$D CFT, HKS \cite{Hartman:2014oaa} determined $\log Z$ when both lie on the same side of $2\pi$, and DPQ \cite{Dey:2024nje} extended this region by alternating modular invariance with a bound on the low-twist contribution. At a sufficiently high twist cutoff, their argument determines the leading free energy everywhere off $\beta\bar\beta=4\pi^2$. We add the boundary of their undetermined region and the self-dual hyperbola $\beta\bar\beta=4\pi^2$. See Theorem~\ref{thm:strong-torus} for details.

Our third result shows that liberation can indeed fail inside the undetermined region. We fix $0<\alpha<1/2$ and take the scalar $\cO=:\!J \cdot \!J\!:$, where $J$ is a generalized free field of dimension $\De/2$ and twist just above $\alpha\De$, so that its spin is $\ell\approx(\frac12-\alpha)\De$. This $\cO$ satisfies the hypotheses of our second result with twist cutoff $2\alpha\De$. At large $\De$, its correlator is dominated by the components of $J$ with spin $\pm\ell$ in the plane of the four operators. Both favor the s-channel pairing below the self-dual line and the t-channel pairing above it, but spin in the plane weighs $z$ and $\bar z$ differently and biases the two components toward different channels. On part of the mixed wedges, the bias overwhelms: one constituent of each operator pairs with its s-channel partner and the other with its t-channel partner, and this connected contraction exceeds the disconnected ones by a factor exponential in $\De$. We call this region \emph{mixed phase}, it lies strictly inside the undetermined region\footnote{As $z\to0$ and $\bar z\to1$ the mixed phase lies between the curves $1-\bar z=z^{(1-\alpha)/\alpha}$ and $1-\bar z=z^{\alpha/(1-\alpha)}$ at leading order in the logarithms. This curve has the same slope as the boundary curve of the undetermined region. As a result, the asymptotic opening angle of the undetermined region cannot be narrowed there.}. The same phase appears in local two-dimensional CFTs, built from composites of currents in symmetric product orbifolds of compact bosons.

Liberation has a bulk meaning. In a free massive bulk theory, the worldline representation has the schematic large-mass form \cite{Maxfield:2017rkn}
\begin{equation}\label{eq:introworldline}
 \langle\cO(x_1)\cO(x_2)\cO(x_3)\cO(x_4)\rangle
 \sim\int_{\partial\Gamma=\{x_1,x_2,x_3,x_4\}}[D\Gamma]\,
 e^{-mL[\Gamma,g]+o(m)},
\end{equation}
where $\Gamma$ is a pair of worldlines ending on the four points and $L$ is its regularized length. In empty AdS, with radius set to one and $m\simeq\De$, the integral localizes on geodesics, and since $e^{-\De L_{ij}}=|x_{ij}|^{-2\De}$, the three pairings give the three terms of \eqref{eq:introGFF}. Liberation reproduces the leading logarithm of this sum. It does not see interactions that change $\log\mathcal G$ by $o(\De)$, and it does not imply that $\cO$ is dual to a single particle, since composites of a large number of heavy particles can also be liberated.

One of the main obstructions to liberation is gravity. Every heavy operator in a holographic CFT interacts gravitationally.
For a compact body of mass $m\simeq\De$ in AdS described by a worldline, the body curves the geometry at the AdS scale by a relative amount of order $G_Nm\sim\De/C_T$, so the action of the other body's worldline changes by order $\De^2/C_T$. This is $o(\De)$ when $\De\ll C_T$,
and the correlator is liberated. When $\De/C_T$ tends to a nonzero constant, gravity contributes at order $\De$, and liberation fails. In two dimensions this follows for Virasoro primaries from Virasoro symmetry alone (due to exchange of stress-energy tensor and its composites), for any limiting $\De/c>0$.

We provide examples to clarify how much information this leading limit discards. For $\cO=:\!\phi^L\!:$ with $\De=L\Df$, fixed $L$ and $\Df\to\infty$ gives liberation with a growing gap, while fixed $\Df$ and $L\to\infty$ fails the self-dual criterion. Taking both large again gives liberation, while the weighted light spectrum saturates the sparseness bound, as is the case for symmetric product orbifold CFTs on the torus \cite{Hartman:2014oaa}. The two liberated cases differ only near the self-dual line. There, in terms of $x=\log\frac{z}{1-z}$ and $\bar x=\log\frac{\bar z}{1-\bar z}$, $\log\mathcal G$ switches from the s-channel value to the t-channel value over a window of width $1/\Df$ in $x+\bar x$. That is of order $1/\De$ for fixed $L$, as for a single heavy particle, but is parametrically wider when $L$ and $\Df$ both grow. The $\gamma$-rescaled potential of Sect.~\ref{ex:gff}, which zooms into a window of width $\De^{-\gamma}$ around the self-dual point, detects this difference.

\subsection{Comments on technique}
To prove these results, we write both the correlator and the partition function as sums of positive Boltzmann weights, with the identity contributing exactly $1$: the correlator as non-negative coefficients times $\rho^h\bar\rho^{\bar h}$ or $z^h\bar z^{\bar h}$, and the partition function as multiplicities times $e^{-\beta h-\bar\beta\bar h}$. In suitable variables $x,\bar x$, defined through a function $\beta(x)$ that plays the role of an inverse temperature, crossing and modular invariance both become the reflection $(x,\bar x)\to(-x,-\bar x)$, up to a factor $e^{(x+\bar x)/\hbar}$ with $\hbar=1/\De$ or $1/c$. We call these data a \emph{bootstrap triple} (Sect.~\ref{sec:triples}). Similar positive representations were used in \cite{Poland:2026uiy,Kim:2015oca} to relate modular invariance and crossing, and spectral moments of the heavy OPE were studied in \cite{Poland:2024hvb,Chiang:2026nmd}. In all of these, positivity and crossing bound the slope of $\hbar\log[\mathcal G(z,\bar z)/\mathcal G(1/2,1/2)]$ between $0$ and $1$ \cite[Lemma~B.4]{Poland:2026uiy}, so a pointwise limit, where it exists, is uniform on compact sets. Assuming that the free energy density vanishes at the self-dual point, \cite[Sect.~3.1.2, Prop.~3.4]{Poland:2026uiy} obtained liberation on the diagonal and a lower bound away from it. We show that this assumption is equivalent to sparseness, the condition used on the torus in \cite{Hartman:2014oaa,Dey:2024nje}, and that it implies liberation on both wedges $z,\bar z\leq1/2$ and $z,\bar z\geq1/2$. Also new are the finite-$\De$ form of this criterion (Proposition~\ref{prop:selfdual-estimate}), the extension of the relay to correlators, a non-liberated phase from spinning composites inside the undetermined region, and the Virasoro obstruction. On the torus, we add the boundary of the undetermined region and the self-dual hyperbola, which follow from the closure argument of Lemma~\ref{lem:lipschitz}. In short, our methods combine and sharpen those of \cite{Poland:2026uiy,Hartman:2014oaa,Dey:2024nje}.

Here is the organization of the rest of the paper. Sect.~\ref{sec:results} states the results for the four-point function and the torus. Sect.~\ref{sec:examples} develops the examples. Sects.~\ref{sec:triples}--\ref{sec:strong} give the common framework and proofs. Remark~\ref{sec:DPQ} compares the relay algorithm arguments, and Sect.~\ref{sec:disc} identifies the remaining open questions and future directions. Proofs for the example analysis are in Appendix~\ref{app:examples}.
\section{Statement of Results}
\label{sec:results}

We state the two theorems for scalar four-point functions at large external dimension and torus partition functions at large central charge. The general statements and proofs are Theorems~\ref{weaklibthm} and \ref{thm2} in Sects.~\ref{sec:triples}--\ref{sec:strong}.

\subsection{The four-point function}
\label{sec:fourpoint}
We consider a sequence of Hermitian scalar primaries $\cO$, in a unitary CFT or in a family of unitary CFTs, whose dimensions $\De$ increase without bound. For each $\cO$, let us consider the reduced four-point function
\begin{equation}
\mathcal{G}(u,v) = \frac{\langle \cO(x_1) \cO(x_2) \cO(x_3) \cO(x_4)\rangle }{\langle \cO(x_1) \cO(x_2)\rangle \langle \cO(x_3) \cO(x_4)\rangle},\qquad
u = \frac{x_{12}^2 x_{34}^2}{x_{14}^2 x_{23}^2},\quad v = \frac{x_{13}^2 x_{24}^2}{x_{14}^2 x_{23}^2},
\end{equation}
and write $u = z \bar{z}$ and $v=(1-z)(1-\bar{z})$.\footnote{This labeling of the insertion points differs from the more common one, in which $u=x_{12}^2x_{34}^2/x_{13}^2x_{24}^2$, by the exchange $x_3\leftrightarrow x_4$. In our labeling, the crossed channel of \eqref{crossingrelation} is $(13)(24)$, and $u^\De$ in \eqref{eq:introGFF} is the contribution of the pairing $(14)(23)$.} We work throughout in the Lorentzian causal diamond $0<z,\bar{z} <1$. Exchanging $x_2 \leftrightarrow x_3$ gives the crossing relation
\begin{equation}
\label{crossingrelation}
\mathcal{G}(z,\bar{z}) = \Big( \frac{z}{1-z}\, \frac{\bar{z}}{1-\bar{z}}\Big)^{\!\De} \mathcal{G}(1-z,1-\bar{z}).
\end{equation}
Each correlator in the sequence obeys \eqref{crossingrelation} with its own $\De$, so $\De\to\infty$ is a limit along the sequence.

The other ingredient is positivity. The reduced correlator has two expansions with non-negative coefficients, in monomials of $z$ and of $\rho=\frac{1-\sqrt{1-z}}{1+\sqrt{1-z}}$:
\begin{equation}
\label{zexp}
\mathcal{G} = \sum_{h,\bar{h}} a_{h,\bar{h}}^{(z)}\, z^h \bar{z}^{\bar{h}},\qquad a_{h,\bar{h}}^{(z)}\geq0,
\end{equation}
and
\begin{equation}
\label{rhoexp}
\mathcal{G} = \sum_{h,\bar{h}} a_{h,\bar{h}}^{(\rho)}\, \rho^h \bar{\rho}^{\bar{h}},\qquad a_{h,\bar{h}}^{(\rho)}\geq0.
\end{equation}

\subsection{Liberation of the four-point function}
\label{sec:liberation4pt}

Since the two expansions obey the same crossing relation, we can state each theorem once for both. For each correlator in the sequence, define the cumulative spectral counts
\begin{equation}
\label{eq:sc}
\mathcal{N}^{(\rho)}(E) = \sum_{h+\bar{h} \leq E} a_{h,\bar{h}}^{(\rho)},\qquad \mathcal{N}^{(z)}(E) =\sum_{h+\bar{h}\leq E}  a_{h,\bar{h}}^{(z)},
\end{equation}
which count, with their weights, the monomials of total weight at most $E$.

\begin{theorem}[Self-dual criterion for four-point functions]
\label{thm:weak-crossing}
Let $\{(\mathcal{G},\De)\}$ be a sequence, with $\De\to\infty$, of reduced four-point functions obeying \eqref{crossingrelation}, each with a non-negative $\rho$- or $z$-expansion.  Define $x(z)=\log\frac{z}{1-z}$. The following statements are equivalent.
\begin{enumerate}[label=(\arabic*)]
    \item \emph{(Global bound)} At every $(z,\bar z)\in(0,1)^2$,
    \begin{equation}
    \begin{aligned}
    \max\Big(0,x(z)+x(\bar{z})\Big)&\leq\liminf_{\De\to\infty}\frac{1}{\De}\log\mathcal{G}(z,\bar z)\\
    &\leq\limsup_{\De\to\infty}\frac{1}{\De}\log\mathcal{G}(z,\bar z)\leq\max\Big(0,x(z) \Big)+\max\Big(0,x(\bar z)\Big).
    \end{aligned}
    \end{equation}
    \item \emph{(Free diagonal)} For $z=\bar{z}$, the limit exists and is
    \begin{equation}
    \lim_{\De\to\infty}\frac{1}{\De}\log\mathcal{G}(z,z) = \max\Big(0,2 x(z)\Big).
    \end{equation}
    \item \emph{(Sparseness)} For some fixed $\epsilon>0$, the cumulative spectral count of the non-negative expansion, defined in \eqref{eq:sc}, obeys
    \begin{equation}
    \begin{aligned}
    \mathcal N^{(\rho)}(E)&\lesssim(3+2\sqrt2)^E,\qquad &&0\leq E\leq\sqrt2\,\De+\epsilon,\\
    \text{or}\qquad\mathcal N^{(z)}(E)&\lesssim2^E, &&0\leq E\leq2\De+\epsilon,
    \end{aligned}
    \end{equation}
    where $\lesssim$ allows a factor $e^{o(\De)}$, uniform below the cutoff.
    \item \emph{(Self-dual limit)} At the crossing symmetric point, $\displaystyle\lim_{\De\to\infty}\frac{1}{\De}\log\mathcal{G}\big(\tfrac12,\tfrac12\big) = 0$.
\end{enumerate}
\end{theorem}

Only statement 3 refers to the expansion; statements 1, 2, and 4 are conditions on the correlator alone. In a unitary CFT both expansions are non-negative, so the theorem applies to each of them, and the two sparseness conditions are therefore equivalent to each other, although the two counts weigh the exchanged spectrum differently and are cut off at different weights.
This is the analog for correlators of the sparseness condition of HKS \cite{Hartman:2014oaa}.

\begin{figure}[t]
    \centering
    \includegraphics[width=\linewidth]{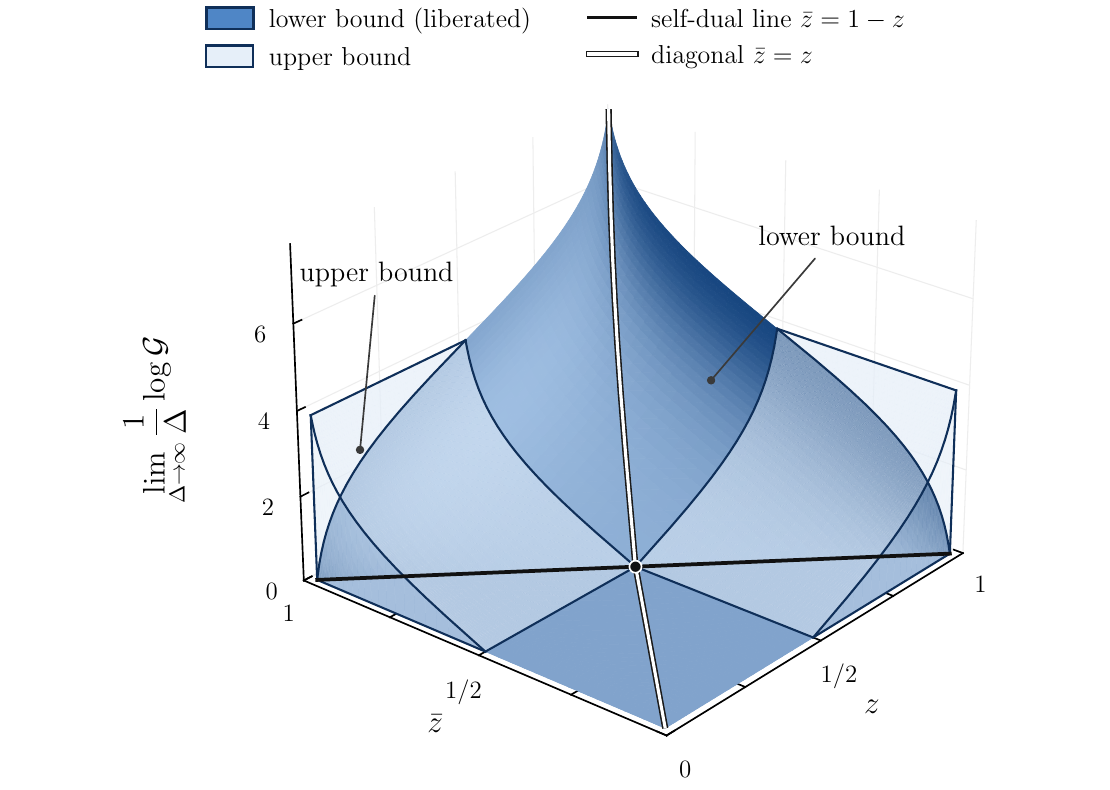}
    \caption{The upper (light, transparent) and lower (dark, solid) bounds of Theorem~\ref{thm:weak-crossing} on the large $\De$ limit of any convergent subsequence satisfying the hypotheses. On the cold wedge $z,\bar{z}  \leq 1/2$ and on the hot wedge $z ,\bar z \geq \frac12$ the bounds coincide, and the limit is unique. On the mixed wedges $z \leq 1/2 \leq \bar{z}$ and $\bar z\leq 1/2 \leq z$, they do not}
    \label{fig:weaklibboundscrossing}
\end{figure}

In practice, the most useful of the four statements is the last. To establish liberation, it suffices to know how fast the correlator grows at the single point $z=\bar z=1/2$. Statement 1 then determines the heavy limit wherever $z$ and $\bar z$ lie on the same side of $1/2$, since its lower and upper bounds coincide there (Fig.~\ref{fig:weaklibboundscrossing}). The limit is $0$, the contribution of the identity, when both are below $1/2$, and $\log\frac{z\bar z}{(1-z)(1-\bar z)}$, the contribution of its crossing image, when both are above. In the mixed region the two bounds differ, and the theorem does not determine the limit. The upper bound there is saturated by the spinning composites of Sect.~\ref{ex:chiral} with any fixed $\tau_J \geq d-2$ and $\De \to \infty$.

The spin plays a crucial role in the mixed region. When $\bar z$ is above $1/2$, the factor $\bar z^{\bar h}$ suppresses an exchanged state weakly, so a state with small $h$ and large $\bar h$,\footnote{The four operators lie in a plane, which is preserved by a subgroup $\mathrm{SL}(2,\mathbf{R})\times\mathrm{SL}(2,\mathbf{R})$ of the conformal group; we call the weights $h,\bar h$ under it \emph{planar weights}.} that is, with large spin and small twist, can compete with the identity even when the spectrum is sparse. To extend liberation into the mixed region we must therefore also control the states of small twist. For each correlator in the sequence and each $\alpha>0$, let $\mathcal{G}^{(\alpha)}_0$ be the part of the expansion \eqref{rhoexp} or \eqref{zexp} with $\min(h,\bar{h})\leq\alpha\De$. We say that the sequence is \emph{$\alpha$-twist sparse} in that expansion if this part does not grow exponentially anywhere below the self-dual line,
\begin{equation}
\label{eq:crossing-twist-sparse}
\lim_{\De\to\infty}\frac{1}{\De}\log\mathcal{G}^{(\alpha)}_0(z,\bar{z}) = 0\qquad\text{for all}\quad z+\bar{z}<1.
\end{equation}
A global primary of twist $\tau$ contributes only monomials with $\min(h,\bar{h})\geq\tau/2$, in either expansion, so this can be achieved by imposing a condition on the primaries of twist below $2\alpha\De$. In particular, a twist gap $\tau>2\alpha\De$ above the identity suffices, and $\alpha=1$ corresponds to the double-twist threshold $\tau = 2\De$ of the GFF. A local CFT has no such gap, since the stress tensor and its first composites have twist of order one; there, twist sparseness asks that their total contribution stay subexponential.

\begin{theorem}[Twist-sparse extension for four-point functions]
\label{thm:strong-crossing}
Let $\{(\mathcal{G},\De)\}$ be a sequence, with $\De\to\infty$, of reduced four-point functions obeying \eqref{crossingrelation} and the equivalent conditions of Theorem~\ref{thm:weak-crossing}. Suppose that the sequence is $\alpha$-twist sparse in the $\rho$-expansion for some $\alpha\in[0,\alpha_*^{(\rho)}]$, or in the $z$-expansion for some $\alpha\in[0,\alpha_*^{(z)}]$, where $\alpha_*^{(\rho)} = 1/\sqrt{2}$ and $\alpha_*^{(z)} = 1$. Then
\begin{equation}
  \lim_{\De\to\infty}\frac{1}{\De}\log\mathcal{G}(z,\bar{z}) = \max\Big(0,\log\frac{z\bar z}{(1-z)(1-\bar{z})}\Big)
  \label{eq:crossing-strong}
\end{equation}
for all $(z,\bar{z})\in(0,1)^2$ outside the corresponding undetermined region, $\mathcal{E}^{(\rho)}_\alpha$ or $\mathcal{E}^{(z)}_\alpha$. With $\hz=\min(z,1-z)$ and $\hzb=\min(\bar{z},1-\bar{z})$, these are
\begin{align}
  \mathcal{E}^{(\rho)}_\alpha &= \bigg\{(z,\bar{z})\ \bigg|\ (1-2z)(1-2\bar{z})<0\ \text{ and }\ \max\Big(\frac{\hz}{1-\hz},\frac{\hzb}{1-\hzb}\Big)<\Big(\big(3+2\sqrt{2}\big)^2\rho(\hz)\,\rho(\hzb)\Big)^{\alpha}\bigg\},
  \label{eq:crossing-enigmatic-rho}\\
  \mathcal{E}^{(z)}_\alpha &= \bigg\{(z,\bar{z})\ \bigg|\ (1-2z)(1-2\bar{z})<0\ \text{ and }\ \max\Big(\frac{\hz}{1-\hz},\frac{\hzb}{1-\hzb}\Big)<\big(4\hz\hzb\big)^{\alpha}\bigg\},
  \label{eq:crossing-enigmatic-z}
\end{align}
\end{theorem}

The undetermined region lies in the mixed wedges and is invariant under crossing and under $z\leftrightarrow\bar{z}$. It is a band around the self-dual line, and the liberated answer \eqref{eq:crossing-strong} holds on its closed complement, which contains the boundary of the band and the self-dual point.

\begin{figure}[t]
    \centering
    \includegraphics[width=\linewidth]{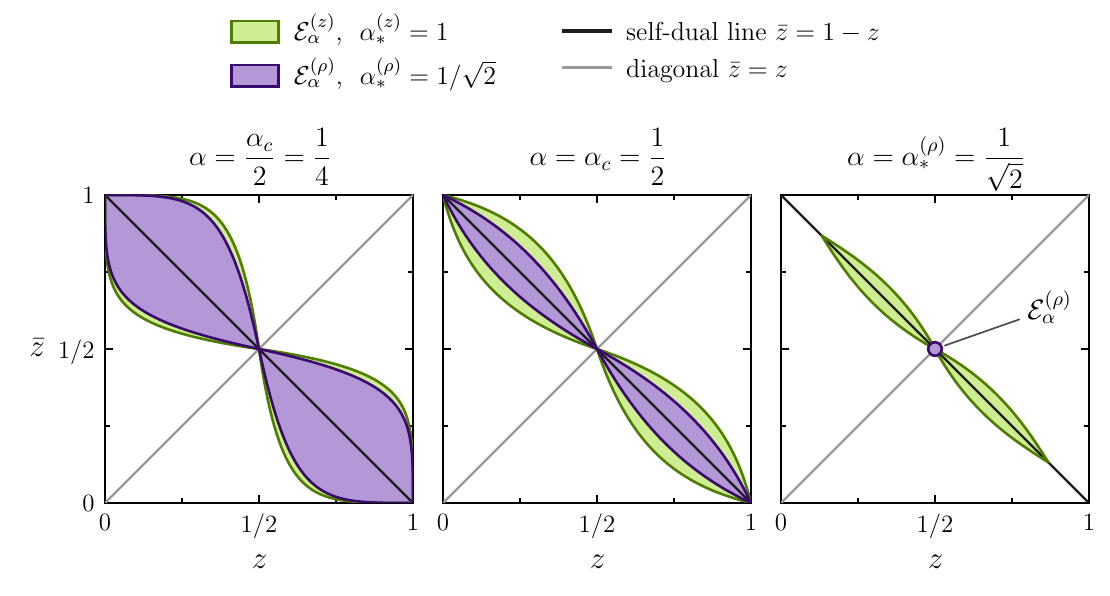}
    \caption{Undetermined regions $\mathcal{E}_\alpha^{(z)}$ (light) and $\mathcal{E}_\alpha^{(\rho)}$ (dark) of four-point functions in the causal diamond, for $\alpha=1/4$, $1/2$, and $\alpha_*^{(\rho)}=1/\sqrt{2}$, assuming twist sparseness in the $z$- or the $\rho$-expansion. At $\alpha = \alpha_*^{(\rho)}$, the region $\mathcal{E}_\alpha^{(\rho)}$ has disappeared, while a sliver of $\mathcal{E}_\alpha^{(z)}$ remains}
    \label{fig:crossenigmatic}
\end{figure}

The undetermined region shrinks as the twist cutoff $\alpha$ grows, and in both expansions its shape changes at $\alpha=1/2$. At $\alpha=0$, it is the whole mixed region, and Theorem~\ref{thm:strong-crossing} reduces to statement 1 of Theorem~\ref{thm:weak-crossing}. For $0<\alpha\leq\frac12$, the band covers the entire self-dual line except the self-dual point, and it extends all the way to the corners $z\to0$, $\bar z\to1$ and $z\to1$, $\bar z\to0$, where both cross ratios approach a lightcone. For $\frac12<\alpha<\alpha_*$, the band stays a finite distance away from these corners, so liberation holds at every fixed point of the mixed region sufficiently close to the lightcones.
At $\alpha=\alpha_*$, the undetermined region is empty, and the liberated answer holds on the whole causal diamond (Fig.~\ref{fig:crossenigmatic}).

An astute reader may notice that unlike Theorem~\ref{thm:weak-crossing}, Theorem~\ref{thm:strong-crossing} depends on the choice of expansion, in both its hypothesis and its conclusion. Sparseness in Theorem~\ref{thm:weak-crossing} is equivalent to a condition on the full correlator at a single point, which does not care how the correlator is expanded. Twist sparseness instead constrains a part of the correlator, the monomials with $\min(h,\bar h)\le\alpha\De$, and which monomials make up this part depends on the expansion, because the two expansions distribute the descendants differently. Each $\rho$-monomial expands into $z$-monomials with non-negative coefficients and larger exponents, so $\mathcal G^{(\alpha,z)}_0\le\mathcal G^{(\alpha,\rho)}_0$. Twist sparseness in the $\rho$-expansion therefore implies twist sparseness in the $z$-expansion at the same $\alpha$, but not conversely. For primaries the two agree, because a primary of twist $\tau$ contributes only monomials with $\min(h,\bar h)\ge\tau/2$ in either expansion, so a twist gap gives both hypotheses with the same $\alpha$.

The conclusions differ as well. The undetermined region is set by comparing the crossed-channel identity with the expansion variable. Since $\log\rho$ varies faster than $\log z$ away from the self-dual point, a given twist cutoff buys more suppression in the $\rho$-expansion. Hence $\mathcal{E}^{(\rho)}_\alpha\subseteq\mathcal{E}^{(z)}_\alpha$, and the $\rho$-region already closes at $\alpha_*^{(\rho)}=1/\sqrt2$.
In particular, a sequence that obeys the conditions of Theorem~\ref{thm:weak-crossing}, and whose non-identity primaries all have twist $\tau>\sqrt{2}\De$, is liberated on the whole causal diamond; with the $z$-expansion alone, the same conclusion would require a gap up to the double-twist threshold $\tau=2\De$. The liberated answer itself is fixed by the identity and crossing, and it is the same for both, so a sequence that is twist sparse in both expansions is liberated outside the intersection of the two undetermined regions.

\begin{remark}
\label{sec:blocks}
Our theorems use the full positive monomial expansion, including descendants. There is nevertheless a useful sufficient condition stated in terms of primary exchanges. Let
\begin{equation}\label{eq:primary-twist-sum}
 \mathcal P_\alpha(z,\bar z)=1+\sum_{\substack{\cO'\ne\mathbf 1\\ \tau_{\cO'}\leq2\alpha\De}}
 a_{\cO'}G_{\cO'}(z,\bar z),
 \qquad \tau_{\cO'}=\De_{\cO'}-\ell_{\cO'} .
\end{equation}
Here the sum is over global conformal primaries and retains their entire conformal blocks. Every descendant of a primary of twist $\tau$ has planar weights with $\min(h,\bar h)\geq\tau/2$. Positivity therefore gives, in either expansion,
\begin{equation}\label{eq:primary-sufficient}
 1\leq\mathcal G^{(\alpha)}_0(z,\bar z)\leq\mathcal P_\alpha(z,\bar z).
\end{equation}
Thus $\De^{-1}\log\mathcal P_\alpha\to0$ throughout $z+\bar z<1$ suffices for $\alpha$-twist sparseness. A gap above $2\alpha\De$ is the special case $\mathcal P_\alpha=1$.

\end{remark}

\subsection{The torus partition function}
\label{sec:torus}
 We consider a sequence of unitary two-dimensional CFTs with $c_L=c_R=c$, a unique normalizable vacuum, and finite torus partition functions at positive left and right inverse temperatures, whose central charges increase without bound. The partition function of each theory on a torus is
\begin{equation}
Z(\tau)  = \mathrm{Tr}_\mathcal{H}\, q^{L_0 - \frac{c}{24} } \bar{q}^{\bar{L}_0 - \frac{c}{24} },\qquad q = e^{2\pi i \tau},
\end{equation}
where $L_0$ and $\bar L_0$ are the zero modes of the two Virasoro algebras. We will let $\beta=-2\pi i\tau$ and $\bar\beta=2\pi i\bar\tau$ vary independently over $(0,\infty)$; they are the inverse temperatures of the left- and right-movers. If $n_{h,\bar h}$ is the degeneracy of states with $L_0=h$ and $\bar L_0=\bar h$, then
\begin{equation}
\label{partfuncdecomp}
Z(\beta,\bar{\beta} ) = \sum_{h,\bar{h}} n_{h,\bar{h}}\, e^{-\beta(h - \frac{c}{24} )-\bar{\beta}(\bar{h} - \frac{c}{24} )} =e^{(\beta +\bar{\beta} ) \frac{c}{24}}\, \widehat{Z}(\beta,\bar{\beta}),
\end{equation}
where the reduced partition function $\widehat{Z}$ is normalized so that the vacuum, with $h=\bar h=0$, contributes $1$. Exchanging the two cycles of the torus, $\tau\to-1/\tau$, gives the modular relation
\begin{equation}
\label{modularrelation}
\widehat{Z}(\beta,\bar{\beta} ) =e^{\frac{c}{24} \left(\frac{4\pi^2}{\beta} + \frac{4\pi^2}{\bar{\beta}} - \beta - \bar{\beta} \right)}\, \widehat{Z}\Big(\frac{4\pi^2}{\beta}, \frac{4\pi^2}{\bar{\beta}}\Big).
\end{equation}
Each member of the sequence obeys \eqref{modularrelation} with its own value of $c$, and $c\to\infty$ is always taken along this sequence. The light spectrum is measured by the cumulative spectral count
\begin{equation}
\label{microcanonicalcountmod}
\mathcal{N}(E) = \sum_{h+\bar{h} \leq E} n_{h,\bar{h}} ,
\end{equation}
the number of states of total weight at most $E$.

\begin{theorem}[Self-dual criterion for torus partition functions]
\label{weaklibtorus}
Let $\{(\widehat{Z},c) \}$ be a sequence of reduced partition functions with $c\to\infty$ obeying \eqref{modularrelation}, with cumulative spectral counts $\{\mathcal{N}\}$. The following statements are equivalent.
\begin{enumerate}[label=(\arabic*)]
    \item \emph{(Global bound)} At every $(\beta,\bar\beta)\in(0,\infty)^2$,
    \begin{equation}
    \begin{aligned}
\max\Big(0 , \frac{4\pi^2}{\beta} +\frac{4\pi^2}{\bar{\beta}}- \beta - \bar{\beta}  \Big) &\leq \liminf_{c\to\infty}\frac{24}{c}\log\widehat Z(\beta,\bar\beta)\\
&\leq \limsup_{c\to\infty}\frac{24}{c}\log\widehat Z(\beta,\bar\beta)\\
&\leq \max\Big(0 , \frac{4\pi^2}{\beta} - \beta \Big)+\max\Big(0 , \frac{4\pi^2}{\bar{\beta}} - \bar{\beta} \Big).
    \end{aligned}
    \end{equation}
    \item \emph{(Free diagonal)} For $\beta = \bar{\beta}$, the limit exists and is
    \begin{equation}
\lim_{c \to \infty} \frac{24}{c} \log \widehat{Z}(\beta,\beta) = \max\Big(0, 2\Big( \frac{4\pi^2}{\beta} - \beta\Big) \Big).
    \end{equation}
    \item \emph{(Sparseness)} For some fixed $\epsilon>0$, $\mathcal N(E)\lesssim e^{2\pi E}$ uniformly for $0\leq E\leq c/12+\epsilon$ as $c\to\infty$.
    \item \emph{(Self-dual limit)} At $\beta = \bar{\beta} = 2 \pi$, $\displaystyle\lim_{c \to \infty} \frac{24}{c} \log\widehat{Z}(2 \pi,2\pi)= 0$.
\end{enumerate}
\end{theorem}

Here $\lesssim$ allows a factor $e^{o(c)}$, uniform below the cutoff. The bound $e^{2\pi E}$ is the inverse of the Boltzmann factor $e^{-2\pi E}$ of a state of total weight $E$ at the self-dual temperature $\beta=\bar\beta=2\pi$, where the mean total weight is $c/12$. In particular, sparseness says that the states below weight $E$ together contribute at most $e^{o(c)}$ to $\widehat Z(2\pi,2\pi)$. HKS \cite{Hartman:2014oaa} established the implication from sparseness to pointwise liberation on the diagonal away from $\beta=2\pi$. Here we include the self-dual point and prove the converse, leading to the above equivalence statement.

The limit in statement 2 is uniform on compact subsets of $(0,\infty)$, including neighborhoods of $\beta=2\pi$, where it has a kink, because modular invariance bounds the derivative of $\frac{24}{c}\log\widehat Z$ uniformly in $c$ (Lemma~\ref{lem:lipschitz}).

\begin{figure}[t]
    \centering
    \includegraphics[width=\linewidth]{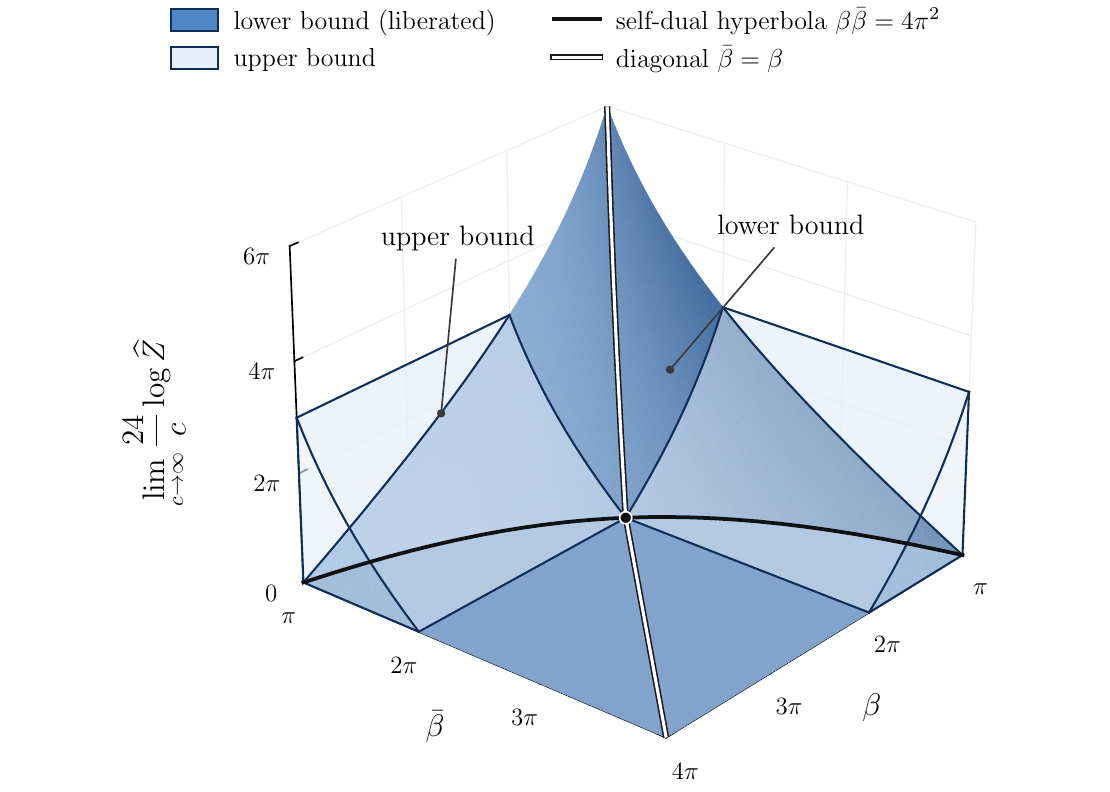}
    \caption{The upper (light, transparent) and lower (dark, solid) bounds of Theorem~\ref{weaklibtorus} on the large $c$ limit of any convergent subsequence. On the cold wedge $\beta ,\bar \beta \geq 2\pi$ and on the hot wedge $\beta ,\bar \beta \leq 2\pi$ the two bounds coincide, and the limit is unique. On the mixed wedges $\beta \leq 2\pi \leq \bar{\beta}$ and $\bar\beta \leq 2\pi \leq \beta$, they do not. To extend the lower bound, which is the liberated answer, into the mixed wedges, one needs an assumption on the twist spectrum}
    \label{fig:weaklibboundstorus}
\end{figure}

Statement 1 also determines the free energy density beyond the diagonal, on the cold wedge $\beta,\bar\beta\geq2\pi$ and on the hot wedge $\beta,\bar\beta\leq2\pi$, where
\begin{equation}
\label{eq:torus-wedges}
\lim_{c\to \infty} \frac{24}{c} \log \widehat{Z}(\beta,\bar{\beta}) = \max\Big(0 , \frac{4\pi^2}{\beta} +\frac{4\pi^2}{\bar{\beta}}- \beta - \bar{\beta}    \Big).
\end{equation}
On the mixed wedges, where one of $\beta$ and $\bar\beta$ is above $2\pi$ and the other below, it only gives a bound (Fig.~\ref{fig:weaklibboundstorus}).

In the mixed wedges, the total weight $h+\bar h$ is no longer the right quantity to control. A state of large spin can be heavy and still dominate, if its larger weight is paired with the higher temperature, where the Boltzmann suppression is weak, and its smaller weight with the lower one. What must be controlled is the twist $2\min(h,\bar h)$. Following \cite{Dey:2024nje}, for each $\alpha>0$, let
\begin{equation}
\label{eq:torus-partial}
  \widehat Z^{(\alpha)}_0(\beta,\bar\beta)
  =\sum_{\min(h,\bar h)\leq\alpha c/24} n_{h,\bar h}\,e^{-\beta h-\bar\beta\bar h}
\end{equation}
be the contribution of the states whose twist lies below a cutoff proportional to $c$. We say that the sequence is \emph{$\alpha$-twist sparse} if these states do not contribute to the free energy density on the side of the self-dual hyperbola $\beta\bar\beta=4\pi^2$ where the vacuum of the direct channel dominates,
\begin{equation}
  \lim_{c\to\infty}\frac{24}{c}\log\widehat Z^{(\alpha)}_0(\beta,\bar\beta)=0
  \qquad\text{for all}\quad \beta\bar\beta>4\pi^2 .
  \label{eq:torus-twist-sparse}
\end{equation}
A sequence that is $\alpha$-twist sparse is $\alpha'$-twist sparse for every $0<\alpha'<\alpha$. At $\alpha=0$ we impose no twist hypothesis.

\begin{theorem}[Twist-sparse extension for torus partition functions]
\label{thm:strong-torus}
Let $\{(\widehat Z,c)\}$ be a sequence of reduced partition functions with $c\to\infty$ obeying \eqref{modularrelation} and the equivalent conditions of Theorem~\ref{weaklibtorus}. If the sequence is $\alpha$-twist sparse for some $\alpha\in[0,1]$, then
\begin{equation}
  \lim_{c\to\infty}\frac{24}{c}\log\widehat Z(\beta,\bar\beta)
  =\max\Big(0,\ \frac{4\pi^2}{\beta}+\frac{4\pi^2}{\bar\beta}-\beta-\bar\beta\Big)
  \label{eq:torus-strong}
\end{equation}
at all temperatures outside the undetermined region $\mathcal{E}_\alpha=\mathcal{E}_\alpha^+\cup \mathcal{E}_\alpha^-$. Here $\mathcal{E}_\alpha^-$ is the image of $\mathcal{E}_\alpha^+$ under $\beta\leftrightarrow\bar\beta$, and
\begin{equation}
  \mathcal{E}_\alpha^+=\Big\{\beta>2\pi>\bar\beta\ \Big|\ \bar\beta<b_\alpha(\beta)\ \text{ and }\
  \frac{4\pi^2}{\beta}<b_\alpha\Big(\frac{4\pi^2}{\bar\beta}\Big)\Big\},
  \label{eq:torus-enigmatic}
\end{equation}
where the boundary function $b_\alpha$ is defined for $\beta\geq 2\pi$ by
\begin{equation}
  b_\alpha(\beta)=
  \begin{cases}
    \dfrac{\alpha(4\pi-\beta)+\sqrt{\alpha^2(4\pi-\beta)^2+16\pi^2(1-\alpha)}}{2},
      & \alpha\leq\tfrac12\ \text{ or }\ \beta<\dfrac{2\pi}{2\alpha-1},\\[14pt]
    \dfrac{4\pi^2}{\beta}, & \text{otherwise}.
  \end{cases}
  \label{eq:torus-boundary}
\end{equation}
In the first case, $b_\alpha(\beta)$ is the positive root of $b^2-\alpha(4\pi-\beta)\,b=4\pi^2(1-\alpha)$.
\end{theorem}

Theorem~\ref{thm:strong-torus} extends the domain of the large-$c$ result of DPQ \cite{Dey:2024nje} at the cost of a weaker error term. There, the free energy was determined on the open set of temperatures outside the closure of $\mathcal{E}_\alpha$ and off the self-dual hyperbola, with an error of order one in $c$ when the sparseness bounds do not depend on $c$. Here it is determined at leading order in $c$, but on the closed complement of $\mathcal{E}_\alpha$. This region also contains the boundary of the undetermined region and the points of the self-dual hyperbola outside it. As in Theorem~\ref{weaklibtorus}, the improvement comes from the Lipschitz bound of Lemma~\ref{lem:lipschitz}.

\begin{figure}[t]
    \centering
    \includegraphics[width=\linewidth]{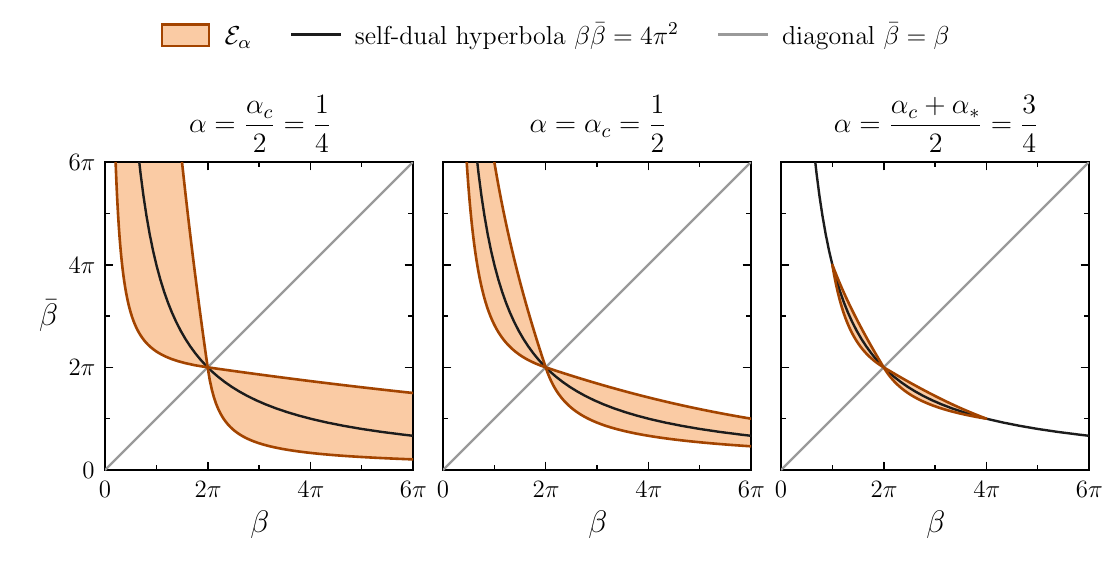}
    \caption{Undetermined regions (shaded) of modular invariant torus partition functions in the $(\beta,\bar{\beta})$ plane, for $\alpha =\alpha_c/2= 1/4$, $\alpha_c = 1/2$, and $(\alpha_c+\alpha_*)/2 = 3/4$}
    \label{fig:modenigmatic}
\end{figure}

The shape of the undetermined region was found in \cite{Dey:2024nje}. It is a band around the self-dual hyperbola, pinched at the self-dual point, which shrinks as the twist cutoff grows (Fig.~\ref{fig:modenigmatic}). At $\alpha=0$, $b_0(\beta)=2\pi$ and $\mathcal{E}_0$ is the union of the two open mixed wedges, so that Theorem~\ref{thm:strong-torus} reduces to statement 1 of Theorem~\ref{weaklibtorus}. For $0<\alpha\leq\frac12$, the band extends to arbitrarily low temperature, where $b_\alpha(\beta)\simeq\frac{1-\alpha}{\alpha}\frac{4\pi^2}{\beta}$, and it contains every point of the self-dual hyperbola except the self-dual point. For $\frac12<\alpha<1$, its closure is compact, and the curve $\bar\beta=b_\alpha(\beta)$ meets the self-dual hyperbola at $\beta=2\pi/(2\alpha-1)$, beyond which the vacuum answer holds on the hyperbola as well. At $\alpha=1$, the undetermined region is empty, and \eqref{eq:torus-strong} holds at all temperatures.

\section{Examples}
\label{sec:examples}
The theorems of Sect.~\ref{sec:results} leave open whether liberation requires a gap in the exchanged spectrum, a single-particle external operator, or the absence of interactions. The examples below show that it requires none of the three. They also show that liberation can fail inside the undetermined region, so the region is not merely a weakness of the technique, and that it fails once $\De$ is comparable to the central charge, rigorously in two dimensions and heuristically in higher dimensions. All proofs for this section are in Appendix~\ref{app:examples}.

\subsection{Composites of a generalized free field}
\label{ex:gff}

The simplest example is the family of correlators
\begin{equation}
\mathcal{G}^{(\Df, L)}(u,v) = \frac{\langle \phi^L(x_1) \phi^L(x_2) \phi^L(x_3) \phi^L(x_4) \rangle}{\langle \phi^L(x_1) \phi^L(x_2)\rangle \langle \phi^L(x_3) \phi^L(x_4) \rangle}
\end{equation}
in a generalized free theory of a scalar $\phi$ of dimension $\Df$. Their decomposition into higher-spin blocks \cite{Alday:2016njk} and their heavy limits have been studied in \cite{Poland:2024hvb,Poland:2026uiy}
(see also \cite{Ferrero:2023gnu} for a general $L$ formula with $\De_\phi =1$, and \cite{Aprile:2025nta}
for the conformal block expansion of free correlators). There are two parameters, the length $L$ of the composite and the dimension $\Df$ of its constituent, and $\De=L\Df$. Fixing $\Df$ and taking $L\to\infty$ gives a sequence with $\De\to\infty$ within a single generalized free theory, while fixing $L$ and taking $\Df\to\infty$ gives a sequence of a single kind of operator in a family of theories. Label a Wick contraction by the numbers $a,b,c$ of propagators in the $(12)(34)$, $(14)(23)$, and $(13)(24)$ channels, with $a+b+c=L$. With $\langle\phi^L\phi^L\rangle$ normalized to $L!$, summing over contractions gives \cite{Poland:2024hvb}
\begin{align}
\label{eq:GFFclosed}
\mathcal{G}^{(\Df, L )}(z,\bar{z}) &= \sum_{k=0}^{L}\binom{L}{k}^2(z\bar{z})^{k\Df}\,{}_2F_1\Big(-k,-k;1;\big((1-z)(1-\bar{z})\big)^{-\Df}\Big) \nonumber\\
&= \sum_{a+b+c=L}\Big(\frac{L!}{a!\,b!\,c!}\Big)^2 (z\bar{z})^{b\Df}\Big(\frac{z\bar{z}}{(1-z)(1-\bar{z})}\Big)^{c\Df}.
\end{align}
The terms are non-negative, and by the multinomial theorem their square roots sum to $\mathcal H^L$, where
\begin{equation}
\mathcal H(z,\bar{z})\equiv1+(z\bar{z})^{\frac{\Df}{2}}+\Big(\frac{z\bar{z}}{(1-z)(1-\bar{z})}\Big)^{\frac{\Df}{2}}.
\end{equation}
Comparing the sum of squares with the square of the sum (for the upper bound), and using the Cauchy--Schwarz inequality for the $\binom{L+2}{2}$ terms in the sum (for the lower bound), one finds
\begin{equation}
\label{eq:GFFsandwich}
\binom{L+2}{2}^{-1}\,\mathcal H(z,\bar{z})^{2L}\ \leq\ \mathcal{G}^{(\Df,L)}(z,\bar{z})\ \leq\ \mathcal H(z,\bar{z})^{2L}.
\end{equation}
Since $\binom{L+2}{2}=e^{o(\De)}$, the leading logarithm of the correlator is that of $\mathcal H^{2L}$, and the heavy limits follow from the behavior of $\mathcal H$.

The three ways of sending $\De\to\infty$ behave differently, and the difference lies in the light spectrum. Sparseness, statement~3 of Theorem~\ref{thm:weak-crossing}, is an upper bound on the light spectrum. We say that a sequence saturates it when the spectral weight attains this bound at leading exponential order, as in HKS \cite[Sect.~5.2 and App.~C.2]{Hartman:2014oaa}.

\begin{proposition}
\label{prop:GFF}
For the correlators \eqref{eq:GFFclosed}, with $(z,\bar{z})\in(0,1)^2$:
\begin{enumerate}[label=(\roman*)]
\item\label{gff:L} If $L$ is fixed and $\Df\to\infty$, the sequence is liberated on the whole causal diamond, and its light spectrum is sparse: $\mathcal{N}^{(\rho)}(\ve\De)=1$ for $\ve<2/L$.
\item\label{gff:Df} If $\Df$ is fixed and $L\to\infty$, the sequence is liberated nowhere:
\begin{equation}
\lim_{L\to\infty}\frac{1}{\De}\log\mathcal{G}^{(\Df,L)}(z,\bar{z})=\frac{2}{\Df}\log \mathcal H(z,\bar{z})>\max\Big(0,\log\frac{z\bar{z}}{(1-z)(1-\bar{z})}\Big).
\end{equation}
\item\label{gff:frag} If $L,\Df\to\infty$ at any relative rate, the sequence is liberated on the whole causal diamond. It saturates the sparseness bound in both expansions: for every fixed $\epsilon>0$,
\begin{equation}\label{eq:GFFsaturation}
\begin{aligned}
 \log\mathcal N^{(\rho)}(E)&=E\log(3+2\sqrt2)+o(\De),
 &&0\leq E\leq\sqrt2\De+\epsilon,\\
 \log\mathcal N^{(z)}(E)&=E\log2+o(\De),
 &&0\leq E\leq2\De+\epsilon,
\end{aligned}
\end{equation}
with errors uniform on the indicated intervals.
\end{enumerate}
\end{proposition}

The proof is in Appendix~\ref{threeheavylimits}. At fixed $L$, every term other than the identity contains $(z\bar z)^{\Df}$, so the exchanged spectrum has a gap proportional to $\De=L\Delta_\phi$, as for the generalized free field itself. At fixed $\Df$, the constituents remain light, and their many ways of pairing across channels make the light spectrum too dense. Indeed, $\frac1\De\log\mathcal G(1/2,1/2)$ tends to $2\log(2+2^{-\Df})/\Df>0$, so the sequence is not liberated. When both parameters grow, the sequence is liberated again, but without a gap, and its light spectrum grows at the largest rate that sparseness allows. Following \cite[Sect.~4.1.1]{Poland:2026uiy}, we call this composite of many heavy constituents a \emph{fragmenton}.

Liberation controls only the leading logarithm, so it does not imply that the identity dominates in relative size. The fragmenton with $\Df=L$, so that $\De=L^2$, makes this concrete. The estimate \eqref{eq:GFFsandwich} gives
\begin{equation}\label{eq:identity-fraction}
 \log\mathcal G\Big(\tfrac12,\tfrac12\Big)
 =2\sqrt{\De}\log2+O(\log\De).
\end{equation}
The self-dual criterion holds, yet the identity supplies a vanishing fraction of the correlator there, and $\mathcal G/\mathcal G^\GFF$ diverges. This is perhaps the cleanest illustration of how coarse the leading logarithm is.

The fragmenton also shows that liberation does not require a strict gap. To see how much freedom it leaves, we ask what the spectrum of a liberated sequence looks like. We work on the diagonal $z=\bar z$.
Since all coefficients are non-negative, $\mathcal N^{(z)}(E)\,z^{E}\le\mathcal G(z,z)$ for every $0<z<1$. For a liberated sequence, the free diagonal of Theorem~\ref{thm:weak-crossing} then gives, with $e=E/\De$,
\begin{equation}\label{eq:hull-bound}
  \limsup_{\De\to\infty}\frac1\De\log\mathcal N^{(z)}(e\De)\;\le\;\hat s(e)
  \equiv\inf_{0<z<1}\Big[\max\Big(0,\,2\log\frac{z}{1-z}\Big)-e\log z\Big].
\end{equation}
When the entropy $s(e)=\lim\frac1\De\log\mathcal N^{(z)}(e\De)$ exists, we have $$\lim\frac1\De\log\mathcal G(z,z)=\sup_e\big[s(e)+e\log z\big]\,.$$ The liberated limit is thus the Legendre transform of $s$ in the variable $-\log z$, and $\hat s$ is the concave hull of $s$.
The infimum in \eqref{eq:hull-bound} is elementary. For $z\le 1/2$ the bracket is $-e\log z$, which decreases in $z$. For $z>1/2$ its derivative is $z^{-1}\big[\,2-e+2z/(1-z)\big]$, which is non-negative for all such $z$ exactly when $e\le4$. Hence
\begin{equation}\label{eq:hull-line}
  \hat s(e)=e\log2\quad\text{for }0\le e\le4,
  \qquad
  \hat s(e)<e\log2\quad\text{for }e>4 .
\end{equation}

Geometrically, the corner of the liberated limit at $z=\frac12$ corresponds to a range of weights over which the hull grows linearly, at the rate $\log2$ allowed by sparseness. Its endpoints correspond to the identity at $E=0$ and to its crossing image at $E=4\De$, twice the mean weight $2\De$ exchanged at the crossing symmetric point (Sect.~\ref{sec:liberation4pt}). The segment lies on the sparseness bound $s=e\log2$ of Theorem~\ref{thm:weak-crossing}. A liberated spectrum must therefore lie on or below this line, and it can reach the line only for $E\le4\De$. Below this weight, liberation is compatible with a gap, with saturation, and with anything in between. The same argument applies to the $\rho$-expansion \eqref{rhoexp}, with $\rho^{E}$ in place of $z^{E}$.
There we have $\hat s(e)=e\log(3+2\sqrt2)$ exactly for $0\le e\le2\sqrt2$, again up to twice the mean weight $\sqrt2\,\De$.

\begin{figure}[t]
  \centering
  \includegraphics[width=.8\linewidth]{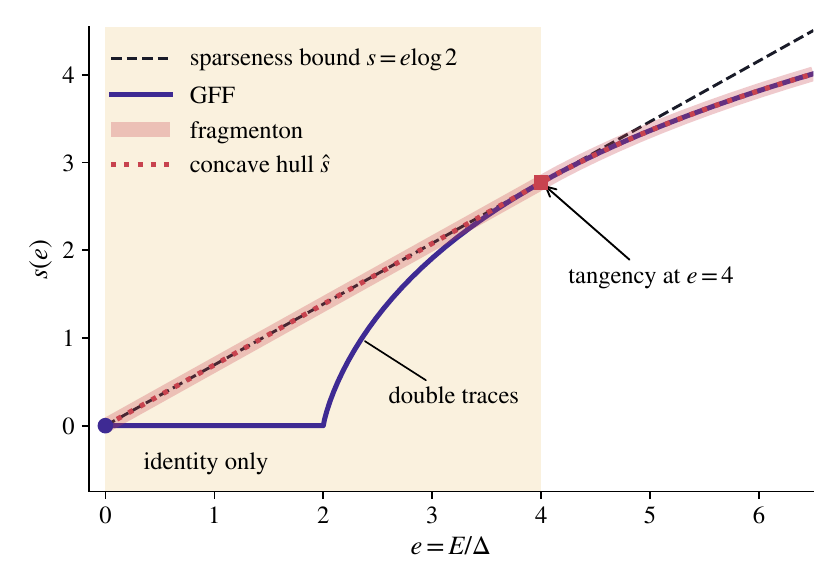}
  \caption{Entropy on the diagonal in the $z$-expansion, $s(e)=\lim\frac1\De\log\mathcal N^{(z)}(e\De)$. Dashed: the sparseness bound $s=e\log2$. Solid: the generalized free field; only the identity contributes for $e<2$, and double traces for $e\ge2$, eq.~\eqref{eq:gff-entropy}. Dotted: the concave hull $\hat s$, the Legendre transform of the liberated limit, eq.~\eqref{eq:hull-line}. It coincides with the bound on $[0,4]$ (shaded), the only range on which a liberated spectrum can saturate sparseness. The generalized free field touches the bound only at $e=0$ and, tangentially, at $e=4$. Red band: the fragmenton ($L,\Df\to\infty$), whose entropy is the hull itself, so that it saturates the bound on the whole shaded range}
  \label{fig:saturation-entropy}
\end{figure}

The generalized free field and the fragmenton populate this segment in opposite ways (Fig.~\ref{fig:saturation-entropy}). On the diagonal, $\mathcal G^\GFF=1+z^{2\De}+z^{2\De}(1-z)^{-2\De}$. The identity sits at $E=0$, the term $u^\De$ is a single monomial, and the crossing image of the identity contributes $\binom{2\De+n-1}{n}$ at total weight $E=2\De+n$ in the original channel. Hence
\begin{equation}\label{eq:gff-entropy}
  s_{\rm GFF}(e)=e\log\frac e2-(e-2)\log\frac{e-2}{2},\qquad e\ge2,
\end{equation}
while $s_{\rm GFF}(e)=0$ for $0\le e<2$, where only the identity contributes. Since $s_{\rm GFF}(4)=4\log2$ and $s_{\rm GFF}'(4)=\log2$, the double-trace curve is tangent to the line at $e=4$ and lies strictly below it elsewhere. Its concave hull is therefore the segment from the identity to the point of tangency, continued by the double-trace curve, and its Legendre transform is again the liberated limit. The generalized free field thus touches the line only at the two ends of the segment, and at the crossing symmetric point the identity and the double-trace band coexist. The fragmenton instead lies on the line, at least up to $E=2\De+\epsilon$ by \eqref{eq:GFFsaturation}, so all intermediate weights contribute at the same exponential order. In fact, the fragmenton saturates the whole segment. At weight $E=e\De$ with $e\le4$, the terms of \eqref{eq:GFFclosed} with $c=\lfloor eL/4\rfloor$ contribute a binomial coefficient $\binom{2c\Df+n-1}{n}$, with $n=E-2c\Df \in \mathbb{N}$. The maximizer comes from $n\approx E/2$,  leading to the entropy growth of $e\log2$. For $e>4$, the terms with $c=L$ give the double-trace curve. Its entropy is therefore the hull $\hat s$, so the fragmenton is liberated with the densest spectrum that liberation allows. The two sequences have the same leading logarithm, and their difference appears only in the subleading behavior near the kink.

The leading rescaled logarithm does not distinguish the two liberated regimes, since both are sparse, although only the fixed-$L$ regime has an exchanged gap proportional to $\De$. A shrinking kinematic window distinguishes them. In the coordinates $x=\log\frac{z}{1-z}$ and $\bar x=\log\frac{\bar z}{1-\bar z}$, crossing is the reflection $(x,\bar x)\mapsto(-x,-\bar x)$ and the liberated value is $\max(0,x+\bar x)$. Write $\mathcal{Z}(x,\bar{x})=\mathcal{G}(z(x),z(\bar{x}))$, and define the \emph{$\gamma$-rescaled potential}
\begin{equation}
\label{eq:alphalocal}
\Lambda^{(\gamma)}(x,\bar{x}) = \lim_{\De \to \infty} \frac{1}{\De^{1-\gamma}} \log \mathcal{Z}\big( x \De^{-\gamma}, \bar{x} \De^{-\gamma} \big),\qquad 0\leq\gamma\leq1 ,
\end{equation}
which probes a window of size $\De^{-\gamma}$ around the self-dual point, and which for $\gamma=0$ is the limit of $\frac1\De\log\mathcal Z$ (closely related quantities were called $\alpha$-local rate functions in \cite{Poland:2026uiy}; we write $\gamma$ since $\alpha$ is our twist cutoff). Set $\Df = b  M^{\gamma_c}$ and $L = \lfloor M^{1-\gamma_c}/b \rfloor$, with $\gamma_c \in (0,1)$ and $b>0$, so that $\De\approx M\to\infty$. Then
\begin{equation}
\label{eq:fragmentonregimes}
\Lambda^{(\gamma)}(x,\bar{x}) =
\begin{cases}
\max(0,x+\bar{x}), & \gamma<\gamma_c,\\[4pt]
\dfrac{2}{b} \log \big(1 +e^{b(x+\bar{x})/2} \big), & \gamma=\gamma_c,\\[10pt]
+\infty, & \gamma>\gamma_c.
\end{cases}
\end{equation}
To see where $\gamma_c$ comes from, note that near the self-dual point \eqref{eq:GFFsandwich} gives
\begin{equation}\label{eq:GFF-near-selfdual}
\log \mathcal{Z}(x,\bar x)=2L\log\big(1+e^{\Df(x+\bar x)/2}\big)+O(\log L)+O(L\,2^{-\Df}).
\end{equation}
This crosses over from $0$ to $\De(x+\bar x)$ in a window $|x+\bar x|\sim\Df^{-1}=b^{-1}\De^{-\gamma_c}$, whose size is set by the dimension of the constituents rather than by $\De$. A window much larger than this, $\gamma<\gamma_c$, sees a sharp kink. A window much smaller, $\gamma>\gamma_c$, sees the self-dual value $2L\log2\propto\De^{1-\gamma_c}$, which diverges in units of $\De^{1-\gamma}$. At $\gamma=\gamma_c$ the two scales coincide. Since $\gamma_c>0$, the smooth crossover is invisible at $\gamma=0$, where $\frac1\De\log\mathcal G$ tends to the kinked function $\max(0,x+\bar x)$. Thus the fragmenton is liberated while saturating the sparseness condition.

\subsection{Heavy scalar with spinning constituents}
\label{ex:chiral}
We consider a heavy scalar built from two constituents. In a generalized free theory, its four-point function is a sum over Wick contractions, and at large $\De$ one contraction dominates at each value of the cross ratios. Scalar constituents all pair in the s-channel ($x_1$ with $x_2$, $x_3$ with $x_4$) below the self-dual line $z+\bar z=1$ and in the t-channel ($x_1$ with $x_3$, $x_2$ with $x_4$) above it, so the composite behaves like a single free particle and is liberated. Spin can break this. A constituent with spin in the plane of the four operators weighs $z$ and $\bar z$ differently, so when $z$ and $\bar z$ lie on opposite sides of $1/2$, constituents of opposite spin can prefer opposite channels. If the spin grows with the dimension, this happens on a whole region of cross ratios inside the undetermined region of Theorem~\ref{thm:strong-crossing}, which we call the \emph{mixed phase}. There the dominant contraction connects all four operators, and liberation fails. We construct the example in generalized free field theory, and then realize it in local two-dimensional CFTs.

This example generalizes the chiral product correlators of \cite{Poland:2026uiy}, in which each constituent is chiral: all of its weight is in $h$ or in $\bar h$. Its twist then vanishes, and so does the twist gap. To produce a twist gap proportional to $\De$, the constituents must carry both large spin and twist, in a ratio that we can choose. A spinning generalized free field is the simplest such constituent. We take a symmetric traceless generalized free field $J^{\mu_1\dots\mu_\ell}$ of spin $\ell$ and dimension $\De_J=\De/2\geq\ell+d-2$, and form the normal-ordered composite $\cO=:J\cdot J:$, a double-twist operator. Holographically, $\cO$ is dual to a composite of two massive fields of opposing spin $\ell$ \cite{Costa:2014kfa}.

The spin enters through the plane that contains the four insertion points. An operator in this plane has weights $(h,\bar h)$ under the conformal transformations of the plane, where $h+\bar h$ is its scaling dimension and $m=h-\bar h$ is its spin under rotations of the plane. We call $m$ the \emph{planar spin}, to distinguish it from the spin $\ell$ of $J$ under $SO(d)$. The components of $J$ have planar spins $m=-\ell,\dots,\ell$. For example, a vector has two components along the plane, $J_w$ and $J_{\bar w}$ in a complex coordinate $w$, with $m=\pm1$, and $d-2$ components normal to it, with $m=0$. The two extreme components, with $m=\pm\ell$, have the most unequal weights, $(h,\bar h)=(\De_J-\frac{\tau_J}2,\frac{\tau_J}2)$ for $m=\ell$ and the reverse for $m=-\ell$, where $\tau_J=\De_J-\ell$ is the twist of $J$. These two components turn out to dominate the heavy limit.

The reduced correlator of $\cO$ is (Appendix~\ref{app:JJ})
\begin{equation}
\label{eq:JJ}
\begin{aligned}
\mathcal{G}={}&1+(z\bar z)^{\De}+(\eta\bar\eta)^{\De}\\
&+\frac{4}{D^2}\Big[(z\bar z)^{\De_J}\Xi_\ell\Big(\frac{z}{\bar z}\Big)+(\eta\bar\eta)^{\De_J}\Xi_\ell\Big(\frac{\eta}{\bar\eta}\Big)+(z\bar z\,\eta\bar\eta)^{\De_J}\Xi_\ell\Big(\frac{1-z}{1-\bar z}\Big)\Big],
\end{aligned}
\end{equation}
where $\eta=z/(1-z)$ and
\[
 \Xi_\ell(\lambda)=\sum_{m=-\ell}^{\ell}N_{\ell,m}\,\lambda^m .
\]
Here $N_{\ell,m}$ is the number of components of $J$ with planar spin $m$, so that $N_{\ell,\pm\ell}=1$, and
\[
 D=\sum_{m=-\ell}^{\ell}N_{\ell,m}=\frac{(2\ell+d-2)(\ell+d-3)!}{\ell!\,(d-2)!}
\]
is the number of independent components of $J$; in $d=3$, every $N_{\ell,m}=1$ and $D=2\ell+1$. The first line of \eqref{eq:JJ} contains the three Wick contractions in which both constituents of each insertion pair with the same partner, as for a generalized free field of dimension $\De$. In each of the other three, the two constituents of each insertion pair with different partners, so the contraction joins all four insertions in a cycle. The term $\lambda^m$ of $\Xi_\ell$ is the contribution of the components of planar spin $m$.

The lowest exchanged twist is $2\tau_J$, that of the double-twist operators $[JJ]$ with all indices symmetrized, so the twist gap is set by the ratio $\ell/\De_J$. For the heavy limit, we hold this ratio fixed. Given $0<\alpha<\frac12$, a twist gap just above $2\alpha\De$ requires $\ell/\De_J\to1-2\alpha$, which fixes $\ell$ up to $o(\De)$. We take $\ell$ to be the largest integer below $\frac{(1-2\alpha)\De}{2}$. Then $\tau_J=\frac\De2-\ell=\alpha_\De\De$ with $\alpha<\alpha_\De\leq\alpha+\frac1\De$, and for large $\De$ both $\ell\geq1$ and the unitarity bound hold.

\begin{proposition}[Composites of a generalized free field of large spin]
\label{prop:JJ}
Let $0<\alpha<\frac12$, and let $\mathcal{G}$ be the correlator \eqref{eq:JJ} with $\tau_J=\alpha_\De\De$, $\alpha<\alpha_\De\leq\alpha+\frac1\De$ and $\De$ large.
\begin{enumerate}[label=(\roman*)]
\item $\mathcal{G}$ obeys \eqref{crossingrelation}, stays bounded at $z=\bar z=\frac12$, and has a non-negative $\rho$-expansion in which every non-identity monomial has $\min(h,\bar h)\geq\alpha_\De\De>\alpha\De$. So the sequence obeys the equivalent conditions of Theorem~\ref{thm:weak-crossing}, and it is $\alpha$-twist sparse in the $\rho$-expansion.
\item The following holds
\begin{equation}
\label{eq:JJrate}
\lim_{\De\to\infty}\frac{1}{\De}\log\mathcal{G}(z,\bar z)=F(z,\bar z)\equiv\max\big(0,\,\alpha x+(1-\alpha)\bar x\big)+\max\big(0,\,(1-\alpha)x+\alpha\bar x\big).
\end{equation}
\item The sequence is liberated exactly outside $\mathcal{V}_\alpha\cup\chi(\mathcal{V}_\alpha)$, where $\chi(x,\bar x)=(\bar x,x)$ and
\begin{equation}
\mathcal{V}_\alpha=\Big\{(x,\bar x)\ \Big|\ x<0<\bar x\ \text{ and }\ \lambda_\alpha|x|<\bar x<\frac{|x|}{\lambda_\alpha}\Big\},\qquad \lambda_\alpha=\frac{\alpha}{1-\alpha}.
\end{equation}
This is a proper subset of $\mathcal{E}^{(\rho)}_\alpha$, and hence of $\mathcal{E}^{(z)}_\alpha$. Near the double lightcone limits, the boundaries of both undetermined regions approach the lines $\bar x=\lambda_\alpha|x|$ and $\bar x=|x|/\lambda_\alpha$ that bound $\mathcal V_\alpha$.
\end{enumerate}
\end{proposition}

The proof is in Appendix~\ref{app:JJ}. Fig.~\ref{fig:bulkchiral} shows where the limit \eqref{eq:JJrate} comes from. Each constituent pairs in the s-channel (vertical line) or the t-channel (horizontal line); the terms containing $z\bar z$, whose lines would run along the diagonals, are smaller because $z<\eta$ and $\bar z<\bar\eta$. The left and right panels are the terms $1$ and $(\eta\bar\eta)^{\De}$, so moving one constituent from vertical to horizontal contributes a factor $(\eta\bar\eta)^{\De_J}$. In the middle panel only one has moved. This is the term $(\eta\bar\eta)^{\De_J}\Xi_\ell(\eta/\bar\eta)$, in which the moved constituent, of planar spin $m$, contributes an extra factor $(\eta/\bar\eta)^{m}$. For $m=\pm\ell$, with $\eta=e^{x}$ and $\bar\eta=e^{\bar x}$, the moved constituent contributes, at leading order in $\De$,
\begin{equation}\label{eq:JJ-spin-term}
 (\eta\bar\eta)^{\De_J}\Big(\frac{\eta}{\bar\eta}\Big)^{\pm\ell}\simeq e^{\De A_\pm},\qquad
 A_\pm=\frac{x+\bar x}{2}\mp\Big(\frac12-\alpha\Big)(\bar x-x).
\end{equation}
The first term of $A_\pm$ is the same as for scalar constituents. The second, the spin term, has opposite signs for $m=\pm\ell$, so it cancels in the right panel. Up to powers of $\De$, $\mathcal G$ has the size of $(1+e^{\De A_+})(1+e^{\De A_-})$. The four terms of this product are the panels: $1$ is the left panel, $e^{\De A_-}$ and $e^{\De A_+}$ are the middle panel as drawn and with the colors exchanged, and $e^{\De(A_++A_-)}$ is the right panel. The factor $1+e^{\De A_\pm}$ describes the constituent of planar spin $\pm\ell$: the $1$ is its s-channel pairing and the exponential its t-channel pairing. So this constituent pairs in the t-channel where $A_\pm>0$ and in the s-channel where $A_\pm<0$, whatever the other constituent does. Since $\frac1\De\log(1+e^{\De A})\to\max(0,A)$, this gives $F=\max(0,A_+)+\max(0,A_-)$.

If $A_+$ and $A_-$ have the same sign, the constituents make the same choice and the sequence is liberated, with the left panel dominant where both are negative (the \emph{cold phase}) and the right panel where both are positive (the \emph{hot phase}). If their signs differ, the spin term wins and the middle panel dominates (the mixed phase), with the colors as drawn in $\mathcal V_\alpha$, where $A_->0>A_+$, and exchanged in $\chi(\mathcal V_\alpha)$. It exceeds the outer panels by a factor $e^{\De\min(|A_+|,|A_-|)+o(\De)}$, so liberation fails, and the leading logarithm detects that $\cO$ is a composite.

\begin{figure}[h]
    \centering
    \includegraphics[width=\linewidth]{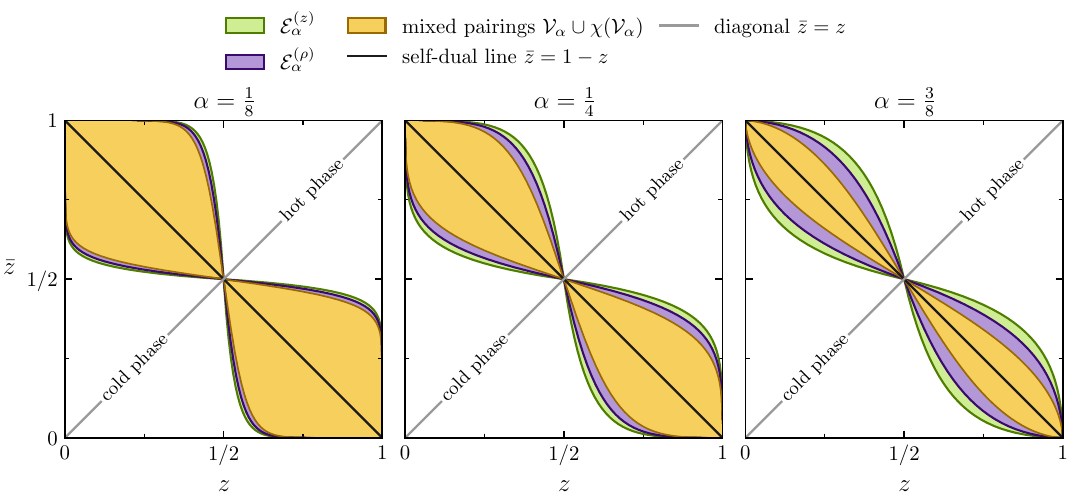}
    \caption{Regions $\mathcal{V}_\alpha\cup\chi(\mathcal{V}_\alpha)$ in which the constituents of planar spin $\pm\ell$ in \eqref{eq:JJ} have different dominant pairings, for $\alpha=1/8,1/4,3/8$. They lie inside the undetermined regions $\mathcal{E}^{(\rho)}_\alpha$ and $\mathcal{E}^{(z)}_\alpha$ at the same cutoff. The inner and outer boundaries have the same asymptotic opening as $z\to0$, $\bar z\to1$, after taking the heavy limit at fixed kinematics. The inner regions shrink to the self-dual line as $\alpha\to1/2$.}
    \label{fig:chiralphases}
\end{figure}

Theorem~\ref{thm:strong-crossing} allows the mixed phase, because a vertical cut through the middle panel, which gives the s-channel OPE, crosses two lines: the double-twist operators $[JJ]$ of twist $2\alpha_\De\De$, just above the cutoff. The theorem confines the non-liberated region to $\mathcal{E}^{(\rho)}_\alpha$, which is wider than the mixed phase at every point, by the largest angle near the self-dual point (Lemma~\ref{lem:position}), and has boundaries of the same slopes near the double lightcone limits (Fig.~\ref{fig:chiralphases}). The same holds in the $z$-expansion, where $\mathcal{G}$ is also $\alpha$-twist sparse.

The coefficient of the spin term is the ratio of spin to dimension, $1/2-\alpha=\lim\ell/\De$, and it sets the size of the mixed phase (Fig.~\ref{fig:chiralphases}). As $\alpha\to 1/2$, the spin term becomes negligible, and the mixed phase closes onto the self-dual line. As $\alpha\to0$, the twist of $J$ becomes negligible compared to its spin, and its extreme components become effectively chiral: $A_+\to x$, which depends only on $z$, and $A_-\to\bar x$, which depends only on $\bar z$. Each constituent then switches to the t-channel when its own cross ratio passes $1/2$, so the mixed phase fills the mixed wedges. In two dimensions, unitarity allows $\tau_J=0$, so this limit is reached exactly, and \eqref{eq:JJ} becomes the chiral product $(1+\eta^{\De}+z^{\De})(1+\bar\eta^{\De}+\bar z^{\De})$ of \cite{Poland:2026uiy}. The same limit, $\max(0,x)+\max(0,\bar x)$, holds in any $d$ if $\tau_J\geq d-2$ is held fixed as $\De\to\infty$. It equals the upper bound of Theorem~\ref{thm:weak-crossing}, so that bound cannot be improved.

\begin{figure}[h]
    \centering
    \includegraphics[width=0.9\linewidth]{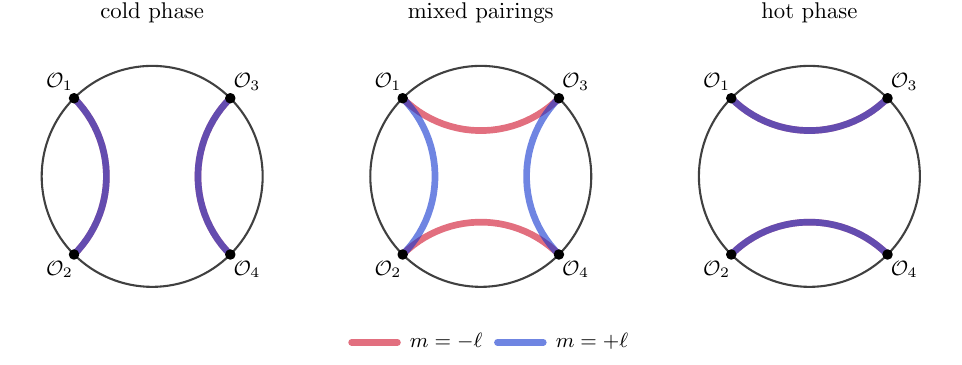}
    \caption{Schematic bulk contraction diagrams for \eqref{eq:JJ}. In the cold and hot phases, both constituent lines connect the same pair of insertions. In $\mathcal{V}_\alpha$, the dominant contractions pair the constituent of planar spin $-\ell$ in the t-channel and that of spin $+\ell$ in the s-channel. Coincident lines in the drawing denote a common pairing, not binding into one particle}
    \label{fig:bulkchiral}
\end{figure}

\paragraph{Local realization} The field $J$ is generalized free, so the example above has no stress tensor, and one may ask whether the mixed phase survives in a local CFT. There, the stress tensor and its composites appear in every OPE $\cO\times\cO$ with low twist, and twist sparseness requires their total contribution to stay subexponential. We show that there is a sequence of two-dimensional local CFTs in which these exchanges are indeed small, so that the correlators are $\alpha$-twist sparse, and the mixed phase is still present.

These exchanges can be suppressed by large-$N$ factorization. With this in mind, it is very natural to consider the symmetric product orbifold of $N$ copies of a CFT, and the $S_N$-invariant operator $U=\frac1{\sqrt N}\sum_{i=1}^NV_i$, where $V_i$ is a primary $V$ of the $i$-th copy. Different copies are uncorrelated and $\langle V\rangle=0$, so in a four-point function the operators sit either pairwise in two different copies, or all in one copy. These two cases give
\begin{equation}
\langle UUUU\rangle=\Big(1-\frac1N\Big)\big[\text{Wick contractions}\big]+\frac1N\,\langle VVVV\rangle .
\end{equation}
Exchanges inside one copy, such as its stress tensor, appear only in the second term. So as $N\to\infty$, $U$ behaves as a generalized free field \cite{Lunin:2000yv,Pakman:2009zz,El-Showk:2011yvt}.

As we have seen in the previous subsection, to ensure the existence of the mixed phase, the constituents had to carry large planar spin, as the components $m=\pm\ell$ of $J$ do. Free-boson currents provide this most simply. Since $j$ has weights $(1,0)$ and $\bar j$ has weights $(0,1)$, a product of $l$ currents $j$ and $k$ currents $\bar j$ of distinct bosons has weights $(l,k)$, and its ratio of twist to dimension, $2l/(l+k)$, plays the role of $\tau_J/\De_J$. We therefore take $U_\pm=\frac1{\sqrt N}\sum_iV_{\pm,i}$, where $V_+$ and $V_-$ are two such products of opposite spin, with weights $(l,k)$ and $(k,l)$, built from different bosons, so that $\cO=U_+U_-$ is a scalar, of dimension $\De=2K$ with $K=l+k$. They play the roles of the components of $J$ of planar spin $-\ell$ and $+\ell$ in $d=2$, with $\De_J=K$ and $\ell=k-l$, and as $N\to\infty$ the four-point function of $\cO$ approaches the two-dimensional case of \eqref{eq:JJ}.

For the heavy limit, we fix $0<\alpha<\frac12$ and let $l/K\to\alpha$ from above, so that, as for $J$, the lowest exchanged twist $4l$ lies just above the cutoff $2\alpha\De=4\alpha K$. The factor $\frac1N$ must also beat the growth of the single-copy term: each of the $K$ currents can contract independently, so $\langle VVVV\rangle$ grows exponentially in $K$. We therefore let the number of copies $N_K$ grow faster than any exponential of $K$, for example $N_K=\lceil e^{K^2}\rceil$. We write $\cO_K$ for $\cO$ in the resulting theory $\mathcal T_K$, the symmetric product orbifold of $N_K$ copies of $2K$ compact bosons, with $\De_K=2K$;

To be precise, let us consider $\mathcal X_K$, the theory of $2K$ compact bosons, and let $\mathcal T_{K,N}=\mathcal X_K^{\otimes N}/S_N$ be the symmetric product orbifold of $N$ copies of $\mathcal X_K$.
This is a local unitary CFT with $c_L=c_R=2NK$. We label the bosons $\phi_{\pm,i,a}$, where $i=1,\dots,N$ labels the copy and $\pm$ and $a=1,\dots,K$ label the bosons within it, and write $j_{\pm,i,a}$ and $\bar j_{\pm,i,a}$ for the holomorphic and antiholomorphic currents $j=i\partial\phi$ and $\bar j=i\bar\partial\phi$ of $\phi_{\pm,i,a}$. For integers $0<l<k$ with $l+k=K$, let
\begin{equation}\label{eq:local-ops}
\begin{gathered}
 V_{+,i}=\prod_{a=1}^{l}j_{+,i,a}\prod_{a=l+1}^{K}\bar j_{+,i,a},\qquad
 V_{-,i}=\prod_{a=1}^{l}\bar j_{-,i,a}\prod_{a=l+1}^{K}j_{-,i,a},\\
 U_\pm=\frac1{\sqrt N}\sum_{i=1}^NV_{\pm,i},\qquad \cO=U_+U_- .
\end{gathered}
\end{equation}
The operators $V_{+,i}$ and $V_{-,i}$ are Virasoro primaries of weights $(l,k)$ and $(k,l)$, so $\cO$ is a scalar Virasoro primary of dimension $\De=2K$. The operators $U_\pm$ and $\cO$ are invariant under permutations of the copies, so they are operators of $\mathcal T_{K,N}$. The constituents $U_+$ and $U_-$ play the roles of the components of $J$ of planar spin $-\ell$ and $+\ell$ in $d=2$, with $\De_J=K$ and $\ell=k-l$.

\begin{proposition}\label{prop:local-JJ}
For every $0<\alpha<\frac12$, the operators $\cO_K$ are unit-normalized Hermitian scalar Virasoro primaries in the local unitary CFTs $\mathcal T_K$, and their reduced correlators $\mathcal G_K$ are $\alpha$-twist sparse in both positive expansions and satisfy
\begin{equation}\label{eq:local-JJ-limit}
 \mathcal G_K(\tfrac12,\tfrac12)\longrightarrow4,\qquad
 \frac{\log\mathcal G_K(z,\bar z)}{\De_K}\longrightarrow F(z,\bar z),
\end{equation}
locally uniformly on $(0,1)^2$, where $F$ is given in \eqref{eq:JJrate}.
\end{proposition}

The proof in Appendix~\ref{app:local-JJ} computes the correlator exactly for every number of copies and bounds its entire low-twist sum. So even for local CFTs, the asymptotic opening of the undetermined region of Theorem~\ref{thm:strong-crossing} cannot be narrowed.

\FloatBarrier

\subsection{The Virasoro obstruction}
\label{ex:virasoro}

We now turn to what prevents liberation. Liberation fails when many exchanged light operators together contribute at the exponential rate $\De$, competing with the identity and its crossing image. The GFF composite with $L\to\infty$ at fixed $\Df$ is one example, but it is a property of a particular sequence of operators rather than a universal obstruction.

A universal candidate is the stress tensor, which appears in every OPE $\cO\times\cO$ of a local CFT, with a coefficient fixed by the Ward identity that grows as $\De/\sqrt{C_T}$. By itself, its contribution grows only polynomially in $\De$ and does not change the leading logarithm. In two dimensions, however, the stress tensor comes with its composites, whose dimensions and OPE coefficients are fixed by Virasoro symmetry. Together they form the vacuum module, which contributes $\mathcal{F}_0(z)\mathcal{F}_0(\bar z)$ to $\mathcal{G}$, where $\mathcal{F}_0$ is the Virasoro vacuum block. For $c>1$, every Virasoro block contributes positively, so $\mathcal{G}(z,\bar z)\geq\mathcal{F}_0(z)\mathcal{F}_0(\bar z)$, and we show in Appendix~\ref{app:virasoro}, keeping only the states $L_{-2}^n|0\rangle$ of the vacuum module, that
\begin{equation}
\label{eq:2dlower}
\mathcal{F}_0(z)\ \geq\ {}_2F_1\Big(\De,\De;\frac c8;\rho(z)^2\Big).
\end{equation}

\begin{proposition}[Virasoro obstruction to liberation]
\label{prop:bodies2d}
Let $\{\cO\}$ be a sequence of scalar Virasoro primaries, with dimensions $\De\to\infty$, in unitary CFT$_2$s with central charges $c>1$.
\begin{enumerate}[label=(\roman*)]
\item If $\De/c\to\nu\in(0,\infty]$, then $\liminf\frac1\De\log\mathcal{G}(z,\bar z)>\max(0,x+\bar x)$ at every point of the causal diamond, so the sequence is liberated nowhere.
\item If $\De^{1+\gamma_c}/c\to k$, with $k>0$ and $0<\gamma_c<1$, then $\liminf\De^{\gamma_c-1}\log\mathcal{G}(\frac12,\frac12)\geq16\rho_*^2\,k$, where $\rho_*=3-2\sqrt2$.
\end{enumerate}
\end{proposition}

The proof is in Appendix~\ref{app:virasoro}. Here $\nu$ and $k$ are normalized by $c$, as is natural in two dimensions; since $C_T=2c$, they are twice the ratios $\De/C_T$ and $\De^{1+\gamma_c}/C_T$ used in Sect.~\ref{ex:bodies}. Statement (i) includes sequences in a single CFT$_2$, at fixed $c$, where $\frac1\De\log\mathcal{G}(1/2,1/2)$ is at least $4\log\frac{1+\sqrt2}{2}\approx0.753$ in the limit, from the large-$\De$ limit of \eqref{eq:2dlower}. A heavy scalar in a unitary CFT$_2$ can therefore be liberated only if $c$ grows faster than $\De$, as in the local realization of Sect.~\ref{ex:chiral}. In statement (ii), the vacuum module alone shifts the self-dual value of $\Lambda^{(\gamma_c)}$ by at least $16\rho_*^2k\approx0.471\,k$.

\subsection{Compact gravitating bodies}
\label{ex:bodies}

In higher dimensions we have no analog of Proposition~\ref{prop:bodies2d}, but holography suggests one. In a holographic CFT, every heavy operator interacts gravitationally through the stress tensor and its composites. We study heavy operators dual to \emph{compact bodies}, whose size is small compared to the AdS radius, which we set to one. The example serves two purposes. It shows that a correlator can be liberated even when classical gravitational interactions change it by an exponentially large factor, since that factor is $e^{o(\De)}$. It also gives a heuristic for obstruction to liberation in $d\geq3$. Assuming the worldline approximation and the decomposition \eqref{eq:HHHH} of the correlator into three channels, we find that such operators are liberated when $\De/C_T\to0$, and that liberation fails when $\De/C_T$ tends to a small positive constant. Graviton exchange between the worldlines exponentiates the stress tensor block \cite{Maxfield:2017rkn}, and at fixed $\De/C_T$ the exponent grows at rate $\De$.

Let us consider our favourite example: a holographic CFT$_d$ with large central charge $C_T$, dual to Einstein gravity in AdS$_{d+1}$,
\begin{equation}
S_{\rm grav}[g]=-\frac{1}{16\pi G_N}\int d^{d+1}y\,\sqrt{g}\,\big(\mathcal{R}+d(d-1)\big),\qquad C_T\propto\frac{1}{G_N},
\end{equation}
and a scalar $\cO$ of dimension $\De$ dual to a compact body of mass $m$, with $\De(\De-d)=m^2$, which couples only to the metric. The worldline path integral \eqref{eq:introworldline} now includes the metric,
\begin{equation}
\label{eq:worldlinePI}
\langle\cO(x_1)\cO(x_2)\cO(x_3)\cO(x_4)\rangle=\int[Dg]\,e^{-S_{\rm grav}[g]}\int_{\partial\Gamma=\{x_1,x_2,x_3,x_4\}}[D\Gamma]\,e^{-mL[\Gamma,g]+\dots},
\end{equation}
where the ellipsis denotes finite-size couplings, such as tidal operators, which respond to the field of the other body. Their suppression requires assumptions on the size and response coefficients of the body, in addition to small $\De/C_T$ \cite{Goldberger:2004jt,Correia:2026fnr}. We neglect them, and return to them at the end of the section. We focus on the s-channel contribution $\mathcal{G}_s(z,\bar z)$ to the reduced correlator, in which $\Gamma$ joins $x_1$ to $x_2$ and $x_3$ to $x_4$. The t- and u-channel contributions are its crossing images, so that
\begin{equation}
\label{eq:HHHH}
\mathcal{G}(z,\bar z)=\mathcal{G}_s(z,\bar z)+\Bigg(\frac{z\bar z}{(1-z)(1-\bar z)}\Bigg)^{\De}\mathcal{G}_s(1-z,1-\bar z)+(z\bar z)^{\De}\mathcal{G}_s\Big(\frac1z,\frac1{\bar z}\Big),
\end{equation}
where the u-channel term follows from $x_2\leftrightarrow x_4$. Its argument lies outside the causal diamond, so there $\mathcal{G}_s$ is continued from the Euclidean section and need not be positive, but its prefactor $(z\bar z)^\De$ is small in the diamond. This decomposition is an assumption, controlled only when $\De/C_T$ is small. Without gravity, $\mathcal{G}_s=1$, and $\mathcal{G}$ is the reduced generalized free field correlator.

With gravity turned on, one of the heavy particles back-reacts and modifies the geometry and the second one follows a geodesic in that geometry. The gravitational interaction is proportional to $G_Nm_2m_1$, giving rise to $\Delta^2/C_T$ factor. Thus we have
\begin{equation}
\log\mathcal{G}_s(z,\bar z)=\frac{\De^{2}}{C_T}\Big(g_1(z,\bar z)+O(\De^{-1})\Big)
\end{equation}
Here in the term $\frac{\De^{2}}{C_T}$, we should think of $\Delta$ as the mass of the second object, while $L[\Gamma,g]$ on the back-reacted geometry is proportional to  $\Delta/C_T$. The error term comes from the loop contribution around the dominating worldline saddle of the path integral. In general, one would expect (at least if we keep $\Delta/C_T$ small enough) that the contributions of the following are present:

\begin{equation}
\label{eq:bodiesexp}
\log\mathcal{G}_s(z,\bar z)\supset \sum_{n\geq1}\frac{\De^{n+1}}{C_T^{\,n}}\Big(g_n(z,\bar z)+O(\De^{-1})\Big).
\end{equation}
Here $g_n$ is the classical on-shell action at order $G_N^n$, and the $O(\De^{-1})$ terms are the loops. Since $\De^{n+1}/C_T^{\,n}=\De\,\nu^n$, at fixed $\nu$ every $g_n$ contributes at rate $\De$, and the loops do not.

As alluded before, at first order, the interaction is the exchange of a single graviton between the geodesics $x_1x_2$ and $x_3x_4$. This geodesic Witten diagram equals the conformal block $G_T$ of the stress tensor \cite{Hijano:2015zsa,Castro:2017hpx,Dyer:2017zef}, with its coefficient fixed by the OPE limit:
\begin{equation}
\label{eq:g1}
g_1(z,\bar z)=\omega_d\,G_T(z,\bar z),\qquad \lambda^2_{\cO\cO T}=\omega_d\frac{\De^2}{C_T},\qquad \omega_d=\frac{d^2}{4(d-1)^2},
\end{equation}
where the last formula is the Ward identity \cite{Osborn:1993cr,Poland:2018epd}, with $C_T$ normalized so that a free scalar has $C_T=\frac{d}{d-1}$, and $G_T\sim z^{\frac{d-2}{2}}\bar z^{\frac{d+2}{2}}$ for $0<z\ll\bar z\ll1$.\footnote{In the conventions of \cite[Eqs.~(52)--(54) and (86)]{Poland:2018epd}, $G_T=4g_{d,2}^{0,0}$. Thus the coefficient multiplying our block is one quarter of the coefficient in their Eq.~(86), giving $\omega_d$ in \eqref{eq:g1}.} To this order, $\mathcal{G}_s=e^{\lambda^2_{\cO\cO T}G_T}$, which is the gravitational attraction of the two bodies; this exponentiation is derived in \cite[Sect.~4.2]{Maxfield:2017rkn}. In general, one might expect that
$\mathcal{G}_s \gtrsim e^{\lambda^2_{\cO\cO T}G_T}$ (recall a similar bound holds in $2$ dimension).

If we keep only this first term and if the u-channel term of \eqref{eq:HHHH} stays small in the causal diamond, then at leading order $\log\mathcal G$ is the larger of the logarithms of the s- and t-channel terms, so
\begin{equation}
\label{eq:bodiesrate}
\frac1\De\log\mathcal{G}(z,\bar z)\approx\max\Big(\nu\,\omega_dG_T(z,\bar z),\ x+\bar x+\nu\,\omega_dG_T(1-z,1-\bar z)\Big),
\end{equation}
up to corrections of order $\nu^2$. If $\nu\to0$, the gravitational terms drop out and the correlator is liberated, even though $\mathcal{G}_s$ can be exponentially large, since its exponent $\omega_d\De^2G_T/C_T$ is small compared to $\De$. If instead $\nu$ tends to a small positive constant, gravity contributes at rate $\De$ and liberation fails: the potential is raised by the stress tensor block of the dominant channel, and it is positive at the self-dual point.

When $\nu\to0$, gravity is visible only in the $\gamma$-rescaled potential. Near the self-dual point, both channels are dressed by the same factor $\mathcal G_s(\frac12,\frac12)$, so if $\De^{1+\gamma_c}/C_T\to k$, with $k>0$ and $0<\gamma_c<1$,
\begin{equation}
\label{eq:bodiesshift}
\Lambda^{(\gamma)}(x,\bar x)=
\begin{cases}
\max(0,x+\bar x), & \gamma<\gamma_c,\\[4pt]
\max(0,x+\bar x)+k\,\omega_d\,G_T(\frac12,\frac12), & \gamma=\gamma_c,\\[4pt]
+\infty, & \gamma>\gamma_c.
\end{cases}
\end{equation}
The kink between the channels therefore remains sharp on every scale $\De^{-\gamma}$ with $\gamma<1$. The scalar GFF composite with $L,\Df\to\infty$ also grows at the self-dual point, but there the mixed pairings of its constituents interpolate between the channels, which gives the smooth function of \eqref{eq:fragmentonregimes} at $\gamma_c$.

In $d=2$, one has $C_T=2c$, $\omega_2=1$, and $G_T=k_4(z)+k_4(\bar z)$ with $k_4(z)=z^2{}_2F_1(2,2;4;z)$, and the multi-stress tensors form the Virasoro vacuum module. If this module dominates and the channel decomposition holds, then $\mathcal{G}_s=\mathcal{F}_0(z)\mathcal{F}_0(\bar z)$, and \eqref{eq:bodiesexp} is the large-$c$ expansion of the vacuum block at fixed $\De/c$ \cite{Hartman:2013mia,Fitzpatrick:2014vua}. Proposition~\ref{prop:bodies2d} establishes the obstruction without these assumptions, and its bound is close to the holographic value: the shift $16\rho_*^2k\approx0.471\,k$ of statement (ii) compares with \eqref{eq:bodiesshift}, which in this normalization is $k_4(\frac12)\,k\approx0.477\,k$.

In higher dimensions the multi-stress tensors have no such symmetry. Finite-size effects appear in the OPE as additional exchanged operators, and since each contributes positively to $\mathcal G(\frac12,\frac12)$, they could restore liberation only by lowering the OPE coefficients of the multi-stress tensors. We do not expect this: in heavy-light correlators the lowest-twist coefficients are universal functions of $\De/C_T$ \cite{Fitzpatrick:2019zqz,Karlsson:2019dbd} and exponentiate in the Regge limit \cite{Kulaxizi:2018dxo,Fitzpatrick:2019efk}. The worldline approximation breaks down when $\De/C_T$ is of order one, which is where we expect liberation to be obstructed universally (Sect.~\ref{sec:disc}).

\section{Bootstrap Triples}
\label{sec:triples}

We now formulate the torus and correlator results using only a positive Laplace representation, the normalized identity contribution, and a crossing reflection. The choice of inverse-temperature function records the differences between the applications.

\subsection{Crossing as a reflection}
\label{sec:onereflection}

We first recall what the two problems are, and then write them in common variables.

\paragraph{The modular bootstrap} The reduced partition function $\widehat{Z}(\beta,\bar{\beta})$ of \eqref{partfuncdecomp} is a sum of Boltzmann weights $e^{-\beta h - \bar{\beta}\bar{h}}$, with non-negative multiplicities $n_{h,\bar{h}}$, over states of weights $h,\bar{h}\geq 0$, in which the vacuum contributes $1$. This ensures that $\widehat{Z}\to 1$ in the limit of zero temperature, $\beta,\bar{\beta}\to\infty$. Moreover, the modular relation \eqref{modularrelation} relates $\widehat{Z}$ at one pair of temperatures to $\widehat{Z}$ at another, up to an explicit exponential prefactor that comes from the vacuum energy, and applying it twice returns the original pair. It is an involution of the temperatures. Note that $2\min(h,\bar{h})$ is the twist of a state.

\paragraph{The correlator bootstrap} Next we consider the reduced four-point function $\mathcal{G}(z,\bar{z})$ of Sect.~\ref{sec:fourpoint}. The crossing relation \eqref{crossingrelation} is again an involution, since $(z,\bar{z})\mapsto(1-z,1-\bar{z})$ squares to the identity. Positivity, with the identity normalized, is present as well, and in two different ways: the $\rho$-expansion \eqref{rhoexp} and the $z$-expansion \eqref{zexp}. Each is a sum of non-negative coefficients $a^{(\rho)}_{h,\bar{h}}$ or $a^{(z)}_{h,\bar{h}}$ multiplying the monomials $\rho^h\bar{\rho}^{\bar{h}}$ or $z^h\bar{z}^{\bar{h}}$, in which the identity contributes $a^{(\rho)}_{0,0} = a^{(z)}_{0,0} = 1$. In both expansions, we again call $2\min(h,\bar{h})$ the twist of a monomial.

\paragraph{The two problems together} Each problem carries an involution, but the two involutions do not look alike. The remedy is to change variables so that each becomes a reflection.

We begin with the modular bootstrap and introduce a variable $x$ through
\begin{equation}\label{eq:betamod}
  \beta(x)=-12x+\sqrt{144x^2+4\pi^2},\qquad\text{so that}\qquad \beta(x)\beta(-x)=4\pi^2,\quad \beta(x)-\beta(-x)=-24x ,
\end{equation}
and write $\mathcal{Z}(x,\bar{x})\equiv\widehat{Z}(\beta(x),\beta(\bar{x}))$. The first identity sends the temperatures on the right-hand side of \eqref{modularrelation} to $(\beta(-x),\beta(-\bar{x}))$, and the second turns the prefactor into an exponential of $x+\bar{x}$. The modular relation becomes
\begin{equation}\label{eq:invmod}
  \mathcal{Z}(x,\bar{x})=e^{\frac{x+\bar{x}}{\hbar}}\,\mathcal{Z}(-x,-\bar{x}),\qquad \hbar=\frac1c .
\end{equation}
Modular invariance has become the reflection $(x,\bar{x})\mapsto(-x,-\bar{x})$, and the vacuum energy has become an explicit exponential. One also checks that $\beta'(0)=-12$, and that $s_\infty\equiv\lim_{x\to-\infty}|\beta'(x)|=24$.

For the correlator, we set $\hbar=1/\De$ and
\begin{equation}
  z(x)=\frac{e^{x}}{1+e^{x}},\qquad\text{that is,}\qquad x=\log\frac{z}{1-z}.
\end{equation}
Then $1-z(x)=z(-x)$, so the prefactor in \eqref{crossingrelation} is $e^{(x+\bar{x})/\hbar}$, and crossing takes the form \eqref{eq:invmod}, now with $\mathcal{Z}(x,\bar{x})\equiv\mathcal{G}(z(x),z(\bar{x}))$. In the same variable, the monomials become Boltzmann weights. For the $z$-expansion, we write $z=e^{-\beta_z(x)}$, with
\begin{equation}\label{eq:betaz}
  \beta_z(x)=\log(1+e^{-x}),\qquad \beta_z(x)-\beta_z(-x)=-x,\qquad
  \beta_z'(0)=-\tfrac12,\qquad s_\infty=1 .
\end{equation}
For the $\rho$-expansion, write $\rho=e^{-\beta_\rho(x)}$, with
\begin{equation}\label{eq:betarho}
\beta_\rho(x)=2\log\big(1+\sqrt{1+e^{x}}\big)-x,\qquad \beta'_\rho(0)=-\frac{1}{\sqrt{2}},\qquad s_\infty=1 .
\end{equation}

The two problems are now written in the same way. In each, there is a positive measure $\mu$ on pairs of exchanged weights $(\delta,\bar{\delta})$, such that $\mu({0,0})=1$, and
\begin{equation}
  \mathcal{Z}(x,\bar{x})=\int e^{-\beta(x)\delta-\beta(\bar{x})\bar{\delta}}\,\dd\mu(\delta,\bar{\delta}) .
\end{equation}
For the torus, $(\delta,\bar{\delta})=(h,\bar{h})$, and $\mu$ counts the multiplicities $n_{h,\bar{h}}$. For the correlator, $\mu$ collects the coefficients $a^{(\rho)}_{h,\bar{h}}$ or $a^{(z)}_{h,\bar{h}}$ of the corresponding expansion. In each problem, crossing symmetry is \eqref{eq:invmod}. Counting the two expansions of the correlator separately, there are three applications of what follows. They differ in the function $\beta$ and in the meaning of $\hbar$, and in nothing else.

The Laplace variable $\beta$ makes positivity explicit, while $x$ makes crossing a reflection. We call $\beta(x)$ an \emph{effective inverse temperature}; this is literal for the torus, but for correlators it is a change of kinematic variables. Throughout, $\mathcal{Z}(x,\bar x)=e^{\Phi(x,\bar x)}$ is the reduced object in these coordinates: $\widehat Z(\beta(x),\beta(\bar x))$ on the torus and $\mathcal G(z(x),z(\bar x))$ for the correlator. We use ``potential'' as shorthand for the logarithm $\Phi$, and explicitly include the factor $\hbar$ when discussing its leading rescaled limit.

\subsection{Definitions}
\label{sec:definitions}

With this background, we can define effective inverse temperatures and bootstrap triples in general.

\begin{definition}[Effective inverse temperature]\label{def:beta}
An \emph{effective inverse temperature} is a continuously differentiable bijection $\beta:\mathbb{R}\to(0,\infty)$ that is strictly decreasing and strictly convex, and that obeys
\begin{equation}\label{eq:sinf}
  s_\infty\equiv\lim_{x\to-\infty}|\beta'(x)|\;\le\;2|\beta'(0)| .
\end{equation}
\end{definition}

Increasing $x$ runs through all positive temperatures. Since $\beta'$ is increasing, the limit in \eqref{eq:sinf} exists. Convexity of $\beta$ makes the function $F_a$ of Sect.~\ref{sec:relay} convex (Lemma~\ref{lem:envelope}), which controls the relay. The bound \eqref{eq:sinf} is a technical condition that makes each step of the relay invertible (Lemma~\ref{lem:envelope}(1)). All three applications satisfy it, with equality for the torus and the $z$-expansion.

\enlargethispage{2\baselineskip}
\begin{definition}[Bootstrap triple]\label{def:triple}
A \emph{bootstrap triple} $(\Phi,\hbar,\beta)$ consists of an effective inverse temperature $\beta$, a scale parameter $\hbar>0$, and a potential $\Phi:\mathbb{R}^2\to(0,\infty)$, with $\mathcal{Z}\equiv e^{\Phi}$, such that:
\begin{enumerate}[label=(\roman*)]
\item\label{it:positivity} \emph{(Positivity)} There is a positive measure $\mu$ on $[0,\infty)^2$, the spectral measure, with
\begin{equation}\label{eq:spectral}
  \mathcal{Z}(x,\bar{x})=\int e^{-\beta(x)\delta-\beta(\bar{x})\bar{\delta}}\,\dd\mu(\delta,\bar{\delta})\qquad\text{for all }(x,\bar{x})\in\mathbb{R}^2 .
\end{equation}
\item\label{it:crossing} \emph{(Crossing symmetry)} Under the involution $(x,\bar{x})\mapsto(-x,-\bar{x})$,
\begin{equation}\label{eq:involution}
\mathcal{Z}(x,\bar{x})=e^{\frac{x+\bar{x}}{\hbar}}\,\mathcal{Z}(-x,-\bar{x}),\qquad\text{or equivalently}\qquad  \Phi(x,\bar{x})=\frac{x+\bar{x}}{\hbar}+\Phi(-x,-\bar{x}) .
\end{equation}
\item\label{it:chiral} \emph{(Parity symmetry)} $\Phi(x,\bar{x})=\Phi(\bar{x},x)$; equivalently, $\mu$ is symmetric under $\delta\leftrightarrow\bar{\delta}$.
\item\label{it:identity} \emph{(Identity)} $\mu(\{(0,0)\})=1$.
\end{enumerate}
\end{definition}

In the applications, $\mu$ is supported on discrete points with positive weights, which are multiplicities or expansion coefficients. The identity condition is equivalent to the boundary condition $\mathcal{Z}(x,x)\to1$ as $x\to-\infty$. Indeed, since $\beta(x)\to\infty$ as $x\to-\infty$, dominated convergence in \eqref{eq:spectral} gives $\lim_{x\to-\infty}\mathcal{Z}(x,x)=\mu(\{\delta+\bar{\delta}=0\})$, which is $\mu(\{(0,0)\})$ because $\mu$ is supported on $[0,\infty)^2$. So $\mu$ has a support at the origin with weight $1$ and a non-negative measure $\mu'$ supported away from it, and $\mathcal{Z}\ge1$, $\Phi\ge0$. In fact $\Phi>0$, since $\mu'=0$ would give $\mathcal{Z}\equiv1$, which is incompatible with \eqref{eq:involution}.

For a four-point function of identical scalars, parity symmetry is the symmetry $z\leftrightarrow\bar z$ under exchange of the two lightcone coordinates. For the torus partition function, it is the symmetry $\widehat{Z}(\beta,\bar{\beta})=\widehat{Z}(\bar{\beta},\beta)$ under exchange of the left- and right-moving temperatures. The latter is an additional spectral symmetry and does not follow from $c_L=c_R$ or modular invariance alone. Parity symmetry enters only in Sect.~\ref{sec:strong}, where it brings the crossing image of a mixed point back into the same quadrant (Lemma~\ref{lem:onestep}), and reduces the proof of Theorem~\ref{thm2} to a single quadrant.

The symmetry assumption is convenient for the relay, but is not a restriction on the applications. Indeed, suppose $\mathcal{Z}$ satisfies positivity, crossing, and the identity condition, without assuming parity symmetry. Then
\begin{equation}\label{eq:symmetrization}
 \mathcal{Z}_{\mathrm{sym}}(x,\bar x)=\tfrac12\big(\mathcal{Z}(x,\bar x)+\mathcal{Z}(\bar x,x)\big)
\end{equation}
defines a bootstrap triple with the same $\hbar$ and $\beta$. Its self-dual value and cumulative spectral counts agree with those of $\mathcal{Z}$. Moreover, its low-twist sum is the average of the two low-twist sums, so twist sparseness on the exchange-invariant region $\mathcal D$ is preserved. The elementary lower bound holds for $\mathcal{Z}$ without parity symmetry, and $\mathcal{Z}\leq2\mathcal{Z}_{\mathrm{sym}}$. Consequently, wherever the symmetric object is liberated, the original one is liberated as well. Theorems~\ref{weaklibtorus}, \ref{thm:strong-torus}, \ref{thm:weak-crossing}, and \ref{thm:strong-crossing} therefore do not require additional parity assumption.

With this symmetrization when needed, the modular bootstrap and the correlator bootstrap are instances of Definition~\ref{def:triple}. The dictionary is collected in Table~\ref{tab:dictionary}, together with several derived constants that appear in the theorems.

\begin{table}[t]
\centering
\small\begin{tabular}{@{}lccc@{}}
\toprule
 & modular & $z$-expansion & $\rho$-expansion \\
\midrule
$\hbar$ & $1/c$ & $1/\De$ & $1/\De$ \\
$\beta(x)$ & $-12x+\sqrt{144x^2+4\pi^2}$ & $\log(1+e^{-x})$ & $2\log(1+\sqrt{1+e^{x}})-x$ \\
$e^{\beta(0)}$ & $e^{2\pi}$ & $2$ & $3+2\sqrt{2}$ \\
$|\beta'(0)|$ & $12$ & $1/2$ & $1/\sqrt2$ \\
$s_\infty$ & $24$ & $1$ & $1$ \\
\midrule
$\hbar\langle E\rangle$ & $1/12$ & $2$ & $\sqrt{2}$ \\
$\alpha_*$ & $1/24$ & $1$ & $1/\sqrt{2}$ \\
$\alpha_c$ & $1/48$ & $1/2$ & $1/2$ \\
\bottomrule
\end{tabular}
\caption{The constants of the three applications. The self-dual mean weight $\langle E\rangle$ is defined in Sect.~\ref{sec:weak}, and the distinguished twist cutoffs $\alpha_*$ and $\alpha_c$ in Sect.~\ref{sec:strong}}
\label{tab:dictionary}
\end{table}

\subsection{Sequences, regions, and symmetries}
\label{sec:conventions}

We are interested in the limit $\hbar\to 0$, which we formalize as follows.

\begin{definition}
\label{def:classicalizing}
Throughout, a sequence of triples means a sequence of bootstrap triples with a common effective inverse temperature $\beta$, indexed by a sequence of values of $\hbar$ that decreases strictly to $0$. Since $\beta$ is shared, we denote it by $\{(\Phi,\hbar)\}$.
\end{definition}

All objects, including $\Phi$, $\mathcal{Z}$, and $\mu$, depend on $\hbar$; we suppress this index. Limits are taken along the sequence. The measures need not converge or share any structure.

It will be convenient to name some regions and symmetries of the kinematic plane.

\begin{definition}
\label{def:regions}
Let
\begin{equation}
  \mathcal{D}=\{x+\bar{x}<0\},\qquad \mathcal{C}=\{x<0,\ \bar{x}<0\},\qquad \mathcal{Q}=\mathcal{D}\cap\{x<0<\bar{x}\} .
\end{equation}
Points of $\mathcal{C}$ are \emph{cold}, and points of $\mathcal{Q}\cup\chi(\mathcal{Q})$, where one coordinate is negative and the other positive, are \emph{mixed}. The \emph{crossing} and \emph{parity} maps are
\begin{equation}
  \varsigma(x,\bar{x})=(-x,-\bar{x}),\qquad \chi(x,\bar{x})=(\bar{x},x) ,
\end{equation}
which generate a group $\mathbb{Z}_2\times\mathbb{Z}_2$, the Klein four-group. The \emph{self-dual point} is $(0,0)$, and the \emph{self-dual line} is $\{x+\bar{x}=0\}$, on which the prefactor in \eqref{eq:involution} is trivial. This line is fixed pointwise by $\varsigma\chi$; crossing alone fixes only the origin.
\end{definition}

In the applications, $\mathcal{C}$ is the cold wedge, $\varsigma(\mathcal{C})$ is the hot wedge, and $\mathcal{D}$ is the side of the self-dual hyperbola $\beta\bar{\beta}=4\pi^2$, or of the self-dual line $z+\bar{z}=1$, that contains the cold wedge. The four images of $\mathcal{Q}$ under the group cover the mixed wedges away from the self-dual line. A statement proved on $\mathcal{Q}$ therefore extends to the rest of the mixed wedges, provided that the hypotheses are invariant under the two maps, and they will be.

\subsection{Three elementary facts}
\label{sec:facts}

Everything that follows rests on the three statements of this subsection. The first two hold for any bootstrap triple, and the third for any sequence of triples.

\begin{lemma}
\label{lem:derivative}
For all $(x,\bar{x})\in\mathbb{R}^2$,
\begin{equation}\label{eq:derivbounds}
  0<\partial_x\Phi<\frac1\hbar,\qquad 0<\partial_{\bar{x}}\Phi<\frac1\hbar .
\end{equation}
Moreover, let $\langle\delta\rangle(x,\bar{x})$ denote the mean of $\delta$ in the probability measure
\[
e^{-\beta(x)\delta-\beta(\bar{x})\bar{\delta}-\Phi(x,\bar{x})}\,\dd\mu(\delta,\bar\delta).
\]
Then
\begin{equation}\label{eq:selfdual}
  \partial_x\Phi(0,0)=\partial_{\bar{x}}\Phi(0,0)=\frac1{2\hbar},\qquad
  \langle\delta\rangle\equiv\langle\delta\rangle(0,0)=\frac{\int\delta\,e^{-\beta(0)(\delta+\bar{\delta})}\,\dd\mu}{\mathcal{Z}(0,0)}
  =\frac{1}{2\hbar\,|\beta'(0)|} .
\end{equation}
We write $\langle\delta\rangle$ without arguments only for this value at the self-dual point; likewise $\langle\bar{\delta}\rangle = \langle\delta\rangle$.
\end{lemma}

\begin{proof}
Differentiating \eqref{eq:spectral} gives $\partial_x\Phi=-\beta'(x)\langle\delta\rangle(x,\bar{x})$, and $\langle\delta\rangle(x,\bar{x})\ge0$. As $-\beta'>0$, we get $\partial_x\Phi\ge0$, with equality only if $\mu$ is supported on $\{\delta=0\}$. In that case $\mathcal{Z}$ would be independent of $x$, which is incompatible with the factor $e^{x/\hbar}$ in \eqref{eq:involution}. Hence $\partial_x\Phi>0$. Differentiating \eqref{eq:involution} gives $\partial_x\Phi(x,\bar{x})=1/\hbar-(\partial_x\Phi)(-x,-\bar{x})<1/\hbar$. At the origin, the same identity reads $\partial_x\Phi(0,0)=1/\hbar-\partial_x\Phi(0,0)$, which is the first equality in \eqref{eq:selfdual}. Combining it with $\partial_x\Phi(0,0)=|\beta'(0)|\langle\delta\rangle$ gives the second. The same argument applies to $\bar{x}$.
\end{proof}

\begin{lemma}
\label{lem:apriori}
For all $(x,\bar{x})\in\mathbb{R}^2$,
\begin{equation}\label{eq:apriori}
  \max\Big(0,\frac{x+\bar{x}}{\hbar}\Big)\ \le\ \Phi(x,\bar{x})\ \le\ \Phi(0,0)+\max\Big(0,\frac{x}{\hbar}\Big)+\max\Big(0,\frac{\bar{x}}{\hbar}\Big).
\end{equation}
\end{lemma}

\begin{proof}
The lower bound is $\Phi>0$ together with $\Phi(x,\bar{x})>(x+\bar{x})/\hbar$, which follows from \eqref{eq:involution} and $\Phi(-x,-\bar{x})>0$. For the upper bound, $\Phi$ is increasing in each variable, with slope less than $1/\hbar$, by \eqref{eq:derivbounds}. So $\Phi(x,\bar{x})\le\Phi(\max(x,0),\max(\bar{x},0))\le\Phi(0,0)+\max(0,x)/\hbar+\max(0,\bar{x})/\hbar$.
\end{proof}

\begin{lemma}
\label{lem:lipschitz}
Let $\{(\Phi,\hbar)\}$ be a sequence of triples, and write $\Plib(x,\bar{x})\equiv\max(0,x+\bar{x})$ for the liberated potential.
\begin{enumerate}[label=(\alph*)]
\item For all $p,q\in\mathbb{R}^2$ and all $\hbar$, $|\hbar\Phi(p)-\hbar\Phi(q)|\le\|p-q\|_1$.
\item The set
\begin{equation}\label{eq:L}
  \mathcal{L}\equiv\Big\{p\in\mathbb{R}^2\ \Big|\ \lim_{\hbar\to0}\hbar\Phi(p)=\Plib(p)\Big\},
\end{equation}
which we call the \emph{liberated set}, is closed.
\item If $\hbar\Phi\to g$ pointwise on a set $S\subset\mathbb{R}^2$, then the convergence is uniform on compact subsets of $S$, and $g$ is $1$-Lipschitz on $S$ with respect to the $L^1$ norm.
\end{enumerate}
\end{lemma}

\begin{proof}
(a) Integrate \eqref{eq:derivbounds} along the two coordinate directions.
(b) Let $p_n\in\mathcal{L}$ with $p_n\to p$. Since $\Plib$ is also $1$-Lipschitz,
\begin{align*}
 |\hbar\Phi(p)-\Plib(p)|&\le|\hbar\Phi(p)-\hbar\Phi(p_n)|+|\hbar\Phi(p_n)-\Plib(p_n)|+|\Plib(p_n)-\Plib(p)|\\
 &\le 2\|p_n-p\|_1+|\hbar\Phi(p_n)-\Plib(p_n)| .
\end{align*}
Fix $n$ and let $\hbar\to0$. The last term vanishes because $p_n\in \mathcal{L}$, so $\limsup_{\hbar\to0}|\hbar\Phi(p)-\Plib(p)|\le2\|p_n-p\|_1$. Now let $n\to\infty$.
(c) The bound in (a) passes to the limit, so $g$ is $1$-Lipschitz on $S$, and a pointwise convergent sequence of functions with a common Lipschitz constant converges uniformly on compact sets.
\end{proof}

\section{The Self-Dual Criterion}
\label{sec:weak}

HKS \cite{Hartman:2014oaa} showed that sparseness of the light spectrum implies pointwise liberation. We now formulate this implication for a bootstrap triple and prove its converse, so that sparseness is equivalent to liberation at the self-dual point.

\subsection{Spectral data}
\label{sec:spectraldata}

\begin{definition}%
\label{def:count}
For a bootstrap triple, the cumulative spectral count is the distribution function of the total weight $E=\delta+\bar{\delta}$ under $\mu$,
\begin{equation}
  \mathcal{N}(E)=\mu\big(\{\delta+\bar{\delta}\le E\}\big).
\end{equation}
It is non-decreasing and right-continuous, it vanishes for $E<0$, and $\mathcal{N}(0)=1$ by the identity condition. On the diagonal, $\mathcal{Z}(x,x)=\int_{[0,\infty)}e^{-\beta(x)E}\,\dd\mathcal{N}(E)$. For $E\ge0$, let $\mathcal{S}^{(E)}_\hbar$ be the largest value, over weights $\sigma\le E$, of $\log\mathcal{N}(\sigma)$ shifted by $-\beta(0)\sigma$, where $\beta(0)$ is the self-dual inverse temperature:
\begin{equation}\label{eq:entropy}
  \mathcal{S}^{(E)}_\hbar=\sup_{\sigma\in[0,E]}\hbar\big(\log\mathcal{N}(\sigma)-\beta(0)\sigma\big)\ \ge\ 0 ,
\end{equation}
where the inequality comes from the term $\sigma=0$.
\end{definition}

Here $E$ is a total exchanged weight; the external dimension of the correlator remains $\De=1/\hbar$. In the applications, $\mathcal{N}$ is the cumulative spectral count \eqref{microcanonicalcountmod} of the torus, or one of the counts $\mathcal{N}^{(\rho)}$ and $\mathcal{N}^{(z)}$ of the correlator. The combination $\log\mathcal{N}(\sigma)-\beta(0)\sigma$ compares the cumulative spectral weight below $\sigma$ with the inverse Boltzmann weight at the self-dual temperature, so the statement $\mathcal{S}^{(E)}_\hbar\to0$ says that below the cutoff $E$ the spectrum is sparse in the sense of HKS. Equivalently, $\mathcal{N}(\sigma)\lesssim e^{\beta(0)\sigma}$ for all $\sigma$ up to the cutoff, where $\lesssim$ denotes an inequality up to a factor $e^{o(1/\hbar)}$, uniform in $\sigma$. This is the form in which statement 3 of Theorems~\ref{weaklibtorus} and \ref{thm:weak-crossing} is written.

\begin{definition}
\label{def:meanweight}
By Lemma~\ref{lem:derivative}, the mean total weight at the self-dual point is
\begin{equation}
  \langle E\rangle\equiv\langle\delta+\bar{\delta}\rangle=\frac{1}{\hbar\,|\beta'(0)|} .
\end{equation}
\end{definition}

\subsection{From sparseness to liberation and back}
\label{sec:weakthm}

We can now state the general form of Theorems~\ref{weaklibtorus} and \ref{thm:weak-crossing}.

\begin{theorem}[Self-dual criterion]\label{weaklibthm}
For any sequence of triples $\{(\Phi,\hbar)\}$ with effective inverse temperature $\beta$, the following statements are equivalent.
\begin{enumerate}[label=(\arabic*)]
\item\label{w:global} \emph{(Global bound)} For every $(x,\bar{x})\in\mathbb{R}^2$,
\begin{equation}\label{eq:globalbounds}
  \max(0,x+\bar{x})\ \le\ \liminf_{\hbar\to0}\hbar\Phi(x,\bar{x})\ \le\ \limsup_{\hbar\to0}\hbar\Phi(x,\bar{x})\ \le\ \max(0,x)+\max(0,\bar{x}).
\end{equation}
\item\label{w:diag} \emph{(Free diagonal)} For $x=\bar{x}$, the limit exists and is
\begin{equation}
\lim_{\hbar\to0}\hbar\Phi(x,x)=\max(0,2x)\qquad \text{for all } x\in\mathbb{R}.
\end{equation}
\item\label{w:sparse} \emph{(Sparseness)} For some fixed $\epsilon>0$,
\begin{equation}
\lim_{\hbar\to0}\mathcal S_\hbar^{(\langle E\rangle+\epsilon)}=0.
\end{equation}
\item\label{w:selfdual} \emph{(Self-dual limit)} $\displaystyle\lim_{\hbar\to0}\hbar\Phi(0,0)=0$.
\end{enumerate}
When they hold, $\cl(\mathcal{C})\cup\varsigma(\cl(\mathcal{C}))\subset\mathcal{L}$. That is, the sequence is liberated on the closed cold wedge and on its crossing image, and the convergence is uniform on compact subsets of these regions.
\end{theorem}

The proof relates sparseness to the value of $\Phi$ at the self-dual point through the following estimate.

\begin{proposition}%
\label{prop:selfdual-estimate}
For any bootstrap triple, put $b_0=\beta(0)$ and $m_\hbar=\langle E\rangle=1/(\hbar|\beta'(0)|)$. For every cutoff $E_\hbar>m_\hbar$,
\begin{equation}\label{eq:selfdual-estimate}
 \mathcal S_\hbar^{(E_\hbar)}\leq\hbar\Phi(0,0)
 \leq\mathcal S_\hbar^{(E_\hbar)}+
 \hbar\log\frac{1+b_0E_\hbar}{1-m_\hbar/E_\hbar}.
\end{equation}
In particular, whenever
\begin{equation}\label{eq:cutoff-margin}
 \hbar\log\frac{1+b_0E_\hbar}{1-m_\hbar/E_\hbar}\longrightarrow0,
\end{equation}
sparseness up to $E_\hbar$ is equivalent to the self-dual limit of Theorem~\ref{weaklibthm}.
\end{proposition}

\begin{proof}
    For $\sigma\ge0$, since $e^{-\beta(0)E}\ge e^{-\beta(0)\sigma}$ for $E \in[0,\sigma]$,
\[
  \mathcal{N}(\sigma)\ \le\ e^{\sigma\beta(0)}\int_{[0,\sigma]}e^{-\beta(0)E}\,\dd\mathcal{N}(E)\ \le\ e^{\sigma\beta(0)}\mathcal{Z}(0,0).
\]
Taking a logarithm, multiplying by $\hbar$, and taking the supremum over $\sigma\in[0,E]$ gives the upper bound
\begin{equation}
\mathcal S_\hbar^{(E_\hbar)}\leq\hbar\Phi(0,0)
\end{equation}
In fact, this is true for all $E_\hbar>0$, and hence for $E_\hbar>m_\hbar$.

For the other inequality,
normalize the spectral measure at the self-dual point:
\[
 \dd\mathbb P_*(E)=\mathcal{Z}(0,0)^{-1}e^{-b_0E}\,\dd\mathcal N(E).
\]
By Lemma~\ref{lem:derivative}, its mean is $m_\hbar$. We have $\mathbb P_*(E>E_\hbar)\leq \langle E \rangle/E_\hbar=m_\hbar/E_\hbar$, and hence
\begin{equation}\label{eq:selfdual-markov}
 \mathcal{Z}(0,0)\leq\frac{\mathcal{Z}_L(E_\hbar)}{1-m_\hbar/E_\hbar},\qquad
 \mathcal{Z}_L(E_\hbar)=\int_{[0,E_\hbar]}e^{-b_0E}\,\dd\mathcal N(E).
\end{equation}
Integration by parts, including the identity contribution, gives
\[
 \mathcal{Z}_L(E_\hbar)=e^{-b_0E_\hbar}\mathcal N(E_\hbar)
       +b_0\int_0^{E_\hbar}e^{-b_0E}\mathcal N(E)\,\dd E
 \leq(1+b_0E_\hbar)e^{\mathcal S_\hbar^{(E_\hbar)}/\hbar}.
\]
Combining these inequalities proves the upper bound.
\end{proof}

\begin{proof}[Proof of Theorem~\ref{weaklibthm}]
We prove
\[
 3\Longleftrightarrow4,\qquad 4\Longrightarrow1\Longrightarrow2\Longrightarrow4.
\]
For $3\Rightarrow4$, set $m_\hbar=\langle E\rangle$ and apply Proposition~\ref{prop:selfdual-estimate} with cutoff $m_\hbar+\epsilon$:
\[
 0\leq\hbar\Phi(0,0)\leq
 \mathcal S_\hbar^{(m_\hbar+\epsilon)}
 +\hbar\log\frac{\bigl(1+\beta(0)(m_\hbar+\epsilon)\bigr)(m_\hbar+\epsilon)}{\epsilon}.
\]
The first term tends to zero by statement 3. Since $m_\hbar=1/(\hbar|\beta'(0)|)$, the correction is $O(\hbar\log(1/\hbar))$ and also vanishes. Conversely, the lower bound in Proposition~\ref{prop:selfdual-estimate} gives $4\Rightarrow3$, for every fixed $\epsilon>0$.

Multiplying the elementary bounds \eqref{eq:apriori} by $\hbar$ and taking the limit gives $4\Rightarrow1$. Setting $x=\bar x$ in \eqref{eq:globalbounds} gives $1\Rightarrow2$, and setting $x=0$ in statement 2 gives $2\Rightarrow4$. For the final statement: on $\cl(\mathcal{C})$, the bounds \eqref{eq:apriori} read $0\le\hbar\Phi\le\hbar\Phi(0,0)\to0=\Plib$, and on $\varsigma(\cl(\mathcal{C}))$ we apply \eqref{eq:involution}. Uniformity on compact subsets is Lemma~\ref{lem:lipschitz}(c).
\end{proof}

Theorems~\ref{thm:weak-crossing} and \ref{weaklibtorus} follow with the constants of Table~\ref{tab:dictionary}, where the cutoff $\langle E\rangle+\epsilon$ becomes $c/12+\epsilon$, $2\De+\epsilon$, or $\sqrt2\De+\epsilon$, with self-dual Boltzmann weight $e^{-2\pi}$, $1/2$, or $(3+2\sqrt2)^{-1}$.

\section{Extension to the Mixed Wedges}
\label{sec:strong}

The self-dual criterion does not give a unique limit on the mixed wedges, where one coordinate is hot and the other is cold. There, low-twist spectral weight at large spin can exceed the identity bound even when the light spectrum satisfies the self-dual criterion, so an additional hypothesis is needed. The natural one is that the low-twist states are sparse too. The conclusion will be liberation everywhere outside an explicit open band around the self-dual line, pinched at the self-dual point, which is the undetermined region.

The proof follows the relay algorithm of \cite{Dey:2024nje}.
In Sect.~\ref{sec:twist} we define twist sparseness, and in Sect.~\ref{sec:relay} we develop the relay step. In Sect.~\ref{sec:enigmatic} we characterize when the relay terminates, which determines the undetermined region. We state and prove the extension theorem in Sect.~\ref{sec:strongthm}, and compare our proof with that of \cite{Dey:2024nje} in Remark~\ref{sec:DPQ}.

\subsection{Twist sparseness}
\label{sec:twist}

We begin by splitting $\mathcal{Z}$ according to the twist.

\begin{definition}[{\cite[eqs.~(2.5) and (2.13)]{Dey:2024nje}}]\label{def:twistsplit}
For a bootstrap triple and $\alpha>0$, write
\begin{equation}\label{eq:split}
  \mathcal{Z}=\mathcal{Z}_0^{(\alpha)}+\mathcal{Z}_1^{(\alpha)},\qquad
  \mathcal{Z}_0^{(\alpha)}(x,\bar{x})=\int_{\min(\delta,\bar{\delta})\le\alpha/\hbar} e^{-\beta(x)\delta-\beta(\bar{x})\bar{\delta}}\,\dd\mu(\delta,\bar{\delta}),
\end{equation}
so that $\mathcal{Z}_0^{(\alpha)}$ collects the identity and all states of twist at most $2\alpha/\hbar$, and $\mathcal{Z}_1^{(\alpha)}$ collects the rest.
\end{definition}

In the applications, $\mathcal{Z}_0^{(\alpha)}$ is the partial sum $\widehat{Z}_0^{(\alpha)}$ or $\mathcal{G}_0^{(\alpha)}$ of Sect.~\ref{sec:results}, up to a normalization of $\alpha$ for the torus, which we discuss below. The split has the following elementary property.

\begin{lemma}
\label{lem:twistineq}
For all $x'\ge x$ and all $\bar{x}$,
\begin{equation}\label{eq:twistineq}
  \mathcal{Z}_1^{(\alpha)}(x,\bar{x})\ \le\ e^{-\frac{\alpha}{\hbar}(\beta(x)-\beta(x'))}\,\mathcal{Z}_1^{(\alpha)}(x',\bar{x})
  \ \le\ e^{-\frac{\alpha}{\hbar}(\beta(x)-\beta(x'))}\,\mathcal{Z}(x',\bar{x}),
\end{equation}
and likewise in the second argument.
\end{lemma}

\begin{proof}
On the support of $\mathcal{Z}_1^{(\alpha)}$, both $\delta$ and $\bar{\delta}$ exceed $\alpha/\hbar$, and $\beta(x)-\beta(x')\ge0$ since $\beta$ is decreasing. Hence $e^{-\beta(x)\delta}=e^{-(\beta(x)-\beta(x'))\delta}e^{-\beta(x')\delta}\le e^{-\frac{\alpha}{\hbar}(\beta(x)-\beta(x'))}e^{-\beta(x')\delta}$ pointwise on the support. Multiplying by $e^{-\beta(\bar{x})\bar{\delta}}\,\mathbf{1}_{\min(\delta,\bar{\delta})>\alpha/\hbar}$ and integrating against $\mu$ gives the first inequality, and $\mathcal{Z}_1^{(\alpha)}\le \mathcal{Z}$ gives the second. The inequality in the second argument follows in the same way, using $\bar\delta>\alpha/\hbar$ in place of $\delta>\alpha/\hbar$.
\end{proof}

Raising the cold coordinate from $x$ to $x'$ thus gains a factor $e^{-\frac\alpha\hbar(\beta(x)-\beta(x'))}$ on the high-twist part, while the crossing involution costs a factor $e^{(x'+\bar{x})/\hbar}$ when $x'+\bar{x}>0$. The relay step below chooses $x'$ so that the gain and the cost cancel.

\begin{definition}[{ \cite[eq.~(2.32) and footnote~9]{Dey:2024nje}}]\label{alphatwistsparseness}
A sequence of triples is \emph{$\alpha$-twist sparse} if
\begin{equation}\label{eq:twistsparse}
  \lim_{\hbar\to0}\hbar\log \mathcal{Z}_0^{(\alpha)}(x,\bar{x})=0\qquad\text{for every }(x,\bar{x})\in\mathcal{D} .
\end{equation}
\end{definition}

This is the only hypothesis on the spectrum beyond those of Theorem~\ref{weaklibthm}. On the cold wedge it follows from statement 4 of that theorem, since $1\le \mathcal{Z}_0^{(\alpha)}\le \mathcal{Z}$ there. Its content is on the mixed wedges, where it bounds the total Boltzmann-weighted low-twist contribution. This is a condition throughout $\mathcal D$, not just at the self-dual point, and it constrains the coefficients as well as the spectrum of low-twist operators. For example, $e^{\beta(0)\delta}$ states with $\bar{\delta}=0$ at a weight $\delta\le\langle E\rangle$ of order $1/\hbar$, such as holomorphic currents in the case of the torus, are compatible with statement 4 of Theorem~\ref{weaklibthm}. But at a mixed point whose hot coordinate is $x>0$, they contribute $e^{(\beta(0)-\beta(x))\delta}$ to $\mathcal{Z}_0^{(\alpha)}$, which is exponentially large, so twist sparseness fails. Only \eqref{eq:twistineq} and \eqref{eq:twistsparse} will be used, so any other decomposition of $\mathcal{Z}$ with these two properties would serve as well. Since $\mathcal{Z}_0^{(\alpha')}\le \mathcal{Z}_0^{(\alpha)}$ for $\alpha'\le\alpha$, a sequence that is $\alpha$-twist sparse is $\alpha'$-twist sparse for every $\alpha'<\alpha$. As in Sect.~\ref{sec:results}, $\alpha=0$ imposes no condition on the twist, and by convention every sequence of triples is $0$-twist sparse. Two values of $\alpha$ play a special role.

\begin{definition}%
\label{def:alphas}
For an effective inverse temperature $\beta$, define
\begin{equation}\label{eq:alphas}
  \alpha_*=\hbar\langle\delta\rangle=\frac{1}{2|\beta'(0)|},\qquad \alpha_c=\frac{1}{2s_\infty} .
\end{equation}
The cutoff $\alpha_*$ is sufficient to eliminate the undetermined region. The value $\alpha_c$ marks a change in the geometry of that region.
\end{definition}

Since $|\beta'(0)|<s_\infty\le2|\beta'(0)|$, one has $\alpha_*/2\le\alpha_c<\alpha_*$. For the correlator, the cutoff $\alpha/\hbar=\alpha\De$ on $\min(\delta,\bar\delta)$ matches the normalization of Sect.~\ref{sec:results}, with $\alpha_*^{(z)}=1$, $\alpha_*^{(\rho)}=1/\sqrt{2}$, and $\alpha_c=\frac12$ for both expansions. In the modular case, $\alpha_*/\hbar=c/24$ and $\alpha_c/\hbar=c/48$, which correspond to twist cutoffs $c/12$ and $c/24$. The first is the twist up to which DPQ assumed sparseness to establish vacuum dominance at all temperatures off the self-dual hyperbola \cite{Dey:2024nje}. In Theorem~\ref{thm:strong-torus} we normalized the cutoff on $\min(h,\bar h)$ as $\alpha c/24$, so the $\alpha$ of that theorem is $24\alpha=\alpha/\alpha_*$ in the present normalization, and the values $\alpha_c=\frac12$ and $\alpha_*=1$ quoted there are the ratios $\alpha_c/\alpha_*$ and $\alpha_*/\alpha_*$.

\subsection{The relay}
\label{sec:relay}

The relay is governed by one auxiliary function of a single variable.

\begin{definition}\label{def:envelope}
For $a\ge0$ and an effective inverse temperature $\beta$, let $F_a(t)=t+a\,\beta(t)$.
\end{definition}

We will only need $F_a$ for $a=\alpha$ and $a=2\alpha$.

\begin{lemma}
\label{lem:envelope}
Let $0<\alpha\le\alpha_*$.
\begin{enumerate}[label=(\arabic*)]
\item $F_\alpha'>0$ on $\mathbb{R}$, so $F_\alpha$ is a strictly increasing bijection of $\mathbb{R}$ onto $(m_\alpha,\infty)$ with continuously differentiable inverse, where $m_\alpha\equiv\lim_{t\to-\infty}F_\alpha(t)\in[-\infty,\infty)$.
\item For $a>0$, $F_a$ is strictly convex. In particular, the maximum of $F_a$ on an interval $[t_1,t_2]$ is $\max\{F_a(t_1),F_a(t_2)\}$.
\item $F_{2\alpha}'(0)=1-\alpha/\alpha_*\ge0$ and $\lim_{t\to-\infty}F_{2\alpha}'(t)=1-\alpha/\alpha_c$. Consequently, the inequality $F_{2\alpha}(x)<F_{2\alpha}(0)$ holds for every $x<0$ if $\alpha\le\alpha_c$; for $\alpha_c<\alpha<\alpha_*$, it holds exactly for $t_\alpha<x<0$, where $t_\alpha$ is the unique negative solution of $F_{2\alpha}(t_\alpha)=F_{2\alpha}(0)$; and it holds for no $x<0$ if $\alpha=\alpha_*$.
\end{enumerate}
\end{lemma}

\begin{proof}
(1) Since $\beta$ is strictly convex, $\beta'$ is strictly increasing, with $\beta'>\lim_{t\to-\infty}\beta'(t)=-s_\infty\ge-2|\beta'(0)|=-1/\alpha_*$, so $\alpha\beta'>-\alpha/\alpha_*\ge-1$ and $F_\alpha'=1+\alpha\beta'>0$. As $F_\alpha(t)\to\infty$ for $t\to\infty$, the range is $(m_\alpha,\infty)$. Since $F_\alpha'$ is continuous and positive, the inverse has the continuous derivative $1/F_\alpha'(F_\alpha^{-1})$. (2) The function $t\mapsto t$ is affine and $a\beta$ is strictly convex, and a convex function on an interval takes its maximum at an endpoint.

(3) In this part, write $F=F_{2\alpha}$, so that $F'=1+2\alpha\beta'$. We use two facts about $F'$. First, its values at $0$ and $-\infty$ are
\[
F'(0)=1-2\alpha|\beta'(0)|=1-\frac{\alpha}{\alpha_*},\qquad \lim_{t\to-\infty}F'(t)=1-2\alpha s_\infty=1-\frac{\alpha}{\alpha_c}.
\]
Second, $F'$ is strictly increasing, since $\beta$ is strictly convex. So $F'$ vanishes at most once, and $F$ decreases to the left of that point and increases to the right of it. The three cases are:
\begin{itemize}
\item $\alpha\le\alpha_c$. Then $F'>1-\alpha/\alpha_c\ge0$ everywhere, so $F$ is increasing, and $F(x)<F(0)$ for all $x<0$.
\item $\alpha=\alpha_*$. Then $F'(0)=0$, so $F'<0$ on $(-\infty,0)$, and $F(x)>F(0)$ for all $x<0$.
\item $\alpha_c<\alpha<\alpha_*$. Then $F'$ is negative near $-\infty$ and positive at $0$, so, being continuous, it vanishes at a single point $t_{\min}<0$. On $[t_{\min},0)$, $F$ is increasing, so $F(x)<F(0)$. On $(-\infty,t_{\min}]$, $F$ is decreasing, from $+\infty$ (its slope tends to a negative constant) down to $F(t_{\min})<F(0)$. So it crosses the value $F(0)$ exactly once there, at a point $t_\alpha$, with $F(x)\ge F(0)$ for $x\le t_\alpha$ and $F(x)<F(0)$ for $t_\alpha<x\le t_{\min}$. Combining the two pieces $(t_\alpha,t_{\min}]$ and $[t_{\min},0)$, $F(x)<F(0)$ for all $t_\alpha<x<0$, while $F(x)\ge F(0)$ for $x\le t_\alpha$; in particular $t_\alpha$ is the only negative solution of $F(t)=F(0)$.\qedhere
\end{itemize}
\end{proof}

\begin{definition}\label{def:M}
For $0<\alpha\le\alpha_*$ and $x<0$, let
\begin{equation}
  M_\alpha(x)=\max_{w\in[x,0]}F_{2\alpha}(w)=\max\{F_{2\alpha}(x),F_{2\alpha}(0)\},
\end{equation}
where the second equality follows from Lemma~\ref{lem:envelope}(2).
\end{definition}

We now make the relay mechanism concrete. Fix a mixed point $p=(x,\bar{x})\in\mathcal{Q}$ and $0<\alpha\le\alpha_*$. The relay steps from $p$ conserve the quantity $Q$ defined below, and the cold coordinates of the successive points form the orbit of a map $B$.

\begin{definition}[{Invariant and orbit; for the invariant, cf.\ \cite[eq.~(3.8)]{Dey:2024nje}}]\label{def:orbit}
For $(x,\bar{x})\in\mathbb{R}^2$, define
\begin{equation}\label{eq:charge}
  Q_\alpha(x,\bar{x})=\alpha\beta(x)+F_\alpha(-\bar{x})=\alpha\big(\beta(x)+\beta(-\bar{x})\big)-\bar{x} ,
\end{equation}
and call $Q\equiv Q_\alpha(p)$ the invariant of $p$. Let $B:[x,\infty)\to\mathbb{R}$ be
\begin{equation}\label{eq:B}
  B(w)=F_\alpha^{-1}\big(Q-\alpha\beta(w)\big),
\end{equation}
and the \emph{orbit} of $p$ is the sequence $w_1=x$, $w_{j+1}=B(w_j)$. By construction, $Q=\alpha\beta(w_j)+F_\alpha(w_{j+1})$ for all $j$. We set
\begin{equation}
  N=\min\{j\ge1\ |\ w_{j+1}\ge0\}\in\{1,2,\dots\}\cup\{\infty\},\qquad p_j=(w_j,-w_{j+1}) .
\end{equation}
It is clear that $Q_\alpha(p_j)=Q$ for all $j$.
\end{definition}

\begin{lemma}
\label{lem:orbit}
$B$ is well defined, continuous, and strictly increasing, $w_2=-\bar{x}$, and $\{w_j\}$ is strictly increasing. For $j<N$, the point $p_j$ lies in $\mathcal{Q}$, with $p_1=p$. If $N<\infty$, then $p_N\in\cl(\mathcal{C})$.
\end{lemma}

\begin{proof}
For $w\ge x$, $Q-\alpha\beta(w)\ge Q-\alpha\beta(x)=F_\alpha(-\bar{x})$ lies in the range of $F_\alpha$, so $B(w)$ is defined, and $B$ is continuous and strictly increasing as a composition of the continuous, strictly increasing maps $-\alpha\beta$ and $F_\alpha^{-1}$. Next, $B(x)=F_\alpha^{-1}(F_\alpha(-\bar{x}))=-\bar{x}$, and $-\bar{x}>x$ because $p\in\mathcal{D}$. If $w_{j+1}>w_j$, then $w_{j+2}=B(w_{j+1})>B(w_j)=w_{j+1}$, so $\{w_j\}$ is strictly increasing by induction. For $j<N$, we have $w_{j+1}<0$, hence $w_j<w_{j+1}<0$, which gives $w_j<0<-w_{j+1}$ and $w_j+(-w_{j+1})<0$, so $p_j\in\mathcal{Q}$. Finally, $w_N<0\le w_{N+1}$ gives $p_N=(w_N,-w_{N+1})\in(-\infty,0)\times(-\infty,0]\subset\cl(\mathcal{C})$.
\end{proof}

\begin{lemma}[{Relay equation, \cite[eq.~(3.6)]{Dey:2024nje}}]\label{lem:relay}
Along an orbit, for every $j$,
\begin{equation}\label{eq:relay}
  w_{j+2}-w_{j+1}=\alpha\big(\beta(w_j)-\beta(w_{j+2})\big),
\end{equation}
and $w_{j+2}$ is the unique solution $w'$ of $w'-w_{j+1}=\alpha(\beta(w_j)-\beta(w'))$.
\end{lemma}

\begin{proof}
Conservation of $Q$ at steps $j$ and $j+1$ gives $\alpha\beta(w_j)+w_{j+1}+\alpha\beta(w_{j+1})=\alpha\beta(w_{j+1})+w_{j+2}+\alpha\beta(w_{j+2})$, which is \eqref{eq:relay}. The equation for $w'$ reads $F_\alpha(w')=\alpha\beta(w_j)+w_{j+1}$, and $F_\alpha$ is injective.
\end{proof}

\begin{lemma}[{Single relay step, \cite[proof of Lemma~3.7]{Dey:2024nje}}]\label{lem:onestep}
For $j<N$,
\begin{equation}\label{eq:onestep}
  \mathcal{Z}(p_j)\ \le\ \mathcal{Z}_0^{(\alpha)}(p_j)+\mathcal{Z}(p_{j+1}) .
\end{equation}
\end{lemma}

\begin{proof}
Split $\mathcal{Z}(p_j)=\mathcal{Z}_0^{(\alpha)}(p_j)+\mathcal{Z}_1^{(\alpha)}(p_j)$, so that only the second term needs work. Apply the twist inequality \eqref{eq:twistineq} in the first argument, raising the cold coordinate from $w_j$ to $w_{j+2}>w_j$, then the involution \eqref{eq:involution}, then $(x,\bar{x})\to (\bar{x},x)$ symmetry:
\begin{align}
  \mathcal{Z}_1^{(\alpha)}(w_j,-w_{j+1})
  &\le e^{-\frac\alpha\hbar(\beta(w_j)-\beta(w_{j+2}))}\,\mathcal{Z}(w_{j+2},-w_{j+1})\nonumber\\
  &= e^{-\frac\alpha\hbar(\beta(w_j)-\beta(w_{j+2}))}\,e^{\frac1\hbar(w_{j+2}-w_{j+1})}\,\mathcal{Z}(-w_{j+2},w_{j+1})\nonumber\\
  &= e^{\frac1\hbar\left[(w_{j+2}-w_{j+1})-\alpha(\beta(w_j)-\beta(w_{j+2}))\right]}\,\mathcal{Z}(w_{j+1},-w_{j+2})
   = \mathcal{Z}(p_{j+1}),
\end{align}
where the exponent vanishes by the relay equation \eqref{eq:relay}.
\end{proof}

\subsection{The undetermined region}
\label{sec:enigmatic}

Iterating Lemma~\ref{lem:onestep} bounds $\mathcal{Z}(p)$ by the low-twist contributions along the orbit, plus the value of $\mathcal{Z}$ at the last point. This is useful only if the orbit reaches the closed cold wedge in finitely many steps, that is, if $N<\infty$, and this is decided by the curve $Q_\alpha(x,\bar x)=M_\alpha(x)$.

\begin{lemma}
\label{uniquenessofbdry}
Let $0<\alpha\le\alpha_*$. For every $x<0$, there is exactly one $f_\alpha(x)\in(0,-x]$ such that
\begin{equation}
  Q_\alpha(x,f_\alpha(x))=M_\alpha(x).
\end{equation}
It is given by
\begin{equation}\label{eq:fclosed}
  f_\alpha(x)=\begin{cases}
  -x, & F_{2\alpha}(x)\ge F_{2\alpha}(0),\\[4pt]
  -F_\alpha^{-1}\big(\alpha(2\beta(0)-\beta(x))\big), & \text{otherwise},
  \end{cases}
\end{equation}
and it is continuous in $x$. Moreover, for every $\bar{x}\in\mathbb{R}$, one has $Q_\alpha(x,\bar{x})>M_\alpha(x)$ if and only if $\bar{x}<f_\alpha(x)$.
\end{lemma}

\begin{proof}
Fix $x<0$ and let $\phi(f)=Q_\alpha(x,f)=\alpha(\beta(x)+\beta(-f))-f$. First, $\phi$ is strictly decreasing, since $\phi'(f)=-(1+\alpha\beta'(-f))=-F_\alpha'(-f)<0$ by Lemma~\ref{lem:envelope}(1). Next, $\phi(0)>M_\alpha(x)$. Indeed, $\phi(0)=\alpha(\beta(x)+\beta(0))>2\alpha\beta(0)=F_{2\alpha}(0)$ since $\beta(x)>\beta(0)$, and $\phi(0)-F_{2\alpha}(x)=F_\alpha(0)-F_\alpha(x)>0$ since $F_\alpha$ is strictly increasing. Finally, $\phi(-x)=x+2\alpha\beta(x)=F_{2\alpha}(x)\le M_\alpha(x)$. So $\phi$ is strictly decreasing with $\phi(0)>M_\alpha(x)\ge\phi(-x)$, and there is a unique $f\in(0,-x]$ with $\phi(f)=M_\alpha(x)$; moreover, $\phi(\bar{x})>M_\alpha(x)$ if and only if $\bar{x}<f$.

When $F_{2\alpha}(x)\ge F_{2\alpha}(0)$, we have $\phi(-x)=M_\alpha(x)$, so $f=-x$. Otherwise $M_\alpha(x)=2\alpha\beta(0)$, and the equation $\phi(f)=2\alpha\beta(0)$ reads $F_\alpha(-f)=\alpha(2\beta(0)-\beta(x))$. The inverse is defined here. By the hypothesis of this case, $x+2\alpha\beta(x)<2\alpha\beta(0)$, which is equivalent to $F_\alpha(x)<\alpha(2\beta(0)-\beta(x))$, while $\alpha(2\beta(0)-\beta(x))<\alpha\beta(0)=F_\alpha(0)$. So $\alpha(2\beta(0)-\beta(x))\in(F_\alpha(x),F_\alpha(0))$, and solving for $f$ gives the second line of \eqref{eq:fclosed}, with $f\in(0,-x)$. Finally, where $F_{2\alpha}(x)=F_{2\alpha}(0)$ we have $\alpha(2\beta(0)-\beta(x))=F_\alpha(x)$, so the two branches agree there, and $f_\alpha$ is continuous.
\end{proof}

\begin{definition}[Undetermined region]\label{enigmaticregion}
For $0<\alpha\le\alpha_*$, let
\begin{equation}
A_\alpha = \{ (x,\bar{x})\ |\ x<0,\ f_\alpha(x) \leq \bar{x} \leq - x\}
\end{equation}
be the region between the boundary curve $\bar{x}=f_\alpha(x)$ and the self-dual line in the quadrant $x<0 < \bar{x}$, and let
\begin{equation}
K_\alpha = A_\alpha \cup \chi(A_\alpha) \cup \varsigma(A_\alpha) \cup \varsigma \chi(A_\alpha)
\end{equation}
be its orbit under the group of Definition~\ref{def:regions}. The \emph{undetermined region} is the interior of $K_\alpha$,
\begin{equation}
\mathcal{E}_\alpha = \inter(K_\alpha) = \mathbb{R}^2 \setminus \cl (\mathbb{R}^2 \setminus K_\alpha),
\end{equation}
where $\cl(S)$ denotes the closure of $S$. For $\alpha=0$ we set $f_0\equiv0$, so that $\mathcal{E}_0$ is the union of the two open mixed wedges.
\end{definition}

The convention at $\alpha=0$ is the limit of \eqref{eq:fclosed}, since $F_\alpha^{-1}(\alpha(2\beta(0)-\beta(x)))\to F_0^{-1}(0)=0$ as $\alpha\to0$. We can now state the termination criterion.

\begin{lemma}[{cf.\ \cite[Remark~2.2(f)]{Dey:2024nje}}]\label{terminationlemma}
Let $0<\alpha\le\alpha_*$ and $p=(x,\bar{x})\in\mathcal{Q}$, with orbit as in Definition~\ref{def:orbit}. Then $N<\infty$ if and only if $\bar{x}<f_\alpha(x)$, that is, if and only if $p\notin K_\alpha$.
\end{lemma}

\begin{proof}
By Lemma~\ref{uniquenessofbdry}, $\bar{x}<f_\alpha(x)$ is equivalent to $Q>M_\alpha(x)$, and hence, by the definition of $M_\alpha$, to $Q>F_{2\alpha}(w)$ for every $w\in[x,0]$. Now let $w\ge x$. Since $F_\alpha$ is strictly increasing and $F_\alpha+\alpha\beta=F_{2\alpha}$, the definition of $B$ gives $B(w)>w$ if and only if $Q-\alpha\beta(w)>F_\alpha(w)$, that is, if and only if $Q>F_{2\alpha}(w)$. Comparing the two statements, $\bar{x}<f_\alpha(x)$ if and only if $B(w)>w$ for every $w\in[x,0]$.

Suppose that $B(w)>w$ on $[x,0]$. If $N$ were infinite, the orbit would be a strictly increasing sequence in $[x,0)$, and its limit would be a fixed point of $B$ in $[x,0]$, by continuity. There is none, so $N<\infty$. Suppose instead that $B(w_*)\le w_*$ for some $w_*\in[x,0]$. Then $w_*>x=w_1$, since $B(x)=-\bar{x}>x$. As $B$ is strictly increasing, $w_j<w_*$ implies $w_{j+1}=B(w_j)<B(w_*)\le w_*$. By induction, $w_j<w_*\le0$ for every $j$, so $N=\infty$. Finally, $\chi(A_\alpha)$ and $\varsigma(A_\alpha)$ lie in $\{\bar{x}<0<x\}$, and $\varsigma\chi(A_\alpha)$ lies in $\{x+\bar{x}\ge0\}$, so the only piece of $K_\alpha$ that meets $\mathcal{Q}$ is $A_\alpha$. Hence $p\notin K_\alpha$ if and only if $\bar{x}<f_\alpha(x)$.
\end{proof}

The next proposition gives the undetermined region in closed form and collects the properties described in Sect.~\ref{sec:results}.

\begin{proposition}[{Shape of the undetermined region; cf.\ \cite[Remark~2.2(d) and Sect.~2.4]{Dey:2024nje}}]\label{enigmaticshape}
Let $0<\alpha\le\alpha_*$.
\begin{enumerate}[label=(\arabic*)]
\item \emph{(Closed form)} $\mathcal{E}_\alpha=\mathcal{E}_\alpha^+\cup\chi(\mathcal{E}_\alpha^+)$, where
\begin{equation}\label{eq:Eclosed}
\begin{aligned}
\mathcal{E}^+_\alpha&\equiv\mathcal{E}_\alpha\cap\{x<0<\bar{x}\}\\
&=\Big\{(x,\bar{x})\ \Big|\ x<0<\bar{x}\ \text{ and }\ \max\big(Q_\alpha(x,\bar{x}),\,Q_\alpha(-\bar{x},-x)\big)<2\alpha\beta(0)\Big\}.
\end{aligned}
\end{equation}
\item \emph{(Monotonicity)} For each $x<0$, $f_\alpha(x)$ is non-decreasing in $\alpha$. Consequently, $\mathcal{E}_\alpha\subseteq\mathcal{E}_{\alpha'}$ whenever $\alpha'\le\alpha$.
\item \emph{(Pinching)} $f_\alpha(x)\to0$ as $x\to0^-$, so the boundary curve starts at the self-dual point.
\item \emph{(Three regimes)} For $\alpha\le\alpha_c$, $\mathcal{E}_\alpha$ contains every point of the self-dual line except the self-dual point. For $\alpha_c<\alpha<\alpha_*$, a point $(x,-x)$ of the self-dual line with $x<0$ lies in $\mathcal{E}_\alpha$ if and only if $x>t_\alpha$, and $\mathcal{E}_\alpha$ is contained in the square $|x|,|\bar{x}|<|t_\alpha|$, so that its closure is compact. For $\alpha=\alpha_*$, the undetermined region is empty.
\end{enumerate}
\end{proposition}

\begin{proof}
(1) The pieces $\chi(A_\alpha)$ and $\varsigma(A_\alpha)$ lie in $\{\bar{x}<0<x\}$, so $K_\alpha\cap\{x<0<\bar{x}\}=A_\alpha\cup\varsigma\chi(A_\alpha)$, and $\mathcal{E}_\alpha$ is the union of $\mathcal{E}_\alpha^+$ and its image under coordinate exchange. We use two facts. First, for $p=(x,\bar{x})\in\mathcal{Q}$, one has $Q_\alpha(\varsigma\chi(p))=Q_\alpha(p)+x+\bar{x}<Q_\alpha(p)$. Second, since $Q_\alpha(x,\cdot)$ is strictly decreasing, $Q_\alpha(p)>Q_\alpha(x,-x)=F_{2\alpha}(x)$. Now take $p\in\mathcal{Q}$. Since $p$ is off the self-dual line and $\varsigma\chi(A_\alpha)\subset\{x+\bar{x}\ge0\}$, $p\in\mathcal{E}_\alpha$ if and only if $p$ is interior to $A_\alpha$, which by continuity of $f_\alpha$ holds if and only if $\bar{x}>f_\alpha(x)$, that is, $Q_\alpha(p)<M_\alpha(x)$. If $Q_\alpha(p)<2\alpha\beta(0)$, then $F_{2\alpha}(x)<Q_\alpha(p)<F_{2\alpha}(0)$, so $M_\alpha(x)=2\alpha\beta(0)$ and $p\in\mathcal{E}_\alpha$. Conversely, if $p\in\mathcal{E}_\alpha$, then $F_{2\alpha}(x)<F_{2\alpha}(0)$, since otherwise $f_\alpha(x)=-x>\bar{x}$; so $M_\alpha(x)=2\alpha\beta(0)$ and $Q_\alpha(p)<2\alpha\beta(0)$. By the first fact, the maximum in \eqref{eq:Eclosed} is $Q_\alpha(p)$, which proves \eqref{eq:Eclosed} on $\mathcal{Q}$. Points with $x<0<\bar{x}$ and $x+\bar{x}>0$ are images $\varsigma\chi(p)$ of points $p\in\mathcal{Q}$, and both $\mathcal{E}_\alpha$ and the condition in \eqref{eq:Eclosed} are invariant under $\varsigma\chi$. Finally, consider a point $(x,-x)$ of the self-dual line with $x<0$, where both arguments of the maximum equal $F_{2\alpha}(x)$. If $F_{2\alpha}(x)<F_{2\alpha}(0)$, then by continuity $f_\alpha(x')<-x'$ for $x'$ near $x$, so a neighborhood of $(x,-x)$ is contained in $A_\alpha\cup\varsigma\chi(A_\alpha)$. If instead $F_{2\alpha}(x)\ge F_{2\alpha}(0)$, then $f_\alpha(x)=-x$, and the points $(x,-x-\epsilon)$ with $\epsilon>0$ lie outside $K_\alpha$. This proves \eqref{eq:Eclosed}.

(2) If $F_{2\alpha}(x)\ge F_{2\alpha}(0)$, then $f_\alpha(x)=-x$, and since $F_{2\alpha}(x)-F_{2\alpha}(0)=x+2\alpha(\beta(x)-\beta(0))$ is increasing in $\alpha$, the same holds for every $\alpha'\geq\alpha$. Otherwise, $y=-f_\alpha(x)\in(x,0)$ solves $G(y,\alpha)\equiv y+\alpha\beta(y)-\alpha(2\beta(0)-\beta(x))=0$. We have $\partial_yG=F_\alpha'(y)>0$ and $\partial_\alpha G=\beta(y)+\beta(x)-2\beta(0)>0$, since $x,y<0$. Since $G$ is continuously differentiable, the implicit function theorem gives $\partial_\alpha y=-\partial_\alpha G/\partial_y G<0$, so $f_\alpha(x)$ increases with $\alpha$ until it reaches $-x$, where the first case takes over by continuity. Since $A_\alpha$ shrinks as $f_\alpha$ grows, so do $K_\alpha$ and $\mathcal{E}_\alpha$.

(3) This is immediate from $0<f_\alpha(x)\le-x$.

(4) By \eqref{eq:Eclosed}, a point $(x,-x)$ with $x<0$ lies in $\mathcal{E}_\alpha$ if and only if $F_{2\alpha}(x)<F_{2\alpha}(0)$. By Lemma~\ref{lem:envelope}(3), this holds for all $x<0$ when $\alpha\le\alpha_c$, exactly for $x>t_\alpha$ when $\alpha_c<\alpha<\alpha_*$, and never when $\alpha=\alpha_*$. For $\alpha_c<\alpha<\alpha_*$, every $p\in\mathcal{E}_\alpha\cap\mathcal{Q}$ has $F_{2\alpha}(x)<Q_\alpha(p)<F_{2\alpha}(0)$, so $t_\alpha<x<0<\bar{x}<-x<|t_\alpha|$, and boundedness follows from the symmetries. For $\alpha=\alpha_*$, the same inequality $F_{2\alpha_*}(x)<F_{2\alpha_*}(0)$ would be required at every point of $\mathcal{E}_{\alpha_*}\cap\mathcal{Q}$, which is impossible for $x\neq 0$. Hence $\mathcal{E}_{\alpha_*}$ is empty.
\end{proof}

\paragraph{The asymptotic cone} When $0<\alpha<\alpha_c$, the band extends to infinity along the self-dual line, and it is natural to ask how wide it is far from the self-dual point. The answer depends only on the slope $s_\infty$ of the effective inverse temperature at low temperature. Define
\begin{equation}\label{eq:lambda}
\lambda_\alpha=\frac{\alpha s_\infty}{1-\alpha s_\infty}=\frac{\alpha}{2\alpha_c-\alpha},
\end{equation}
which increases from $0$ to $1$ as $\alpha$ runs from $0$ to $\alpha_c$.

\begin{proposition}%
\label{enigmaticasymptotics}
Let $0<\alpha<\alpha_c$. Then
\begin{equation}
\lim_{x\to-\infty}\frac{f_\alpha(x)}{|x|}=\lambda_\alpha .
\end{equation}
Consequently, at large $|x|$ and $|\bar{x}|$, the component $\mathcal{E}_\alpha^+$ is asymptotic to the open cone $\lambda_\alpha|x|<\bar{x}<|x|/\lambda_\alpha$. That is, for every $m>0$ with $m\neq\lambda_\alpha,1/\lambda_\alpha$, the point $(x,m|x|)$ lies in $\mathcal{E}_\alpha$ for all sufficiently large $|x|$ if $\lambda_\alpha<m<1/\lambda_\alpha$, and lies outside $\mathcal{E}_\alpha$ for all sufficiently large $|x|$ otherwise. The same holds for $\chi(\mathcal{E}_\alpha^+)$, with the roles of $x$ and $\bar{x}$ exchanged.
\end{proposition}

\begin{proof}
For $\alpha\le\alpha_c$, Lemma~\ref{lem:envelope}(3) gives $F_{2\alpha}(x)<F_{2\alpha}(0)$ for every $x<0$, so the second branch of \eqref{eq:fclosed} applies, and $y\equiv-f_\alpha(x)\in(x,0)$ solves
\begin{equation}
y+\alpha\beta(y)=\alpha\big(2\beta(0)-\beta(x)\big).
\end{equation}
First, $\beta(t)/|t|\to s_\infty$ as $t\to-\infty$, and $\beta'(t)\to-s_\infty$. Second, $y\to-\infty$ as $x\to-\infty$: the right-hand side tends to $-\infty$, while the left-hand side is $F_\alpha(y)$, which is increasing and hence bounded below by $F_\alpha(t_0)$ whenever $y\ge t_0$. Now divide both sides by $|x|$:
\begin{equation}
\frac{|y|}{|x|}\Big(\alpha\frac{\beta(y)}{|y|}-1\Big)=\frac{2\alpha\beta(0)}{|x|}-\alpha\frac{\beta(x)}{|x|}.
\end{equation}
As $x\to-\infty$, the right-hand side tends to $-\alpha s_\infty$, and the bracket on the left tends to $-(1-\alpha s_\infty)$, which is non-zero even at $\alpha=\alpha_c$, since $\alpha s_\infty\leq\frac12$. Hence $|y|/|x|\to\alpha s_\infty/(1-\alpha s_\infty)=\lambda_\alpha$, which is the first statement. The same argument gives $f_{\alpha_c}(x)/|x|\to1$ at the critical cutoff.

For the second, take $p=(x,m|x|)$ with $x<0$. If $m<1$, then $p\in\mathcal{Q}$, and by the proof of Proposition~\ref{enigmaticshape}(1), $p\in\mathcal{E}_\alpha$ if and only if $m>f_\alpha(x)/|x|$. Since the right-hand side tends to $\lambda_\alpha$, this holds for all large $|x|$ if $m>\lambda_\alpha$, and fails for all large $|x|$ if $m<\lambda_\alpha$. If $m>1$, then $\varsigma\chi(p)=(-m|x|,|x|)\in\mathcal{Q}$, and since $\mathcal{E}_\alpha$ is $\varsigma\chi$-invariant, $p\in\mathcal{E}_\alpha$ if and only if $1/m>f_\alpha(-m|x|)/(m|x|)$, whose right-hand side again tends to $\lambda_\alpha$. Finally, if $m=1$, then $p$ lies on the self-dual line, and it is in $\mathcal{E}_\alpha$ by Proposition~\ref{enigmaticshape}(4). The statement for $\chi(\mathcal{E}_\alpha^+)$ follows by coordinate-exchange symmetry.
\end{proof}

The opening of the asymptotic cone depends only on $\alpha s_\infty$, and the example of Proposition~\ref{prop:JJ} shows that it cannot be narrowed. It closes onto the self-dual line as $\alpha\to\alpha_c$, where $\lambda_\alpha\to1$. At $\alpha=\alpha_c$ itself, the band still extends along the whole self-dual line, but its width grows sublinearly, and its shape depends on the subleading behavior of $\beta$ at low temperature. The apex of the band is governed instead by the slope of $\beta$ at the self-dual point, as the following lemma shows.

\begin{lemma}
\label{lem:position}
Let $0<\alpha<\alpha_c$. On $(-\infty,0)$, the ratio $f_\alpha(x)/|x|$ is strictly increasing in $|x|$, with limit $\lambda^*_\alpha=\frac{\alpha}{2\alpha_*-\alpha}$ as $x\to0^-$ and $\lambda_\alpha=\frac{\alpha}{2\alpha_c-\alpha}$ as $x\to-\infty$. In particular, $\lambda^*_\alpha|x|<f_\alpha(x)<\lambda_\alpha|x|$ for all $x<0$.
\end{lemma}

\begin{proof}
For $\alpha\leq\alpha_c$, Lemma~\ref{lem:envelope}(3) and \eqref{eq:fclosed} give $f_\alpha(-s)=-y(s)$ for $s>0$, with
\[
y(s)=F_\alpha^{-1}\big(\alpha(2\beta(0)-\beta(-s))\big).
\] Since $\beta$ is strictly convex, the argument of $F_\alpha^{-1}$ is a strictly concave function of $s$; since $F_\alpha$ is strictly increasing and convex, $F_\alpha^{-1}$ is increasing and concave. So $y$ is strictly concave, with $y(0)=0$, and $y(s)/s$ is strictly decreasing. Its limit as $s\to0^+$ is $y'(0)=-\alpha|\beta'(0)|/(1-\alpha|\beta'(0)|)$, by the chain rule and Lemma~\ref{lem:envelope}(1), which is $-\lambda^*_\alpha$, and its limit as $s\to\infty$ is $-\lambda_\alpha$, by Proposition~\ref{enigmaticasymptotics}.
\end{proof}

Thus the boundary of the undetermined region leaves the self-dual point at the slope $\lambda^*_\alpha$, which is set by $|\beta'(0)|$, and turns monotonically toward the asymptotic slope $\lambda_\alpha$, which is set by $s_\infty$. The two ends of the band are controlled by two different numbers: the slope of the effective inverse temperature at the self-dual point, and its slope at zero temperature.

For the correlator, $s_\infty=1$ in both expansions, so $\lambda_\alpha=\alpha/(1-\alpha)$. Since $x\simeq\log z$ and $\bar x\simeq-\log(1-\bar z)$ as $z\to0$, $\bar z\to1$, both undetermined regions reach this double lightcone limit along the band
\begin{equation}
z^{\frac{1-\alpha}{\alpha}}\ \lesssim\ 1-\bar{z}\ \lesssim\ z^{\frac{\alpha}{1-\alpha}},
\end{equation}
at leading order in the logarithms. Near the self-dual point they differ: $\lambda^*_\alpha=\alpha/(\sqrt2-\alpha)$ for the $\rho$-expansion and $\alpha/(2-\alpha)$ for the $z$-expansion. For the torus, $s_\infty=24$, and in the normalization of Theorem~\ref{thm:strong-torus} the same slope gives the large-$\beta$ behavior $b_\alpha(\beta)\simeq\frac{1-\alpha}{\alpha}\frac{4\pi^2}{\beta}$ quoted in Sect.~\ref{sec:torus}.

\subsection{The extension theorem}
\label{sec:strongthm}

We can now state the general form of Theorems~\ref{thm:strong-torus} and \ref{thm:strong-crossing}.

\begin{theorem}[{Extension under twist sparseness; cf.\ \cite[Corollary~2.5]{Dey:2024nje}}]
\label{thm2}
Let $\{(\Phi,\hbar)\}$ be a sequence of triples that obeys the equivalent conditions of Theorem~\ref{weaklibthm}. If the sequence is $\alpha$-twist sparse for some $\alpha \in [0, \alpha_*]$, where $\alpha_* = \hbar \langle \delta \rangle = 1/(2|\beta'(0)|)$, then
\begin{equation}
\lim_{\hbar \to 0} \hbar \Phi(x,\bar{x}) = \max(0,x+\bar{x}) \qquad \text{for all } (x,\bar{x}) \in \mathbb{R}^2 \setminus\mathcal{E}_\alpha,
\end{equation}
where $\mathcal{E}_\alpha \subset\mathbb{R}^2$ is the undetermined region of Definition~\ref{enigmaticregion}. The convergence is uniform on compact subsets of $\mathbb{R}^2\setminus\mathcal{E}_\alpha$.
\end{theorem}

\begin{proof}
For $\alpha=0$, the complement of $\mathcal{E}_0$ is $\cl(\mathcal{C})\cup\varsigma(\cl(\mathcal{C}))$, and the claim is the final statement of Theorem~\ref{weaklibthm}. So let $0<\alpha\le\alpha_*$. We first prove liberation on $\mathbb{R}^2 \setminus K_\alpha$, in three cases, and then extend it to $\mathbb{R}^2 \setminus \mathcal{E}_\alpha$ by closure.

\emph{Cold points and their images.} On $\cl(\mathcal{C})\cup\varsigma(\cl(\mathcal{C}))$, which contains the self-dual point, the claim is the final statement of Theorem~\ref{weaklibthm}.

\emph{Mixed points of $\mathcal{D}$ outside $K_\alpha$.} Since both the potential and $K_\alpha$ are invariant under $\chi$, we may take $p\in\mathcal{Q}$. Lemma~\ref{terminationlemma} gives $N<\infty$, and iterating Lemma~\ref{lem:onestep} along the points $p_1,\dots,p_N$ of Lemma~\ref{lem:orbit} gives a chain of $N$ positive terms,
\begin{equation}
\mathcal{Z}(p) \leq \sum_{j=1}^{N-1} \mathcal{Z}_0^{(\alpha)}(p_j) + \mathcal{Z}(p_N).
\end{equation}
Bounding the sum by $N$ times its largest term,
\begin{equation}
\hbar \Phi(p) \leq \hbar \log N + \max\Big\{ \max_{j < N} \hbar \log \mathcal{Z}_0^{(\alpha)}(p_j),\ \hbar \log \mathcal{Z}(p_N) \Big\}.
\end{equation}
Now let $\hbar \to 0$. The number $N$ and the orbit depend only on $p$, $\alpha$, and $\beta$, and not on $\hbar$. So $\hbar \log N \to 0$, and the maximum is over a fixed finite list. The $N-1$ terms $\hbar\log \mathcal{Z}_0^{(\alpha)}(p_j)$ tend to $0$ by $\alpha$-twist sparseness, which applies since $p_j\in\mathcal{Q}\subset\mathcal{D}$ for $j<N$ (Lemma~\ref{lem:orbit}), and the last term tends to $0$ by the first case, since $p_N\in\cl(\mathcal{C})$. Therefore $\limsup_{\hbar\to 0} \hbar \Phi(p) \leq 0$, and with the lower bound of Lemma~\ref{lem:apriori}, $\lim_{\hbar\to 0 } \hbar \Phi(p) = 0=\max(0,x+\bar{x})$.

\emph{Points with $x+\bar{x}>0$ outside $K_\alpha\cup\varsigma(\cl(\mathcal{C}))$.} Since $K_\alpha$ is $\varsigma$-invariant, $\varsigma(p)$ also lies outside $K_\alpha$, and it is a mixed point of $\mathcal{D}$, to which the previous case applies. The involution then gives $\hbar \Phi(p) = (x+\bar{x}) + \hbar \Phi( \varsigma(p)) \to x+ \bar{x} = \max(0,x+\bar{x})$.

Every point of $\mathbb{R}^2 \setminus K_\alpha$ falls into one of these cases, since the self-dual line, apart from the self-dual point, lies in $A_\alpha\cup\chi(A_\alpha)\subset K_\alpha$. Therefore $\mathbb{R}^2 \setminus K_\alpha \subset \mathcal{L}$. The liberated set $\mathcal{L}$ is closed by Lemma~\ref{lem:lipschitz}(b), so
\begin{equation}
\mathbb{R}^2 \setminus \mathcal{E}_\alpha = \mathbb{R}^2 \setminus \inter(K_\alpha) = \cl( \mathbb{R}^2 \setminus K_\alpha ) \subset \mathcal{L}.
\end{equation}
Uniform convergence on compact subsets then follows from Lemma~\ref{lem:lipschitz}(c).
\end{proof}

The closure step is where Theorem~\ref{thm2} improves on the result of DPQ, as we discuss in Remark~\ref{sec:DPQ}.

\begin{remark}
   Theorem~\ref{thm2} reproduces Theorems~\ref{thm:strong-torus} and \ref{thm:strong-crossing} using the appropriate $\beta$ variables, see Table~\ref{tab:dictionary}.
\end{remark}

\begin{remark}[Comparison with the proof of DPQ \cite{Dey:2024nje}]
\label{sec:DPQ}
Theorem~\ref{thm2} reorganizes the constructive proof of DPQ. Their algorithm applies the twist inequality, crossing, and coordinate exchange, choosing the next point to cancel the exponential prefactor, which is Lemma~\ref{lem:onestep}. To characterize the starting points from which the algorithm terminates, DPQ filter them by the number of steps and follow individual trajectories through the filtration. These trajectories conserve the invariant $Q$ of Definition~\ref{def:orbit}, and they are our orbits read backwards. The boundary of the terminating set is then a limit of the boundaries of the filtration, with two branches, and defining this limit and its branches requires control of the starting points of the trajectories. Our proof instead fixes the starting point and uses the invariant to define the map $B$. By monotonicity, the orbit terminates exactly when $B$ has no fixed point in $[x,0]$, that is, when $Q_\alpha(p)>M_\alpha(x)$ (Lemma~\ref{terminationlemma}), and the two branches of DPQ are the two terms of $M_\alpha(x)=\max\{F_{2\alpha}(x),F_{2\alpha}(0)\}$.

The additional boundary points come from the Lipschitz bound. In either formulation the relay excludes the non-terminating set $K_\alpha$, including its boundary, and Lemma~\ref{lem:lipschitz}(b) makes the liberated set closed, so that it contains these boundary points as well.
\end{remark}
\FloatBarrier
\section{Discussion}
\label{sec:disc}
In this work, we asked when the four-point function of identical scalars of dimension $\De$ behaves like a GFF, in particular, obeys $\log\mathcal G=\log\mathcal G^\GFF+o(\De)$ as $\De\to\infty$ at fixed cross ratios, as if they were describing heavy particles propagating along free bulk geodesics. We call this behavior liberation in the heavy limit.

We showed that sparseness in the low-lying spectra is equivalent to liberation at the self-dual point, and that this forces liberation wherever $z$ and $\bar z$ lie on the same side of $\frac12$. Sparseness in the low-twist spectrum extends liberation to the mixed wedges ($z<1/2<\bar{z}$ and $\bar{z}<1/2<z$), except in an undetermined region whose boundary we found in closed form. Within this undetermined region, we exhibited a mixed phase that arises in correlators of scalar composites of large-spin operators. The same arguments we used to prove the main theorems sharpen the torus results of \cite{Hartman:2014oaa,Dey:2024nje}.

In two dimensions, Proposition~\ref{prop:bodies2d} shows that liberation fails once $\De$ is comparable to $c$: the vacuum module alone grows too fast and positivity prevents the other modules from canceling it. In AdS$_3$ this is the statement that heavy particles attract gravitationally at order $\De\times(\De/c)$. The reference \cite{Fitzpatrick:2014vua} derived the same attraction from the bootstrap in the heavy-light limit, where the vacuum block produces the deficit-angle spectrum of large-spin double twists, assuming that this block dominates at large spin. Proposition~\ref{prop:bodies2d} uses neither assumption.

In higher dimensions we expect the same, but we have only the worldline argument of Sect.~\ref{ex:bodies}, which breaks down when $\De\sim C_T$: finite-size effects \cite{Goldberger:2004jt,Porto:2016pyg} delocalize the body, the expansion \eqref{eq:bodiesexp} is no longer useful, and the two bodies can collapse into a black hole. Each of these breakdowns seems only to make matters worse for liberation. A body that is not localized on a worldline, such as a gas cloud or a large black hole, can exchange its constituents or microstates, which contribute positively to the correlator, as for the composites of Sect.~\ref{ex:gff} at fixed $\Df$ and for black-hole microstates in two dimensions. What a proof needs is control of the multi-stress tensors $[T^n]$, whose OPE coefficients, unlike that of $T$, are not fixed by symmetry in $d>2$. A bootstrap argument that the stress tensor block approximately exponentiates, through the large-spin trajectories of its composites, would close this gap.

Instead, one could look for a counterexample: a correlator of operators with $\De\sim C_T\to\infty$ that \emph{is} liberated. No such four-point function is known in higher dimensions, even holographically. Two- and three-point functions of such operators have been computed from the geometries they source \cite{Abajian:2023jye,Abajian:2023bqv}, and for half-BPS operators in $\mathcal N=4$ SYM from matrix models \cite{Holguin:2023orq,Kazakov:2024ald,Anempodistov:2025maj}. Extending these methods to four points would test the obstruction directly.

\looseness=-1 When the undetermined region first appeared, for the torus \cite{Dey:2024nje}, it was unclear whether any theory populates it. For four-point functions the spinning composites answer this far from the self-dual point; for the torus it remains open. Near the self-dual point it is open for both: the mixed phase leaves it with slope $\lambda_\alpha$, the undetermined region with the smaller slope $\lambda^*_\alpha$ (Lemma~\ref{lem:position}), and nothing we know fills the wedge between them. For $\alpha_c<\alpha<\alpha_*$ the undetermined region is bounded, and we have no example at all. Our composites cannot enter this window: with two constituents of equal dimension, the exchanged operators that produce the mixed phase have twist at most $\De$, so the composites are twist sparse only for $\alpha<\alpha_c$, and we do not know whether the full cutoff $\alpha_*$ is needed. We expect any failure of liberation in the undetermined region to work as ours does, through exponentially large OPE weight in low-twist states of large spin, whose constituents could be large-spin double-twist operators or, holographically, spinning strings. Interactions between them should matter most near the self-dual point, where the competing pairings are closest in weight, and might widen the mixed phase.

Identical scalars are only the simplest setting. For $\langle\cO_1\cO_1\cO_2\cO_2\rangle$ with dimensions growing at different rates, one can ask when the leading logarithm is given by the geodesics joining $x_1$ to $x_2$ and $x_3$ to $x_4$. The answer should depend on what is assumed about the three OPEs and on the ratio of the dimensions. Positivity also ties this correlator to $\langle\cO_1\cO_1\cO_1\cO_1\rangle$ and $\langle\cO_2\cO_2\cO_2\cO_2\rangle$, and it would be interesting to know whether liberation of the latter two implies liberation of the first. External operators of large spin are another direction, in which the polarizations become new parameters on which the phases can depend.

Our proofs use only positivity, a reflection, a parity symmetry, and an identity contribution normalized to one, which is why the correlator and the torus could be treated together. Any observable with these properties is a bootstrap triple, to which our theorems apply unchanged. Since Proposition~\ref{prop:selfdual-estimate} holds at finite $\hbar$, one can also hope to combine it with Tauberian methods to control the heavy spectrum beyond the leading logarithm. Finally, in the flat-space limit of gapped theories in AdS \cite{vanRees:2022zmr}, liberation is the leading-order condition for the disconnected terms to dominate, a first step toward extracting the scattering amplitude.

\section*{Acknowledgement}
We thank Tom Hartman, Shiraz Minwalla, Onkar Parikkar, David Poland, Balt van Rees, and Pedro Vieira for discussions. ChatGPT Astra 6 and Claude Opus 5.5 were used to build figures, give extensive adversarial reviews of the draft and revision suggestions, as well as develop the current composite example in Appendix~\ref{app:local-JJ} and related content in section $3$. The work of GR is supported by DOE grant DE-SC0017660. The work of SP is supported by the Department of Atomic Energy, Government of India, under Project Identification Number RTI-4012 and the Infosys Endowment for the study of the Quantum Structure of Spacetime. SP would also like to acknowledge his debt to the people of India for their steady support of the study of basic science.

\appendix
\titleformat{\section}{\normalfont\bfseries}{Appendix \thesection.}{0.5em}{}
\renewcommand{\thesubsection}{\thesection.\arabic{subsection}}

\section{Proofs for Sect.~\ref{sec:examples}}
\label{app:examples}

We use the notation of Sects.~\ref{sec:examples} and \ref{sec:triples}: $x=\log\frac{z}{1-z}$ and $\bar x=\log\frac{\bar z}{1-\bar z}$, $\mathcal{Z}(x,\bar x)=\mathcal G(z(x),z(\bar x))$, $\hbar=1/\De$, the effective inverse temperatures $\beta_\rho$ and $\beta_z$ of Table~\ref{tab:dictionary}, and the regions and maps of Definition~\ref{def:regions}.

\subsection{Composites of a generalized free field}
\label{threeheavylimits}

\begin{proof}[Proof of Proposition~\ref{prop:GFF}]
The two sides of \eqref{eq:GFFsandwich} differ by the factor $\binom{L+2}{2}=e^{o(\De)}$, so $\frac1\De\log\mathcal{G}^{(\Df,L)}=\frac{2}{\Df}\log \mathcal H+o(1)$. This is \ref{gff:Df}; the inequality there is strict because $\mathcal H$ exceeds each of its terms. When $\Df\to\infty$, since $z\bar z<1$,
\[
\frac{2}{\Df}\log \mathcal H\to\max\Big(0,\log\frac{z\bar z}{(1-z)(1-\bar z)}\Big),
\]
which gives liberation in \ref{gff:L} and \ref{gff:frag}. For the gap in \ref{gff:L}, note that every term of \eqref{eq:GFFclosed} other than the identity, $a=L$, contains $(z\bar z)^{\Df}$, which is $(\rho\bar\rho)^{\Df}$ times a power series in $\rho,\bar\rho$, since $z=4\rho/(1+\rho)^2$ and $1-z=\big(\frac{1-\rho}{1+\rho}\big)^2$. So every non-identity monomial has $h+\bar h\geq2\Df=2\De/L$.

For saturation in \ref{gff:frag}, treat both expansions at once. Let $\beta$ be $\beta_\rho$ or $\beta_z$, and $\mathcal N$ the corresponding count, so that $\mathcal{Z}(x,x)=\int e^{-\beta(x)E}\,\dd\mathcal N(E)$ (Definition~\ref{def:count}). By Table~\ref{tab:dictionary}, $e^{\beta(0)}$ is $3+2\sqrt2$ or $2$, and the cutoff in \eqref{eq:GFFsaturation} is $\langle E\rangle+\epsilon$, where $\langle E\rangle=\De/|\beta'(0)|$, which is $\sqrt2\De$ or $2\De$, is the mean weight at the self-dual point (Definition~\ref{def:meanweight}).

For the upper bound, drop all states above $E$ in $\mathcal Z(0,0)$. Since $\mathcal Z(0,0)=e^{o(\De)}$ by the first step,
\begin{equation}\label{eq:fragmenton-upper}
 \log\mathcal N(E)\leq\beta(0)E+\log \mathcal{Z}(0,0)=\beta(0)E+o(\De)\qquad\text{for every }E\ge0.
\end{equation}
The proof for the lower bound is, on the other hand, nontrivial. It requires us to zoom into the self-dual point. In particular, we weigh the spectrum at an inverse temperature slightly away from the self-dual value, so that the weighted sum is made almost entirely of states of one chosen weight $E$. This is possible because, in a window of width about $1/\Delta_\phi$ around the self-dual point, the fragmenton’s correlator is smooth rather than kinked, and as the temperature moves across this window, the weight that dominates sweeps from 0 to twice the mean weight. Then we note that the weighted sum is at least 1, since the identity alone contributes 1. Finally we argue that each of the dominant states contributes only about $e^{-\beta(0)E}$, so to add up to order one there must be about $e^{\beta(0)E}$ of them.

To execute the plan, weigh the spectrum at an inverse temperature slightly away from $\beta(0)$,
\[
 \mathcal W(s)=\int e^{-\left(\beta(0)+s/\Df\right)E}\,\dd\mathcal N(E).
\]
This is $\mathcal Z(x,x)$ at $x=-s/(|\beta'(0)|\Df)+O(\Df^{-2})$, inside the window of width $1/\Df$ around the self-dual point, so \eqref{eq:GFF-near-selfdual} gives
\[
 \frac{\log\mathcal W(s)}{L}\ \longrightarrow\ \Psi(s)=2\log\big(1+e^{-s/|\beta'(0)|}\big).
\]
The limit $\Psi$ is smooth, which is what the kink at scale $\De$ lacked. In fact, the smoothness plays the pivotal role for the proof to work. Its slope tells us which weight dominates $\mathcal W(s)$: as for any partition function, the slope of $\log\mathcal W$ in the inverse temperature is minus the average weight, here $-\Psi'(s)\,\De$. As $s$ runs over $\mathbb R$, $-\Psi'(s)$ runs over $(0,2/|\beta'(0)|)$, from zero up to twice the mean weight. So, given a target weight $e\De$ with $0<e<2/|\beta'(0)|$, choose $s$ with $-\Psi'(s)=e$. It remains to check that at this temperature the average weight is also the typical one, that is, that states far from $e\De$ contribute little to $\mathcal W(s)$. Consider first the states above $(e+\delta)\De$. Lowering the inverse temperature by $t/\Df$, with $t>0$, multiplies each of them by at least $e^{t(e+\delta)L}$, since $\De/\Df=L$. Hence
\[
 \int_{E>(e+\delta)\De}e^{-(\beta(0)+s/\Df)E}\,\dd\mathcal N(E)\ \le\ e^{-t(e+\delta)L}\,\mathcal W(s-t).
\]

Divided by $L$, the logarithm of the right-hand side tends to $\Psi(s-t)-t(e+\delta)$. Since $\Psi(s-t)=\Psi(s)+te+O(t^2)$, this equals $\Psi(s)-t\delta+O(t^2)$, which is below $\Psi(s)$ for small $t$. So these states supply an exponentially small fraction of $\mathcal W(s)$. Raising the inverse temperature instead, the same argument with $\mathcal W(s+t)$ shows that the states below $(e-\delta)\De$ also supply an exponentially small fraction. Hence, for every $\delta>0$ and all large $L$ and $\Df$, the states with $(e-\delta)\De\le E\le(e+\delta)\De$ supply at least half of $\mathcal W(s)$:
\begin{equation}\label{eq:fragmenton-window}
 \int_{(e-\delta)\De}^{(e+\delta)\De}e^{-(\beta(0)+s/\Df)E}\,\dd\mathcal N(E)\ \ge\ \tfrac12\,\mathcal W(s).
\end{equation}

Since the identity contributes $1$ to $\mathcal W(s)$, the right-hand side of \eqref{eq:fragmenton-window} is at least $\frac12$.

 Each state in the window has $E\ge(e-\delta)\De$, so it contributes at most $e^{-(\beta(0)+s/\Df)(e-\delta)\De}$ to the left-hand side of \eqref{eq:fragmenton-window}, and there are at most $\mathcal N((e+\delta)\De)$ such states. To reach $\frac12$, there must be at least
\[
 \mathcal N\big((e+\delta)\De\big)\ \geq\ \tfrac12\,e^{(\beta(0)+s/\Df)(e-\delta)\De}=\tfrac12\,e^{\beta(0)(e-\delta)\De+s(e-\delta)L}.
\]
The temperature shift costs a term of order $L=o(\De)$ in the exponent, so
\begin{equation}\label{eq:fragmenton-lower}
 \liminf_{\De\to\infty}\frac1\De\log\mathcal N\big((e+\delta)\De\big)\ \ge\ \beta(0)(e-\delta),
\end{equation}
which matches the upper bound \eqref{eq:fragmenton-upper}.

Applying \eqref{eq:fragmenton-lower} at $e-\delta$ in place of $e$ and letting $\delta\to0$ gives $\liminf\frac1\De\log\mathcal N(e\De)\ge\beta(0)e$ for $0<e\le1/|\beta'(0)|$. Since $\mathcal N$ is monotone and the limit is continuous, the bound is uniform on $0\le E\le\langle E\rangle$, and for $\langle E\rangle\le E\le\langle E\rangle+\epsilon$ monotonicity gives $\mathcal N(E)\ge\mathcal N(\langle E\rangle)$, which changes the exponent by at most $\beta(0)\epsilon$. Together with the uniform upper bound \eqref{eq:fragmenton-upper}, this proves \eqref{eq:GFFsaturation}.

\end{proof}

\subsection{Composites of a generalized free field of large spin}
\label{app:JJ}

In this subsection we derive the correlator \eqref{eq:JJ} of the singlet $\cO=\,:\!J_{\mu_1\dots\mu_\ell}J^{\mu_1\dots\mu_\ell}\!:$ of Sect.~\ref{ex:chiral}, and then prove Proposition~\ref{prop:JJ}. A reader willing to accept \eqref{eq:JJ} can skip directly to the proof of Proposition ~\ref{prop:JJ}.

The derivation runs in several steps. We begin by writing $\cO$ in terms of components of $J$ and computing its two-point function. Second, we sort the Wick contractions of $\langle\cO\cO\cO\cO\rangle$ into six types and evaluate each type up to the trace of a rotation matrix. Then we compute these rotation matrices for a convenient choice of conformal frame. The final step is to compute the traces.

As the first step, we consider $V_\ell$, the space of symmetric traceless tensors of rank $\ell$ on $\mathbb R^d$, with the inner product $\epsilon\cdot\epsilon'=\epsilon^{\mu_1\dots\mu_\ell}\epsilon'_{\mu_1\dots\mu_\ell}$, and let $D=\dim V_\ell$. Fix an orthonormal basis $\epsilon_1,\dots,\epsilon_D$ of $V_\ell$, and let $J_a=\epsilon_a^{\mu_1\dots\mu_\ell}J_{\mu_1\dots\mu_\ell}$. Since $J$ is symmetric and traceless, $J=\sum_a\epsilon_aJ_a$, and therefore
\begin{equation}\label{eq:JJ-components}
\cO=\sum_{a=1}^{D}J_aJ_a .
\end{equation}
A matrix $M\in O(d)$ acts on $V_\ell$ by acting on every index. Let $R(M)$ be the $D\times D$ matrix of this action, $R(M)_{ab}=\epsilon_a\cdot(M\epsilon_b)$. It is orthogonal, and $R(MM')=R(M)R(M')$. In index-free notation, with $J(x,\zeta)=\zeta^{\mu_1}\cdots\zeta^{\mu_\ell}J_{\mu_1\dots\mu_\ell}(x)$ for a null polarization vector $\zeta$, the two-point function of $J$ is \cite{Costa:2011mg}
\begin{equation}
\label{eq:spinning2pt}
\langle J(x_1,\zeta_1)J(x_2,\zeta_2)\rangle=\frac{\big(\zeta_1\cdot I(x_{12})\cdot\zeta_2\big)^{\ell}}{(x_{12}^2)^{\De_J}},\qquad I_{\mu\nu}(x)=\delta_{\mu\nu}-\frac{2x_\mu x_\nu}{x^2}.
\end{equation}
In tensor form, this states that $\langle J_{\mu_1\dots\mu_\ell}(x)J_{\nu_1\dots\nu_\ell}(y)\rangle$ is $I_{\mu_1\nu_1}\cdots I_{\mu_\ell\nu_\ell}/((x-y)^2)^{\De_J}$, with $I=I(x-y)$, up to trace terms that vanish against traceless tensors. Contracting with $\epsilon_a$ and $\epsilon_b$ gives
\begin{equation}\label{eq:JJ-comp2pt}
\langle J_a(x)J_b(y)\rangle=\frac{R\big(I(x-y)\big)_{ab}}{\big((x-y)^2\big)^{\De_J}} .
\end{equation}
Here $I(x)$ is the reflection in the hyperplane orthogonal to $x$, so $I(x)^2=1$ and $I(-x)=I(x)$. In $\langle\cO(x)\cO(y)\rangle$, the two constituents at $x$ contract with the two at $y$ in two ways, and each way gives $((x-y)^2)^{-2\De_J}$ times $\sum_{a,b}R_{ab}R_{ab}=\operatorname{tr}(RR^{T})=D$. With $\De=2\De_J$,
\begin{equation}\label{eq:JJ-O2pt}
\langle\cO(x)\cO(y)\rangle=\frac{2D}{\big((x-y)^2\big)^{\De}} .
\end{equation}

Then we perform the Wick contractions. The correlator $\langle\cO(x_1)\cdots\cO(x_4)\rangle$ is a sum over the ways of pairing the eight constituents, two at each point, with no pair at a single point, since $\cO$ is normal ordered. Draw each pairing as a graph with vertices $1,2,3,4$ and one edge for each propagator. Every vertex has two edges and no edge returns to its own vertex, so the graph is of one of two kinds:
\begin{enumerate}[label=(\alph*)]
\item two double edges, $(12)(34)$, $(13)(24)$, or $(14)(23)$. Both constituents of each insertion go to the same partner, as for a generalized free field of dimension $\De$;
\item a single cycle, $(1234)$, $(1243)$, or $(1324)$, where $(1234)$ has the edges $12$, $23$, $34$, $41$, and so on. The two constituents of each insertion go to different partners, and the graph connects all four insertions.
\end{enumerate}
For $(12)(34)$, the sum over contractions is $\langle\cO(x_1)\cO(x_2)\rangle\langle\cO(x_3)\cO(x_4)\rangle$, which contributes $1$ to the reduced correlator. In the same way, by \eqref{eq:JJ-O2pt}, $(13)(24)$ and $(14)(23)$ contribute
\[
\Big(\frac{x_{12}^2x_{34}^2}{x_{13}^2x_{24}^2}\Big)^{\De}=\Big(\frac uv\Big)^{\De},\qquad
\Big(\frac{x_{12}^2x_{34}^2}{x_{14}^2x_{23}^2}\Big)^{\De}=u^{\De}.
\]

For the cycle $(1234)$, one constituent at each point contracts toward the next point of the cycle and the other toward the previous point. There are two ways to choose which is which at each point, so $2^4=16$ contractions, all of equal weight. Both constituents of $\cO(x_i)$ carry the same index $a_i$ in \eqref{eq:JJ-components}, so the indices are contracted around the cycle. Writing $R^{(ij)}=R(I(x_{ij}))$, each contraction gives
\[
\sum_{a_1,\dots,a_4}\frac{R^{(12)}_{a_1a_2}R^{(23)}_{a_2a_3}R^{(34)}_{a_3a_4}R^{(41)}_{a_4a_1}}{(x_{12}^2x_{23}^2x_{34}^2x_{41}^2)^{\De_J}}
=\frac{\operatorname{tr}R(\Theta_{1234})}{(x_{12}^2x_{23}^2x_{34}^2x_{41}^2)^{\De_J}},\qquad \Theta_{1234}=I(x_{12})I(x_{23})I(x_{34})I(x_{41}).
\]
As a product of four reflections, $\Theta_{1234}$ lies in $SO(d)$.
Multiplying by $16$ and dividing by $\langle\cO(x_1)\cO(x_2)\rangle\langle\cO(x_3)\cO(x_4)\rangle=4D^2(x_{12}^2x_{34}^2)^{-2\De_J}$ gives $\frac4{D^2}u^{\De_J}\operatorname{tr}R(\Theta_{1234})$. The other two cycles are treated in the same way. Altogether,
\begin{equation}\label{eq:JJ-wick}
\mathcal G=1+u^{\De}+\Big(\frac uv\Big)^{\De}+\frac4{D^2}\Big[u^{\De_J}\operatorname{tr}R(\Theta_{1234})+\Big(\frac uv\Big)^{\De_J}\operatorname{tr}R(\Theta_{1243})+\Big(\frac{u^2}v\Big)^{\De_J}\operatorname{tr}R(\Theta_{1324})\Big],
\end{equation}
with $\Theta_{1243}=I(x_{12})I(x_{24})I(x_{43})I(x_{31})$ and $\Theta_{1324}=I(x_{13})I(x_{32})I(x_{24})I(x_{41})$. In terms of $z$, $\bar z$, and $\eta=z/(1-z)$, the cross ratios are $u=z\bar z$, $u/v=\eta\bar\eta$, and $u^2/v=z\bar z\,\eta\bar\eta$.

Since $\cO$ is a scalar primary, $\mathcal G$ depends only on the cross ratios, so it suffices to evaluate \eqref{eq:JJ-wick} with the four points on a two-plane. Choose orthonormal vectors $e_1,e_2$ spanning it, and identify the plane with $\mathbb C$: the point $x_i$ corresponds to $w_i\in\mathbb C$, and $w_{ij}=w_i-w_j$. In this labeling of the points, $z=w_{12}w_{34}/(w_{14}w_{32})$, and $\bar z$ is its complex conjugate. In terms of the null vectors $\zeta_\pm=(e_1\pm ie_2)/\sqrt2$, $x_{ij}=(\bar w_{ij}\,\zeta_++w_{ij}\,\zeta_-)/\sqrt2$, so $x_{ij}\cdot\zeta_+=w_{ij}/\sqrt2$, $x_{ij}\cdot\zeta_-=\bar w_{ij}/\sqrt2$, and $x_{ij}^2=w_{ij}\bar w_{ij}$. The definition of $I$ in \eqref{eq:spinning2pt} (note $I(x)l=l-2x\frac{(x\cdot l)}{x^2}$) then gives
\[
I(x_{ij})\,\zeta_\pm=-\Big(\frac{w_{ij}}{\bar w_{ij}}\Big)^{\pm1}\zeta_\mp ,
\]
while $I(x_{ij})$ fixes the directions orthogonal to the plane. For $\lambda\neq0$, let $g_\lambda$ be the linear map that multiplies $\zeta_+$ by $\lambda$ and $\zeta_-$ by $1/\lambda$, and fixes the directions orthogonal to the plane. Applying the formula above four times, the four signs cancel, and the cycle $(ijkl)$ gives
\[
\Theta_{ijkl}=I(x_{ij})I(x_{jk})I(x_{kl})I(x_{li})=g_\lambda,\qquad \lambda=\frac{w_{jk}\,w_{li}}{w_{ij}\,w_{kl}}\bigg/\frac{\bar w_{jk}\,\bar w_{li}}{\bar w_{ij}\,\bar w_{kl}} .
\]
By the identity $w_{13}w_{24}=w_{12}w_{34}+w_{14}w_{23}$, the ratio $w_{jk}w_{li}/(w_{ij}w_{kl})$ equals $1/z$, $-1/\eta$, and $1/(1-z)$ for the three cycles, so
\begin{equation}\label{eq:cycle-rotations}
\Theta_{1234}=g_{\bar z/z},\qquad
\Theta_{1243}=g_{\bar\eta/\eta},\qquad
\Theta_{1324}=g_{(1-\bar z)/(1-z)} .
\end{equation}

It remains to compute $\Xi_\ell(\lambda)\equiv\operatorname{tr}R(g_\lambda)$. Complete $e_1,e_2$ to an orthonormal basis $e_1,\dots,e_d$ of $\mathbb R^d$. In the basis $\zeta_+$, $\zeta_-$, $e_3,\dots,e_d$, the map $g_\lambda$ is diagonal with eigenvalues $\lambda,1/\lambda,1,\dots,1$, so a symmetrized product of $k_+$ copies of $\zeta_+$, $k_-$ copies of $\zeta_-$, and vectors $e_i$ with $i\geq3$ is multiplied by $\lambda^{k_+-k_-}$. The exponent $k_+-k_-$ is the planar spin $m$ of the corresponding component of $J$, defined in Sect.~\ref{ex:chiral}. Since $g_\lambda$ preserves $V_\ell$, it is diagonalizable on $V_\ell$ with eigenvalues $\lambda^m$, $|m|\leq\ell$, and
\[
\Xi_\ell(\lambda)=\sum_{m=-\ell}^{\ell}N_{\ell,m}\lambda^m ,
\]
where $N_{\ell,m}\geq0$ is the dimension of the eigenspace of planar spin $m$, so that $\sum_mN_{\ell,m}=D$. The extreme multiplicities are $N_{\ell,\pm\ell}=1$, since the only tensor of planar spin $\pm\ell$ is $\zeta_\pm^{\otimes\ell}$, which is traceless because $\zeta_\pm\cdot\zeta_\pm=0$. And $N_{\ell,-m}=N_{\ell,m}$, since the reflection $e_2\to-e_2$ exchanges $\zeta_+$ and $\zeta_-$ and preserves $V_\ell$, so $\Xi_\ell(1/\lambda)=\Xi_\ell(\lambda)$.

Besides these properties, the proof of Proposition~\ref{prop:JJ} uses only that $D$ grows polynomially in $\ell$. This holds because $V_\ell$ lies in the space of symmetric tensors of rank $\ell$, so $D\leq\binom{\ell+d-1}{d-1}\leq(\ell+1)^{d-1}$.
In $d=2$ there are no vectors $e_i$ with $i\geq3$, so the symmetric tensors of planar spin $m$ are the multiples of the symmetrized product of $k_+$ copies of $\zeta_+$ and $k_-$ copies of $\zeta_-$, with $k_++k_-=\ell$ and $k_+-k_-=m$. Since $\zeta_+\cdot\zeta_-=1$, its trace is nonzero unless $k_+$ or $k_-$ vanishes. So only $m=\pm\ell$ occur, $D=2$, and $\Xi_\ell(\lambda)=\lambda^\ell+\lambda^{-\ell}$.

By \eqref{eq:cycle-rotations}, $\operatorname{tr}R(\Theta)=\Xi_\ell(\lambda)$ for the three cycles, with $\lambda=\bar z/z$, $\bar\eta/\eta$, and $(1-\bar z)/(1-z)$. Inserting these into \eqref{eq:JJ-wick}, and using $\Xi_\ell(1/\lambda)=\Xi_\ell(\lambda)$, gives \eqref{eq:JJ}. Since each $\Xi_\ell$ is a finite sum of powers of $\lambda$, \eqref{eq:JJ} is a finite sum of products of powers of $z$, $\bar z$, $1-z$, and $1-\bar z$, so it defines the correlator for independent $z$ and $\bar z$. In the causal diamond $0<z,\bar z<1$, the three arguments of $\Xi_\ell$ are positive. For $\lambda>0$, $\Xi_\ell(\lambda)$ is a sum of $D$ positive terms, counted with multiplicity, each at most $\max(\lambda,1/\lambda)^\ell$, and the terms $m=\pm\ell$ have coefficient one. Hence
\begin{equation}\label{eq:Xi-bounds}
\max\big(\lambda,\tfrac1\lambda\big)^{\ell}\ \leq\ \Xi_\ell(\lambda)\ \leq\ D\max\big(\lambda,\tfrac1\lambda\big)^{\ell},\qquad\lambda>0.
\end{equation}

In $d=2$, with $\Xi_\ell(\lambda)=\lambda^\ell+\lambda^{-\ell}$, $D=2$, and $p=\tau_J=\De_J-\ell$ and $q=\De-\tau_J=\De_J+\ell$, and writing $\eta=z/(1-z)$ in the third cycle term, the correlator becomes
\begin{equation}\label{eq:JJ-2d}
\mathcal G=\big(1+\eta^p\bar\eta^q+z^p\bar z^q\big)\big(1+\eta^q\bar\eta^p+z^q\bar z^p\big),
\end{equation}
as one checks by expanding the product. The first factor is the four-point function of the component of $J$ of planar spin $-\ell$, with weights $(\frac p2,\frac q2)$, and the second is that of the component of planar spin $+\ell$.

\begin{proof}[Proof of Proposition~\ref{prop:JJ}]
(i) Both monomial expansions are non-negative because $\cO$ is a scalar primary of a unitary theory. To find their lowest powers, write $\eta=z(1-z)^{-1}$ and expand $\Xi_\ell(\lambda)=\sum_mN_{\ell,m}\lambda^m$ with $|m|\leq\ell$. Using $\De=2\De_J$, each term of $\mathcal G-1$ in \eqref{eq:JJ} is then a coefficient times powers of $z$, $\bar z$, $(1-z)^{-1}$, and $(1-\bar z)^{-1}$:
\begin{center}
\small
\renewcommand{\arraystretch}{1.35}
\begin{tabular}{@{}lccccc@{}}
\toprule
Term of $\mathcal G-1$ & Coefficient & \multicolumn{4}{c}{Power of}\\
\cmidrule(l){3-6}
 & & $z$ & $\bar z$ & $(1-z)^{-1}$ & $(1-\bar z)^{-1}$\\
\midrule
$(z\bar z)^{\De}$ & $1$ & $\De$ & $\De$ & $0$ & $0$\\
$(\eta\bar\eta)^{\De}$ & $1$ & $\De$ & $\De$ & $\De$ & $\De$\\
$(z\bar z)^{\De_J}(z/\bar z)^m$ & $\frac4{D^2}N_{\ell,m}$ & $\De_J+m$ & $\De_J-m$ & $0$ & $0$\\
$(\eta\bar\eta)^{\De_J}(\eta/\bar\eta)^m$ & $\frac4{D^2}N_{\ell,m}$ & $\De_J+m$ & $\De_J-m$ & $\De_J+m$ & $\De_J-m$\\
$(z\bar z\,\eta\bar\eta)^{\De_J}\big(\frac{1-z}{1-\bar z}\big)^m$ & $\frac4{D^2}N_{\ell,m}$ & $\De$ & $\De$ & $\De_J-m$ & $\De_J+m$\\
\bottomrule
\end{tabular}
\end{center}
Since $|m|\leq\ell$, we have $\De_J\pm m\geq\De_J-\ell=\tau_J>0$, and also $\De=2\De_J>\tau_J$. So every power of $z$ and $\bar z$ in the table is at least $\tau_J$, and every power of $(1-z)^{-1}$ and $(1-\bar z)^{-1}$ is non-negative. The value $\tau_J$ is reached at $m=\pm\ell$ in the third and fourth rows, by the double-twist operators $[JJ]$ of twist $2\tau_J$.
In the $\rho$-expansion, $z=4\rho/(1+\rho)^2$ is $4\rho$ times a power series that starts at $1$, and $(1-z)^{-1}=\big(\frac{1+\rho}{1-\rho}\big)^2$ is a power series that starts at $1$, so each term starts at the same powers of $\rho$ and $\bar\rho$.  Hence, in both expansions, every non-identity monomial has $\min(h,\bar h)\geq\tau_J=\alpha_\De\De>\alpha\De$, so $\mathcal G^{(\alpha)}_0=1$. At $z=\bar z=\frac12$, every argument of $\Xi_\ell$ is $1$ and $\Xi_\ell(1)=D$, so $\mathcal G(\frac12,\frac12)=2+4^{-\De}+\frac4D(1+2^{1-\De})$ is bounded, and statement~4 of Theorem~\ref{thm:weak-crossing} holds.

(ii) Let
\[
M_*=(\eta\bar\eta)^{\De_J}\max\Big(\frac{\eta}{\bar\eta},\frac{\bar\eta}{\eta}\Big)^{\ell}=e^{\De_J(x+\bar x)+\ell|x-\bar x|},
\]
and bound the three cycle terms of \eqref{eq:JJ}, with their factor $\frac4{D^2}$, by \eqref{eq:Xi-bounds}. The second term lies between $\frac4{D^2}M_*$ and $\frac4DM_*$. The first is at most $\frac4D$, since $(z\bar z)^{\De_J}(z/\bar z)^{\pm\ell}=z^{\De_J\pm\ell}\bar z^{\De_J\mp\ell}<1$, both exponents being at least $\tau_J>0$. The third is at most $\frac4DM_*$, since $\frac{1-z}{1-\bar z}=\frac z{\bar z}\,\frac{\bar\eta}\eta$ gives
\[
(z\bar z\,\eta\bar\eta)^{\De_J}\Big(\frac{1-z}{1-\bar z}\Big)^{\pm\ell}=z^{\De_J\pm\ell}\bar z^{\De_J\mp\ell}\,(\eta\bar\eta)^{\De_J}\Big(\frac{\bar\eta}{\eta}\Big)^{\pm\ell}<M_* .
\]
Since also $(z\bar z)^{\De}<1$,
\[
\max\Big(1,\,e^{\De(x+\bar x)},\,\frac4{D^2}M_*\Big)\ \leq\ \mathcal G\ \leq\ 2+\frac4D+e^{\De(x+\bar x)}+\frac8D\,M_* .
\]
As $D\geq2$ grows polynomially in $\ell\leq\De$, the logarithm of each side is $\log\max\big(1,e^{\De(x+\bar x)},M_*\big)+O(\log\De)$. Now $\De_J=\frac\De2$ and $\ell=\De_J-\tau_J=(\frac12-\alpha_\De)\De$ with $|\alpha_\De-\alpha|\leq\frac1\De$, so $\log M_*=\De\max(A_+,A_-)+O(|x|+|\bar x|)$, with $A_\pm$ as in \eqref{eq:JJ-spin-term}; indeed $\max(A_+,A_-)=\frac{x+\bar x}2+(\frac12-\alpha)|x-\bar x|$. Therefore
\[
\frac1\De\log\mathcal G=\max(0,\,A_++A_-,\,A_+,\,A_-)+O\Big(\frac{\log\De+|x|+|\bar x|}{\De}\Big).
\]
The maximum is $A_++A_-$ when both are positive, the positive one when only one is, and $0$ otherwise. That is, it equals $\max(0,A_+)+\max(0,A_-)=F$.

(iii)  Since $A_++A_-=x+\bar x$, the limit $F=\max(0,A_+)+\max(0,A_-)$ exceeds the liberated value $\max(0,x+\bar x)$ exactly where $A_+$ and $A_-$ have opposite signs. As $A_+$ and $A_-$ are weighted averages of $x$ and $\bar x$, this requires $x$ and $\bar x$ to have opposite signs. For $x<0<\bar x$, $A_--A_+=(1-2\alpha)(\bar x-x)>0$, so the condition is $A_->0>A_+$, that is, $\lambda_\alpha|x|<\bar x<|x|/\lambda_\alpha$, which defines $\mathcal V_\alpha$. Exchanging $x$ and $\bar x$ exchanges $A_+$ and $A_-$, so for $\bar x<0<x$ the region is $\chi(\mathcal V_\alpha)$. By (i), Theorem~\ref{thm:strong-crossing} applies in the $\rho$-expansion, so the sequence is liberated outside $\mathcal E^{(\rho)}_\alpha$, and therefore $\mathcal V_\alpha\cup\chi(\mathcal V_\alpha)\subset\mathcal E^{(\rho)}_\alpha$. The boundary function satisfies $f_\alpha(x)<\lambda_\alpha|x|$ (Lemma~\ref{lem:position}) and $f_\alpha(x)/|x|\to\lambda_\alpha$ as $x\to-\infty$ (Proposition~\ref{enigmaticasymptotics}), so the inclusion is proper and the two regions have the same asymptotic cone. Since $\mathcal E^{(\rho)}_\alpha\subseteq\mathcal E^{(z)}_\alpha$ (Sect.~\ref{sec:liberation4pt}), the union is also a proper subset of $\mathcal E^{(z)}_\alpha$, and Proposition~\ref{enigmaticasymptotics} gives the same $\lambda_\alpha$ in the $z$-expansion, so $\mathcal E^{(z)}_\alpha$ has the same asymptotic cone.

\end{proof}

\subsection{Local realization by current composites}
\label{app:local-JJ}

We compute the four-point function of the operators $\cO_K$ of Proposition~\ref{prop:local-JJ} (Lemma~\ref{lem:local-exact}) and then prove the proposition.

On the plane, correlators of untwisted operators that are invariant under permutations of the copies, such as $U_\pm$ and $\cO$ of \eqref{eq:local-ops}, are the same in $\mathcal T_{K,N}$ as in $\mathcal X_K^{\otimes N}$, so we compute them there. Let $w_i$ be the complex coordinate of the $i$-th insertion, as in Appendix~\ref{app:JJ}; in this labeling, $w_{12}w_{34}/(w_{13}w_{24})=-\eta$ and $w_{12}w_{34}/(w_{14}w_{23})=-z$. The currents are Hermitian. Currents of different bosons are uncorrelated, those of a single boson obey Wick's theorem, and different copies are uncorrelated with $\langle V_{\pm,i}\rangle=0$, so
\begin{gather*}
\langle jj\rangle=w_{12}^{-2},\qquad \langle\bar j\bar j\rangle=\bar w_{12}^{-2},\\
\langle U_+U_+\rangle=w_{12}^{-2l}\bar w_{12}^{-2k},\qquad \langle U_-U_-\rangle=w_{12}^{-2k}\bar w_{12}^{-2l},\qquad \langle\cO\cO\rangle=|w_{12}|^{-2\De}.
\end{gather*}
Reduced four-point functions are divided by these two-point functions for the pairs $(12)$ and $(34)$.

We first compute the correlators within a single copy. By Wick's theorem, the reduced four-point function of $j$ is the sum over the pairings $(12)(34)$, $(13)(24)$, and $(14)(23)$,
\begin{equation}\label{eq:current-f}
 f(z)=1+\eta^2+z^2 ,
\end{equation}
and that of $\bar j$ is $f(\bar z)$. Let $S_\pm$ be the reduced four-point function of a single $V_{\pm,i}$. The currents in $V_{\pm,i}$ belong to different bosons, so $S_\pm$ is the product of the reduced four-point functions of its currents:
\begin{equation}\label{eq:local-seed}
 S_+=f(z)^lf(\bar z)^k,\qquad S_-=f(z)^kf(\bar z)^l .
\end{equation}
$\eta^2$ for $(13)(24)$, and $z^2$ for $(14)(23)$. So expanding $S_\pm$ assigns one of the three pairings to each of the $K$ currents, which gives $3^K$ terms. Let $\mathcal G_\pm$ be the sum of the three terms in which all $K$ currents have the same pairing:
\begin{equation}\label{eq:local-base}
 \mathcal G_+=1+\eta^{2l}\bar\eta^{2k}+z^{2l}\bar z^{2k},
 \qquad \mathcal G_-(z,\bar z)=\mathcal G_+(\bar z,z).
\end{equation}
These are the four-point functions of generalized free fields with weights $(l,k)$ and $(k,l)$. Multiplying them out gives \eqref{eq:JJ} in $d=2$, where $\Xi_\ell(\lambda)=\lambda^\ell+\lambda^{-\ell}$ and $D=2$, with $\De_J=K$, $\ell=k-l$, and $\tau_J=2l$.

\begin{lemma}\label{lem:local-exact}
For all $K$, $N$, and $0<l<k$ with $l+k=K$, the reduced four-point function of $\cO=U_+U_-$ is
\begin{equation}\label{eq:local-exact}
 \mathcal G=H_+H_-,\qquad
 H_\pm=\Big(1-\frac1N\Big)\mathcal G_\pm+\frac1N\,S_\pm .
\end{equation}
\end{lemma}

\begin{proof}
$U_+$ and $U_-$ are built from disjoint sets of bosons, so every correlator of $\cO$ is the product of those of $U_+$ and $U_-$, and $\mathcal G=H_+H_-$, where $H_\pm$ is the reduced four-point function of $U_\pm$. To compute $H_+$, expand
\[
 \langle U_+(w_1)\cdots U_+(w_4)\rangle=\frac1{N^2}\sum_{i_1,\dots,i_4=1}^N\langle V_{+,i_1}(w_1)\cdots V_{+,i_4}(w_4)\rangle .
\]
Each term is a product of correlators within single copies. A term vanishes if some copy appears at exactly one or exactly three points, since the one- and three-point functions of $V_{+,i}$ vanish. The remaining terms are of two kinds. In $N$ of them all four indices are equal, and each gives the single-copy correlator, with reduced value $S_+$. In the others the indices take two values, each twice, following one of the pairings $(12)(34)$, $(13)(24)$, or $(14)(23)$, with $N(N-1)$ terms for each. Each such term is a product of two two-point functions of $V_{+,i}$, which has weights $(l,k)$, so its reduced value is $1$, $(w_{12}w_{34}/w_{13}w_{24})^{2l}(\bar w_{12}\bar w_{34}/\bar w_{13}\bar w_{24})^{2k}=\eta^{2l}\bar\eta^{2k}$, or $(w_{12}w_{34}/w_{14}w_{23})^{2l}(\bar w_{12}\bar w_{34}/\bar w_{14}\bar w_{23})^{2k}=z^{2l}\bar z^{2k}$, where the signs drop out because the exponents are even. Since $U_+$ and $V_{+,i}$ have the same two-point function,
\[
H_+=\frac1{N^2}\Big[N\,S_++N(N-1)\big(1+\eta^{2l}\bar\eta^{2k}+z^{2l}\bar z^{2k}\big)\Big]=\frac1N\,S_++\Big(1-\frac1N\Big)\mathcal G_+ ,
\]
and likewise for $H_-$.
\end{proof}

By Lemma~\ref{lem:local-exact}, each constituent's correlator $H_\pm$ has two parts.
However, we want $H_\pm\approx\mathcal G_\pm$, so that $\mathcal G\approx\mathcal G_+\mathcal G_-$ is the correlator \eqref{eq:JJ} of Sect.~\ref{ex:chiral} in $d=2$. This fixes two choices, for given $0<\alpha<\frac12$.

The first is how the currents are split. With $l$ holomorphic and $k$ antiholomorphic currents, $U_+$ has weights $(l,k)$ and twist $\tau_J=2l$, so $\tau_J/\De_K=l/K$, where $\De_K=2K$. This ratio enters in two ways. It sets the rate, since $\eta^{2l}\bar\eta^{2k}=e^{\De_K(\frac lKx+\frac kK\bar x)}$, so the limit $F$ of \eqref{eq:JJrate} requires $l/K\to\alpha$. It also sets the lowest exchanged weight, $\min(h,\bar h)=2l$, and $\alpha$-twist sparseness requires this to lie above the cutoff $\alpha\De_K=2\alpha K$, that is, $l>\alpha K$. Both hold for the smallest integer above $\alpha K$, so we take $l=\lfloor\alpha K\rfloor+1$ and $k=K-l$. Then $|l/K-\alpha|\leq1/K$, and $0<l<k$ for large $K$, since $\alpha<\frac12$.

The second is the number of copies. Averaging over copies suppresses $S_\pm$ only by $\frac1N$, because all four insertions must sit in the same copy. But $S_+$ exceeds $\mathcal G_+$ exponentially: at large $K$, $S_+/\mathcal G_+\approx e^{bK}$, with
\[
b=\alpha\log f(z)+(1-\alpha)\log f(\bar z)-2\max\big(0,\,\alpha x+(1-\alpha)\bar x\big)>0 .
\]
For the generalized free part to win at every point, $N$ must therefore grow faster than any exponential of $K$, so we take $N=N_K$ with $\log N_K/K\to\infty$, for example $N_K=\lceil e^{K^2}\rceil$. This is more than sufficient; in fact, any $N_K$ with $\log N_K/K\to \infty$ works, and we have made no attempt to optimize.

\begin{proof}[Proof of Proposition~\ref{prop:local-JJ}]
The properties of the operators were shown above. Expanding \eqref{eq:local-exact} gives
\begin{equation}\label{eq:local-positive-mixture}
 \mathcal G_K=\Big(1-\frac1N\Big)^2\mathcal G_+\mathcal G_-+R,\qquad
 R=\frac1N\Big(1-\frac1N\Big)\big(S_+\mathcal G_-+\mathcal G_+S_-\big)+\frac1{N^2}\,S_+S_- .
\end{equation}
All the statements follow from $R$ being negligible. We trivially have $\mathcal G_\pm\geq1$ and $S_\pm\geq1$. From above, each of the three terms of $\mathcal G_+$ is at most $e^{2l|x|+2k|\bar x|}$: the first is $1$, the second is $e^{2lx+2k\bar x}$ since $\eta=e^x$, and the third is below $1$ since $0<z,\bar z<1$. The same holds for $\mathcal G_-$ with $l$ and $k$ exchanged. Since $2l+2k=\De_K$, this gives $\mathcal G_\pm\leq3e^{\De_K\max(|x|,|\bar x|)}$. And $S_\pm$ is a product of $K$ factors $f(z)$ or $f(\bar z)$, so $S_\pm\leq\max\big(f(z),f(\bar z)\big)^K$.

Now fix a compact set $C\subset(0,1)^2$. On $C$, $z$ and $\bar z$ stay away from $0$ and $1$, so
\[
B_C=\max_{(z,\bar z)\in C}\Big[\log3+\max\big(|x|,|\bar x|\big)+\frac12\log\max\big(f(z),f(\bar z)\big)\Big]
\]
is finite and independent of $K$. Since $K=\De_K/2$ and $\De_K\geq1$, the bounds above give that $\mathcal G_\pm$ and $S_\pm$ lie between $1$ and $e^{B_C\De_K}$ on $C$. So each product in $R$ lies between $1$ and $e^{2B_C\De_K}$. Their coefficients in \eqref{eq:local-positive-mixture} add up to $\frac2N-\frac1{N^2}<\frac2N$, so $0\leq R\leq\frac2{N_K}e^{2B_C\De_K}$, which tends to zero uniformly on $C$, since $\log N_K/\De_K\to\infty$. Here $B_C$ grows without bound as $C$ approaches the edges of $(0,1)^2$, which is why $C$ is fixed first and $N_K$ must grow faster than any exponential.

For the limit, dividing \eqref{eq:local-positive-mixture} by $\mathcal G_+\mathcal G_-\geq1$ gives $\big|\mathcal G_K/(\mathcal G_+\mathcal G_-)-1\big|\leq\frac2{N_K}+R$, which tends to zero uniformly on $C$. So $\frac1{\De_K}\log\mathcal G_K$ and $\frac1{\De_K}\log(\mathcal G_+\mathcal G_-)$ have the same limit. At $z=\bar z=\frac12$, $\mathcal G_\pm=2+4^{-K}$, so $\mathcal G_K(\frac12,\frac12)\to4$. Elsewhere, $z^{2l}\bar z^{2k}<1$, so $\max(1,\eta^{2l}\bar\eta^{2k})\leq\mathcal G_+\leq3\max(1,\eta^{2l}\bar\eta^{2k})$. With $\log(\eta^{2l}\bar\eta^{2k})=2lx+2k\bar x$, $\De_K=2K$, and $|l/K-\alpha|\leq1/K$, this gives $\frac1{\De_K}\log\mathcal G_+\to\max\big(0,\alpha x+(1-\alpha)\bar x\big)$ uniformly on $C$. In the same way, $\frac1{\De_K}\log\mathcal G_-\to\max\big(0,(1-\alpha)x+\alpha\bar x\big)$, and the sum of the two limits is $F$.

For the twist content, we first check that $\mathcal G_\pm$ and $S_\pm$ have non-negative $z$- and $\rho$-expansions.
It is easy to verify that every non-identity monomial of $\mathcal G_+\mathcal G_-$ has $\min(h,\bar h)\geq2l>\alpha\De_K$, and its low-twist part is exactly $1$.
 Keeping only the monomials with $\min(h,\bar h)\leq\alpha\De_K$ in \eqref{eq:local-positive-mixture} therefore gives $\mathcal G_{K,0}^{(\alpha)}=(1-\frac1N)^2+R_0$, where $R_0$ is the same part of $R$. The coefficients of $R$ are non-negative and its identity coefficient is $\frac2N-\frac1{N^2}$, so $\frac2N-\frac1{N^2}\leq R_0\leq R$. Hence $1\leq\mathcal G_{K,0}^{(\alpha)}\leq1+R$, which tends to one uniformly on $C$. So the sequence is $\alpha$-twist sparse in both expansions.
\end{proof}

\subsection{The Virasoro obstruction}
\label{app:virasoro}

\begin{lemma}\label{lem:vaclower}
Let $c>1$, and let $\mathcal{F}_0$ be the Virasoro vacuum block of a scalar primary $\cO$ of dimension $\De$, with weights $(\frac\De2,\frac\De2)$, normalized so that the vacuum module contributes $\mathcal{F}_0(z)\mathcal{F}_0(\bar z)$ to $\mathcal{G}$. For $z\in(0,1)$, $\mathcal{F}_0(z)\geq{}_2F_1\big(\De,\De;\frac c8;\rho(z)^2\big)$, which is \eqref{eq:2dlower}.
\end{lemma}

\begin{proof}
We work with the holomorphic half; the antiholomorphic half is identical. In radial quantization, place the operators at $\pm\rho$ and $\pm1$, and let $\psi\propto\cO(1)\cO(-1)|0\rangle$, normalized so that $\langle0|\psi\rangle=1$. Then
\begin{equation}
\label{eq:F0squares}
\mathcal{F}_0(z)=\sum_{N\geq0}\rho^N\,\|\Pi_N\psi\|^2,
\end{equation}
where $\Pi_N$ is the orthogonal projection onto level $N$ of the vacuum module, which is unitary for $c>1$. Although $\psi$ is not normalizable, each $\Pi_N\psi$ lies in a finite-dimensional space, and its components $\langle\xi|\psi\rangle$ follow from the Ward identity
\begin{equation}
\label{eq:wardLn}
[L_n,\cO(w)]=\Big(w^{n+1}\partial_w+(n+1)\frac\De2\,w^n\Big)\cO(w)
\end{equation}
and $L_n|0\rangle=0$ for $n\geq-1$, applied to the two-point function. For any nonzero $\xi$ at level $N$, the Cauchy--Schwarz inequality gives
\begin{equation}
\label{eq:CSlevel}
\|\Pi_N\psi\|^2\geq\frac{|\langle\xi|\psi\rangle|^2}{\langle\xi|\xi\rangle}.
\end{equation}
At level $N=2n$, take $\xi=L_{-2}^n|0\rangle$. Since $[L_2,L_{-2}]=4L_0+\frac c2$,
\begin{equation}
L_2L_{-2}^n|0\rangle=\sum_{j=0}^{n-1}\Big(8j+\frac c2\Big)L_{-2}^{n-1}|0\rangle=4n\Big(n-1+\frac c8\Big)L_{-2}^{n-1}|0\rangle,
\end{equation}
and by induction $\langle\xi|\xi\rangle=4^n\,n!\,(\frac c8)_n$. For the overlap $\langle\xi|\psi\rangle=\langle0|L_2^n|\psi\rangle$, note that by \eqref{eq:wardLn} $L_2$ moves the insertions along $w^3\partial_w$, whose flow is $w_t=w(1-2tw^2)^{-1/2}$, with $w_t(\pm1)=\pm(1-2t)^{-1/2}$ and $w_t'(\pm1)=(1-2t)^{-3/2}$. Since $L_2|0\rangle=0$,
\begin{equation}
\langle0|e^{tL_2}|\psi\rangle=\big(w_t'(1)\,w_t'(-1)\big)^{\frac\De2}\Big(\frac{w_t(1)-w_t(-1)}{2}\Big)^{-\De}=(1-2t)^{-\De},
\end{equation}
so $\langle\xi|\psi\rangle=2^n(\De)_n$. By \eqref{eq:CSlevel}, the coefficient of $\rho^{2n}$ in \eqref{eq:F0squares} is at least $(\De)_n^2/\big(n!\,(\frac c8)_n\big)$, which is the coefficient of $\rho^{2n}$ in ${}_2F_1(\De,\De;\frac c8;\rho^2)$. The odd levels contribute non-negatively, which proves the lemma.
\end{proof}

\begin{proof}[Proof of Proposition~\ref{prop:bodies2d}]
Keeping only the vacuum module in either channel, unitarity and crossing give
\begin{equation}
\label{eq:twobounds}
\mathcal{G}\geq\max\Big(\mathcal{F}_0(z)\mathcal{F}_0(\bar z),\ \Big(\frac{z\bar z}{(1-z)(1-\bar z)}\Big)^{\De}\mathcal{F}_0(1-z)\mathcal{F}_0(1-\bar z)\Big),
\end{equation}
All terms of the series expansion of the lower bound in Lemma~\ref{lem:vaclower} are positive, so for $\rho=\rho(z)$ and every $n$,
\begin{equation}
\mathcal{F}_0(z)\geq T_n(\rho)\equiv\frac{(\De)_n^2}{n!\,(\frac c8)_n}\,\rho^{2n}.
\end{equation}

At large $\De$, a single term already captures the leading growth of the series. The terms grow with $n$ until the ratio
\begin{equation}
\frac{T_{n+1}(\rho)}{T_n(\rho)}=\frac{\rho^2(\De+n)^2}{(n+1)(\frac c8+n)}
\end{equation}
falls to one. This happens at $n$ of order $\De$ when $c$ is at most of order $\De$, which is statement (i), and at $n\approx8\rho^2\De^2/c$ when $\De\ll c\ll\De^2$, which is statement (ii).

(i) We want to probe the regime where $n$ is order of $\De$ and $c$ is order of $\De$. Hence let us
fix $t>0$ and set $n=\lfloor t\De\rfloor$. We have
\begin{equation}
    \log T_n(\rho)\geq 2\log \frac{\Gamma(\De+n)}{\Gamma\De}-\log \Gamma(n+1)+2n\log\rho -\log \frac{\Gamma(c/8+n)}{\Gamma (c/8)}
\end{equation}
Since $\De/c \to \nu$, for large enough $\De$, $\De/c>\nu-\delta$ for any given $\delta>0$. Define $b=\frac{1}{8(\nu-\delta)}$ and we have
(since $(y)_n=\Gamma(y+n)/\Gamma(y)$ increases with $y$),
\begin{equation}
\log T_n(\rho)\geq2\log\frac{\Gamma(\De+n)}{\Gamma(\De)}-\log\Gamma(n+1)-\log\frac{\Gamma(b\De+n)}{\Gamma(b\De)}+2n\log\rho.
\end{equation}
Stirling's formula gives $\log\Gamma(u\De+O(1))=u\De\log u+u\De\log\De-u\De+o(\De)$ for fixed $u>0$.
Using this we have
\begin{equation}
\frac1\De\log T_n(\rho)\geq B_b(t)+o(1),
\end{equation}
where
\begin{equation}
B_b(t)=2(1+t)\log(1+t)-t\log t-(b+t)\log(b+t)+b\log b+2t\log\rho.
\end{equation}
Letting $\De\to\infty$, then $\delta\to0$, and then taking the supremum over $t$,
\begin{equation}
\label{eq:Rnu}
\liminf_{\De\to\infty}\frac1\De\log\mathcal{F}_0(z)\geq R_\nu(\rho)\equiv\sup_{t>0}B_{1/(8\nu)}(t),
\end{equation}
Let $a=1/(8\nu)$. The supremum is positive for every $\rho\in(0,1)$, because $B_a(t)\to0$ as $t\to0$, while $B_{a}'(t)=\log\frac{\rho^2(1+t)^2}{t(a+t)}\to+\infty$. Inserting \eqref{eq:Rnu} into both terms of \eqref{eq:twobounds}, and using $\log\frac{z\bar z}{(1-z)(1-\bar z)}=x+\bar x$,
\begin{equation}
\liminf_{\De\to\infty}\frac1\De\log\mathcal{G}(z,\bar z)\geq\max\Big(R_\nu(\rho)+R_\nu(\bar\rho),\ x+\bar x+R_\nu\big(\rho(1-z)\big)+R_\nu\big(\rho(1-\bar z)\big)\Big).
\end{equation}
Since $R_\nu>0$, the first argument exceeds $0$ and the second exceeds $x+\bar x$, so the right-hand side exceeds $\max(0,x+\bar x)$ at every point of the causal diamond.

For $\nu=\infty$, we have $b=0$, and $B_0$ is concave, with its maximum where $\rho(1+t)=t$. This gives $t=\rho/(1-\rho)$ and $R_\infty(\rho)=-2\log(1-\rho)$. At the self-dual point, the two arguments coincide, and they equal $2R_\infty(\rho_*)=4\log\frac{1+\sqrt2}{2}$.

(ii) The first nontrivial term of the series is the exchange of a single stress tensor,
\begin{equation}
T_1(\rho)=\frac{8\rho^2\De^2}{c}\equiv m.
\end{equation}
Since $\De^{1+\gamma_c}/c\to k$ and $\gamma_c<1$, we have $m\to\infty$, with $m$ of order $\De^{1-\gamma_c}$. For $n\ll\De,c$, we have $(\De)_n\approx\De^n$ and $(\frac c8)_n\approx(\frac c8)^n$, so $T_n(\rho)\approx m^n/n!$. Thus the stress tensor exchanges exponentiate to $e^m$, and the largest term is at $n\approx m$.

To make this precise, keep $n=\lfloor m\rfloor$. Then $n/c\leq m/c=8\rho^2\De^2/c^2\to0$, since $\gamma_c>0$. Using $(\De)_n\geq\De^n$ and $(\frac c8)_n\leq(\frac c8+n)^n$,
\begin{equation}
T_n(\rho)\geq\frac{m^n}{n!}\Big(1+\frac{8n}{c}\Big)^{-n}.
\end{equation}
Since $m\geq n$ and $\log n!=n\log n-n+O(\log n)$, we have $\log\frac{m^n}{n!}\geq n-O(\log n)$. Moreover, $n\log(1+\frac{8n}{c})\leq\frac{8n^2}{c}=o(n)$. Therefore $\log T_n(\rho)\geq n(1+o(1))=m(1+o(1))$, that is,
\begin{equation}
\log\mathcal{F}_0(z)\geq8\rho^2\frac{\De^2}{c}\big(1+o(1)\big).
\end{equation}
At $z=\bar z=\frac12$, where $\rho=\bar\rho=\rho_*$, the first term of \eqref{eq:twobounds} gives $\log\mathcal{G}(\frac12,\frac12)\geq16\rho_*^2\frac{\De^2}{c}(1+o(1))$. Since $\De^{\gamma_c-1}\frac{\De^2}{c}=\De^{1+\gamma_c}/c\to k$, this proves (ii).
\end{proof}

\begingroup
\small
\let\originalthebibliography\thebibliography
\renewcommand{\thebibliography}[1]{%
  \originalthebibliography{#1}%
  \setlength{\itemsep}{0pt plus .1pt}%
  \setlength{\parskip}{0pt}}
\bibliographystyle{unsrturl}
\bibliography{liberation}
\endgroup
\end{document}